\documentclass[twocolumn]{aastex62}
\usepackage{graphicx}
\usepackage{gensymb}

\newcommand{\textsw}[1]{\textsc{#1}}
\submitjournal{ApJ}
\shorttitle{halo counts-in-cells}
\shortauthors{Wen et al.}

\begin{document}

\title{Halo Counts-in-cells for Cosmological Models with Different Dark Energy}

\correspondingauthor{Di Wen}
\email{diwen2@illinois.edu}
\author[0000-0003-2812-8607]{Di Wen}
\affil{Department of Astronomy,\\ University of Illinois at Urbana-Champaign,\\ 1002 W. Green Street, Urbana, IL 61801, USA}
\author[0000-0001-6233-8347]{Athol J. Kemball}
\affil{Department of Astronomy,\\ University of Illinois at Urbana-Champaign,\\ 1002 W. Green Street, Urbana, IL 61801, USA}
\author{William C. Saslaw}
\affil{Department of Astronomy,\\ University of Virginia,\\ 530 McCormick Road, Charlottesville, VA 22904, USA}

\begin{abstract}
We examine the counts-in-cells probability distribution functions that describe dark matter halos in the Dark Energy Universe Simulations (DEUS). We describe the measurements between redshifts $z=0$ to $z=4$ on both linear and non-linear scales. The best-fits of the gravitational quasi-equilibrium distribution (GQED), the negative binomial distribution (NBD), the Poisson-Lognormal distribution (PLN), and the Poisson-Lognormal distribution with a bias parameter (PLNB) are compared to simulations. The fits agree reasonably consistently over a range of redshifts and scales. To distinguish quintessence (RPCDM) and phantom ($w$CDM) dark energy from $\Lambda$ dark energy, we present a new method that compares the model parameters of the counts-in-cells probability distribution functions. We find that the mean and variance of the halo counts-in-cells on $2-25h^{-1}$Mpc scales between redshifts $0.65<z<4$ show significant percentage differences for different dark energy cosmologies. On $15-25h^{-1}$Mpc scales, the $g$ parameter in NBD, $\omega$ parameter in PLN, $b$ and $C_b$ parameters in PLNB show larger percentage differences for different dark energy cosmologies than on smaller scales. On $2-6h^{-1}$Mpc scales, kurtosis and the $b$ parameter in the GQED show larger percentage differences for different dark energy cosmologies than on larger scales. For cosmologies explored in the DEUS simulations, the percentage differences between these statistics for the RPCDM and $w$CDM dark energy cosmologies relative to $\Lambda$CDM generally increases with redshift from a few percent to significantly larger percentages at $z=4$. Applying our method to simulations and galaxy surveys can provide a useful way to distinguish among dark energy models and cosmologies in general.
\end{abstract}


\section{Introduction} \label{sec:intro}
Two decades of observational evidence strongly suggest that an unknown form of dark energy with negative pressure dominates our universe. Luminosity distance measurements of high-redshift Type Ia supernovae (SNe) indicate an accelerating expansion of the universe with a positive vacuum energy density (i.e. positive cosmological constant $\Lambda$) \citep{riess98, perlmutter99}. The observed cosmic microwave background (CMB) power spectrum shows very good agreement with spatially-flat cold dark matter models and a cosmological constant \citep[$\Lambda$CDM;][]{planckxiii16,planckvi18}. Combined analyses of the baryon acoustic oscillation (BAO) signal in galaxy surveys and the CMB acoustic scale have found consistency with the results from CMB alone \citep{cole05,eisenstein05,alam17}. The standard flat $\Lambda$CDM model is consistent with constraints from the combined datasets of SNe, CMB and BAO \citep{giannantonio08, kowalski08,abbott19a}. The recent Planck CMB power spectra combined with lensing reconstruction and BAO give a best-fit $\Lambda$CDM model with $\Omega_{\Lambda} = 0.6889 \pm 0.0056$ and a dark energy equation of state $w = -1.03 \pm 0.03$ \citep{planckvi18}. Despite little physical understanding, it is widely accepted that the energy density in the universe today is mostly dark energy. Observational probes of cosmic acceleration and the dark energy equation of state have known shortcomings \citep[see][for reviews]{albrecht06, frieman08, weinberg13, huterer18}. The broadest set of probes for dark energy are necessary to further constrain the dark energy equation of state and to distinguish dark energy models. In this paper we develop a technique using the counts-in-cells (CiC) distribution of dark matter halos to distinguish among dark energy models. Here we apply this technique to N-body cosmological simulations (to be described in Section \ref{sec:deus}) in the first instance to measure the differences for different dark energy models and also as a precursor to applying the method to observed galaxy survey samples with greater measurement uncertainties. In addition, as future simulations become available, this approach can be extended to test their consistency with future observational surveys.

\subsection{Dark Energy}
The nature of dark energy may not be as simple as a vacuum energy represented by constant $\Lambda$ and $w$. Scalar-field models introduce quintessence \citep[see][for reviews]{wetterich88, ratra88, frieman95, copeland06} and allow a time-varying equation of state parametrized by a Taylor expansion $w(a) = w_{0} + (1-a)w_{a}$, where $a$ is the scale factor \citep{planckvi18}. A cosmological constant $\Lambda$ corresponds to $w_{0}=-1$ and $w_{a}=0$. A constant $\Lambda$ and $w=-1$ is consistent with CMB+BAO+SNe+$H_0$ measurements within approximately  10\% uncertainty with 95\% confidence \citep{planckvi18}. However, $w_0$ and $w_a$ for a dynamical dark energy equation of state are less constrained by combinations of different probes \citep[Fig. 4 in][]{planckxiv16}. The case when $w<-1$ is "phantom" dark energy, in which the sum of the pressure and energy density is negative, causing a "big rip" universe \citep{caldwell03}. For a weakly coupled scalar field, $w$ can be parametrised by the slope parameter, $\epsilon_{s}$, whose negative values correspond to phantom models. The value of $\epsilon_{s}$ is not well constrained even combining multiple probes.

\subsection{Clustering Statistics}
Statistics of large scale structure in the CMB, in galaxy surveys and large cosmological simulations can test theoretical models of cosmic evolution. The description of the CMB largely relies on the measurement of angular power spectra \citep[e.g.][]{planckvi18}. Power spectra are also widely used to validate results of dark-matter-only cosmological N-body simulations \citep[e.g.][]{springel05, alimi10, klypin11, skillman14}. To compare simulations with galaxy surveys, halo finders \citep{davis85, behroozi13, knebe13} generate halo catalogs and the halos are then populated with galaxies using theoretical models, such as halo abundance matching \citep{kravtsov04, tasitsiomi04, vale04, conroy06}. Another method to quantify large scale structure is the two-point correlation function; this is the Fourier transform of the power spectrum \citep{peebles80, saslaw00}. The power spectrum is often more convenient for continuous matter density, and the two-point correlation functions for discrete distributions of objects. Higher order spectra or correlation functions are usually more difficult to estimate accurately and are computationally challenging \citep{smith11, munshi11}.

Nonlinear clustering produces higher order correlations \citep[e.g.][]{peebles80, bernardeau02} for galaxies, halos and the underlying density field. The hierarchy of correlation functions, in principle, provides complete information about the statistical distribution of structures, but it is impractical to measure accurate correlation functions of high orders. Moreover, two-point correlation functions and power spectra also do not contain any phase information. Two spatial density distributions can have the same power spectrum, but have very different clustering morphology caused by phase differences \citep{chiang01}. Primordial non-Gaussianity in the initial density field from inflation can leave signatures of spatial phases. Nonlinear evolution can produce non-random phases \citep{coles03}. Useful statistical measures of the spatial phase distribution are yet to be discovered.

\subsection{Counts-in-Cells Distribution}
The number counts of discrete objects in an ensemble of cells of a fixed size and shape gives a CiC probability distribution function (PDF) \citep{saslaw00}. We compare the CiC PDFs in N-body dark matter cosmological simulations for different dark energy models. This method is an extension of galaxy cluster abundance, which can be considered the first moment, i.e. mean, of the CiC distribution with a high mass cut at $\sim 10^{14} M_{\odot}$. If halos greater than galaxy masses are included, the sample is much larger and nonlinear clustering is also probed. CiC distributions contain much more information from higher moments and higher order correlations \citep{saslaw00}. Galaxy-mass halo distributions can be related to the galaxy distributions in galaxy surveys through a mass-luminosity relation. Halos and galaxies are both biased tracers of the underlying dark matter density, but simulations have not found any qualitative difference between the mean bias relations for galaxies and halos \citep{uhlemann18}. Measuring the count-in-cells distributions of halos is a crucial first step to measure the differences of different dark energy models, before similar measurements are conducted with observed galaxy survey samples with greater uncertainties. 

It is possible that the halo and galaxy distributions are very similar if each halo is usually occupied by one galaxy. Even so, tests over a wide range of scales and redshifts are necessary to confirm the validity of the analytical PDF for structure formation and clustering. If these distribution functions are roughly similar at low redshift, then high redshift samples could provide tighter constraints on their functional form. The effect of dark energy on the distribution functions of halo and galaxy clustering has yet to be explored. The analytical models of CiC statistics considered in this paper (Section \ref{models}) have not explicitly included evolving dark energy. To distinguish the difference between galaxy and halo distribution, the low mass halos should be included, although they may not host any galaxies. However, this may be difficult because the resolution of simulations across a wide range of halo masses is limited. The scale-dependence of halo bias \citep{dalal08} and galaxy bias \citep{weinberg04} indicate that the distribution functions of halos and galaxies due to clustering are scale-dependent. While the scale-dependence of the moments of the CiC distribution has been measured in surveys and simulations, the analytical form of the CiC PDF as a function of scale and redshift has not been extensively studied. The halo number density and halo CiC PDF can be influenced differently by different dark energy models, but difficult to measure due to limited simulation volume \citep[e.g.][]{shi16}. The CiC distributions for different equations of states with dark energy might show measurable differences for halos at certain mass ranges, length scales and redshifts.

The main goals of this paper are: 

a) to demonstrate the utility of CiC for distinguishing different dark energy models;

b) to find the best-fit analytical model of the halo CiC PDF; and,

c) to make predictions of the scales and redshifts of CiC statistics that show the largest deviations from $\Lambda$CDM.

In Section \ref{sec:method}, we describe our CiC algorithm and the analytical models used for model fitting. In Section \ref{sec:results}, we give results for the residuals of CiC PDFs, for the moments of the CiC PDFs, and for the best-fit analytical models as functions of cell size and redshift for various dark energy models. In Section \ref{sec:discussion}, we discuss how our method describes the differences in dark energy models and the sources of uncertainties. In Section \ref{sec:conclusions}, we summarize our findings.\\

\section{Method} \label{sec:method}
Our framework for distinguishing cosmological models with CiC statistics is as follows: i) select dark matter halos by mass in the simulation halo catalogs; ii) count the number of halos in the enumerated cell sizes and calculate the associated CiC PDF for each cell size; iii) fit analytical models to the measured CiC PDFs; and, iv) compare the differences in CiC moments and fitted model parameters over the different cosmologies. Before we discuss our method in detail, we first provide some background on the application of the CiC method in surveys and simulations and the analytical models proposed for the CiC PDF.

\subsection{Background}
CiC statistics contain more information than the two-point correlation function. The CiC PDF shows the ensemble average of clustering properties and combines information from a range of length scales up to the cell size \citep{wall12}. The PDF moments can be readily computed (in both two and three dimensions), and be related to the correlation functions \citep{peebles80}. It has been shown that the CiC statistics do not depend on cell shapes strongly \citep{saslaw91, szapudi96, szapudi98}.

The mean of the CiC at a given redshift (e.g. galaxy cluster abundance) constrains models of structure formation and clustering \citep{mana13}. It depends on the evolution of the halo mass function \citep{tinker08}. For a Gaussian-distribution density field or Poisson-distributed halos or galaxies, i.e. without clustering, the mean contains complete information about the statistical properties \citep{wall12}. A clustered distribution has an increased variance and a non-zero skewness compared to a Poisson distribution. These indicate nonlinear gravitational clustering \citep{wall12}. The second order moment $\sigma_8$ is the matter density fluctuation at a scale of 8 Mpc/h. It has been readily measured for the CMB \citep{planckvi18} and used as a cosmological parameter in simulations \citep[e.g.][]{klypin11}. To constrain cosmological models, higher order moments of the CiC as a function of redshift and scale are necessary.

The moments of the galaxy CiC PDF have been used to quantify galaxy clustering in the CFHTLS-Wide survey \citep{wolk13} and the spectroscopic Sloan Digital Sky Survey (SDSS) \citep{bel14}. CiC has also been used to measure galaxy bias for submillimeter galaxies observed with the Atacama Large Millimeter Array (ALMA) \citep{ono14}, galaxy samples from the ALHAMBRA survey \citep{lopez15}, the VIPERS survey \citep{diporto16} and the Dark Energy Survey (DES) \citep{salvador19}. The third and fourth order moments of CiC have been measured for simulated galaxies in smoothed particle hydrodynamic simulations \citep{weinberg04} and dark matter halos in different simulated cosmologies as a function of scale and redshift \citep{casas03, angulo08}. All these CiC measurements only assumed the $\Lambda$CDM model, except for \citet{bel14} who also included a varied $w$.

There are advantages in fitting an analytic form to the measured discrete CiC PDF rather than computing the first few moments numerically. An analytical form inherently contains moments of all orders and sheds light on the stochastic processes underlying the clustering statistics and it provides hints about the physical processes behind structure formation and clustering. The CiC PDF from galaxy surveys and N-body simulations have been compared with theoretical models of structure formation and clustering, but there is no clear consensus on the detailed functional form of the CiC PDF of galaxies, halos or dark matter densities, other than clear evidence for non-Poisson distributions. The analytical forms of CiC PDF models are described in Section \ref{models}. Some examples in the literature of fitting an analytic CiC PDF to survey data or simulations include studies using the 2dF Galaxy Redshift Survey (2dFGRS) and dark matter in the Hubble Volume simulation \citep{croton04}, the DEEP2 Galaxy Redshift Survey and mock galaxies from simulations \citep{conroy05}, SDSS galaxies \citep{yang11,hurtado-gil17}, VIPERS galaxies \citep{bel16}, DES galaxies and MICE mock galaxies \citep{clerkin17} and dark matter halos in other simulations \citep{neyrinck14,ahn15}.

\subsection{DEUS Halo Catalog}\label{sec:deus}
In this study, we measure the CiC statistics of simulated dark matter halos in three cosmologies and find the best-fit parameters of model probability distribution functions in a range of length scales and redshifts. We make use of the publicly available Dark Energy Universe Simulations halo position catalogs \citep[DEUS;][]{alimi12}. A set of three simulations in comoving space with box-lengths of 648 $h^{-1}$Mpc and $2048^{3}$ dark matter particles \citep{alimi12} are selected for comparison between three dark energy models, namely, the cosmological constant ($\Lambda$CDM), Ratra-Peebles potential for quintessence scalar field (RPCDM) and phantom dark energy model ($w$CDM) with constant equation of state. Table \ref{table:DEUSSALL1} lists the cosmological parameters used for these simulations \citep[see][for details on the choice of cosmologies in DEUS]{bouillot15}. In summary, the cosmological parameters for RPCDM and $w$CDM were chosen by DEUS to be statistically consistent with the best-fit flat $\Lambda$CDM model to the WMAP-7yr data while being indistinguishable from $\Lambda$CDM at the significance level of the WMAP-7yr data. In our work we use the simulations as provided by DEUS. It is worth noting that the values of $\Omega_m$, $w_0$ and $w_1$ chosen at the time of DEUS are not fully consistent with recent constraints combining multiple observational probes \citep{abbott19a, abbott19b}. New simulations adopting more tightly constrained cosmologies would produce smaller differences in CiC between different dark energy models, but should still be distinguishable with a subset of similar CiC measurements. We plan to use a wider range of simulations that systematically vary the cosmological parameters in future work to test our proposed framework for distinguishing dark energy models. Here we use the halo catalogs from DEUS with smaller box-lengths than the DEUS-FUR, but with the same cosmological parameters. The smaller box-length and fewer number of particles allow a higher mass resolution so that galaxy-sized halos are included in the halo catalogs. Each halo contains at least 100 particles, found using the Friends-of-Friends algorithm \citep{roy14}, which gives a mass cut at $M_{halo}>2.4\times10^{11} M_{\odot}$ for all three cosmologies. The halos are from 9 snapshots at redshifts spaced between $0\leqslant z\leqslant4$.

\begin{table}
\begin{center}
\begin{tabular}{cccc}
\hline
Parameters & $\Lambda$CDM & RPCDM & $w$CDM \\
\hline
\hline
$\Omega_m$  & 0.2573 & 0.23 & 0.275 \\
$\Omega_b h^2$  & 0.02258 & 0.02273 & 0.02258 \\
$\sigma_8$& 0.801 & 0.66  & 0.852 \\
$w_0$& -1 & -0.87  & -1.2 \\
$w_1$ & 0 & 0.08 & 0 \\
$m_p ($h$^{-1}$ M$_\odot$)  & $2.26 \times 10^{9}$ & $2.02 \times 10^{9}$  & $2.42 \times 10^{9}$ \\
\hline
\end{tabular}
\caption{Cosmological parameter values of the DEUS simulated cosmologies. We use a set of smaller simulations with a smaller box-length and higher mass resolution compared to the DEUS-FUR, but with the same cosmological parameter values \citep{bouillot15}. For all models the scalar spectral index is set to $n_s=0.963$ and the Hubble parameter $h=0.72$. We list the values of a linear equation of state parameterization $w(a) = w_0 + w_1(1- a)$ for the different models and the particle mass $m_p$. For all three simulations the box-length is $L_{box}=648h^{-1}$Mpc and the number of dark matter particles is $2048^3$.}
\label{table:DEUSSALL1}
\end{center}
\end{table}

\subsection{Counts-in-cells algorithm}
The CiC distribution sampled here is the PDF of a discrete random variable, $N$, the number count of dark matter halos with $M_{halo}>2.4\times10^{11} M_{\odot}$ inside spherical cells of a given radius. We make 216 CiC measurements using the 3 halo catalogs for 8 different cell sizes of radii $R=$ 2, 4, 6, 8, 10, 15, 20, 25 $h^{-1}$Mpc, at 9 different redshifts $z$ = 0, 0.1, 0.25, 0.4, 0.65, 1, 1.5, 2.3 and 4. Each side of the cubic simulation box is divided into 512 sections and the centers of the $512^3$ cubes are the centers of the $512^3$ spherical cells used for the CiC measurements (see Appendix \ref{appendix}). Since the cells can be overlapping, our CiC is oversampling the simulation box. The oversampling method is inspired by the infinite sampling CiC method \citep{szapudi98}, which transforms counting discrete objects in sampling cells into calculating the ratios of overlapping area/volume for cells around discrete objects. The infinite sampling CiC method is argued to be free of measurement errors. The densely populated cell centers in our method can be viewed as sampling the overlapping regions in the infinite sampling CiC method. When the number of cells tends to infinity, the ratios of overlapping volumes are then exactly recovered. Our choice of the cell number is based on a resolution study (see Appendix \ref{appendix}). The number of halos in each cell is found by comparing the cell radius and the distance between the cell center and halo center. The occurrence of each count $N$ is then counted for cells that are completely within the simulation box, excluding incomplete cells on the edges of the simulation box. Lastly, the histogram is normalized to produce a PDF, called the CiC distribution. The CiC algorithm is parallelized with \textit{Message Passing Interface} \citep[MPI;][]{gropp14} in C and is run on the Blue Waters supercomputer at the National Center for Supercomputing Applications (NCSA).

\subsection{Model Fitting} \label{models}
Four models are fitted to the resulting CiC distributions. The first is the gravitational quasi-equilibrium distribution (GQED) given by
\begin{equation}\label{GQED}
f_{GQED}(N) = \frac{\overline{N}(1-b)}{N!}
\left(\overline{N}(1-b)+Nb\right)^{N-1} e^{-\overline{N} (1-b)-Nb}
\label{eq:gqed}
\end{equation}where $\overline{N}$ is the average number of halos in a given cell volume and $b$ is the ensemble average ratio of the gravitational correlation energy to twice the kinetic energy of peculiar velocities \citep{saslaw84}.

The second is the negative binomial distribution (NBD) derived by \citet{elizalde92}. We use a formulation with the Gamma function \citep{yang11}, 
\begin{equation}
f_{NBD}(N) =
\frac{\Gamma\left(N+\frac{1}{g}\right)}{\Gamma\left(\frac{1}{g}\right)N!} \frac{n^N \left(\frac{1}{g}\right)^\frac{1}{g}} {\left(n+\frac{1}{g}\right)^{N+\frac{1}{g}}}
\label{eq:nbd}
\end{equation}
where $n$ is the average number of halos in a given cell volume and $g$ is a clustering parameter.
The third model is a log-normal distribution, commonly used to describe the matter density distribution. The Poisson sampled log-normal distribution includes the shot noise as halos are discretely sampled from a log-normally distributed continuous matter density field \citep{clerkin17},
\begin{equation}
	f_{PLN}(N) = \int^\infty_{-1} \frac{\bar{N}^N (1+\delta_g)^N}{N!} e^ {-\bar{N}(1+\delta_g)} f(\delta_g) d\delta_g 
	\label{eq:pln}
\end{equation}where
\begin{equation} 
	f(\delta_g) \mathrm{d}\delta_g = \frac{1}{w \sqrt{2 \pi}}exp\left( \frac{-x^2}{2w^2}\right) \mathrm{d}x
\end{equation} $x = \mathrm{ln} (1+\delta_g) + w^2/2$ and $w^2$ is the variance of the corresponding normal distribution $f[\mathrm{ln}(1+\delta_g)]$.
The fourth model is the log-normal distribution plus a halo bias parameter $b$ and a matter density variance parameter $C_b$ \citep{hurtado-gil17},
\begin{equation}\label{lnbpdf}
f_{PLNB}(\Delta) = \frac{1}{\sqrt{2\pi H_0}}\frac{\exp{(-\frac{1}{2}\frac{y^2}{H_0})}}{\Delta+b-1}
\end{equation}
where 
\begin{equation}
\Delta = N/\bar{N}
\end{equation}
\begin{equation}
H_0 = \log{(1+C_b)}
\end{equation}
\begin{equation}
y = \log{\Big((\Delta+b-1)\frac{\sqrt{1+C_b}}{b}\Big)}
\end{equation} 

Appendix \ref{appendix} shows two examples of the best-fit models for different cell radii and cosmologies. Of these four models, the GQED model has a physical basis in the thermodynamics \citep[and earlier references]{saslaw00} and statistical mechanics \citep{saslaw10} of cosmological gravitational many-body systems. The physics behind the other three models seems more obscure. All four models are Lagrangian distributions. The NBD is essentially a Poisson sampled gamma distribution and the GQED is a Poisson sampled truncated Borel distribution (also called a convolved or compound distribution in some literature) \citep{saslaw00}. One interpretation for the Poisson component of these distribution functions is that the formation of galaxies or dark matter halos around galaxies out of the underlying density field is a Poisson process. The question then is what statistical distribution the underlying dark matter density field follows. Another interpretation is a halo model, where the distribution of dark matter halos that host galaxy clusters follows a Poisson distribution, and the distribution of galaxies in galaxy clusters can be modeled by a halo occupation distribution, essentially governed by the halo mass and mass profile as well as biasing effects \citep{sheth94, fry11}. With suitable values of the bias factors, this full halo model shows good agreement with the first few moments of the dark matter halos and subhalos CiC from an adaptive mesh refinement cosmological simulation of dark matter particles \citep{fry11}. However, the physics behind this phenomenological halo model is not clear. The assumption that the validity of the emergence of cluster-sized halos is a Poisson process requires verification.

We find the best-fit model parameter with the least squares weighted uniformly for all counts $N$
\begin{equation} \label{eres-lsqdist}
x = \sum_{N=0}^{N_{max}} \left(\frac{f_{model}(N) - f_{CiC}(N)}{E}\right)^2
\end{equation} where $E=0.0005$ and $f_{CiC}(N)$ is the measured CiC PDF. The uniform weighting factor $E$ allows the cells with underdensity and overdensity to have the same weighting as the peak of the CiC distribution function. The least-squares is scaled larger with a small $E$ for easier comparison and smaller round-off errors during fitting. We use \textsw{MPFIT\footnote{https://pages.physics.wisc.edu/{$\sim$}craigm/idl/cmpfit.html} \citep{markwardt09}}, a least squares fitting library in C using the Levenberg-Marquardt technique \citep{press02}, to iteratively search for the best-fit parameters and the least squares for all four models. Both PLN and PLNB are integral functions. The CQUAD doubly-adaptive integration in the GNU Scientific Library\footnote{https://www.gnu.org/software/gsl/doc/html/integration.html \citep{galassi09}} is used to evaluate the integral functions. Due to the presence of factorials of large integers as well as very small or large exponential functions in all four models, double precision floating-point numbers in C/C++ programming language are insufficient for evaluating our model over the desired range of parameters. We make use of the GNU \textsw{MPFR} library\footnote{https://www.mpfr.org/} \citep{fousse07}, a C library for multiple-precision floating-point computations with correct rounding, for function evaluations. At least 80-bits significand must be kept to ensure the smoothness of model functions, the success of integration and the convergence of the least squares fitting.

\section{Results} \label{sec:results}
\subsection{Jackknife Error of CiC\label{jackknife}}
The jackknife method \citep{shao95} is used to estimate the uncertainties of the CiC distributions for two test cases, cell radii $R=2h^{-1}$ Mpc and $R=10h^{-1}$ Mpc at $z=0$ in $\Lambda$CDM. The sample of $512^3$ cells are divided into $n$ equal volume subsamples by coordinates, then one subsample is deleted at a time to obtain a jackknife CiC subsample. For a given $n$, there are therefore $n$ different jackknife CiC subsamples, and every cell is deleted once. The jackknife error is given by
\begin{equation}
    \sigma_{JK} = \sqrt{\frac{n-1}{n}\sum_{i=1}^{n}\left(\delta_{i} - \delta_{JK}\right)^2}
\end{equation}
where
\begin{equation}
    \delta_{JK} = \frac{1}{n}\delta_{i}
\end{equation}
and $\delta_{i}$ denotes the rms error when all the cells except the $i$-th are used. 
The sum of the squared jackknife errors for a CiC distribution may vary with delete fraction. For the two test cases, we choose $n$ between 2 and 2048.

\begin{figure}
\plotone{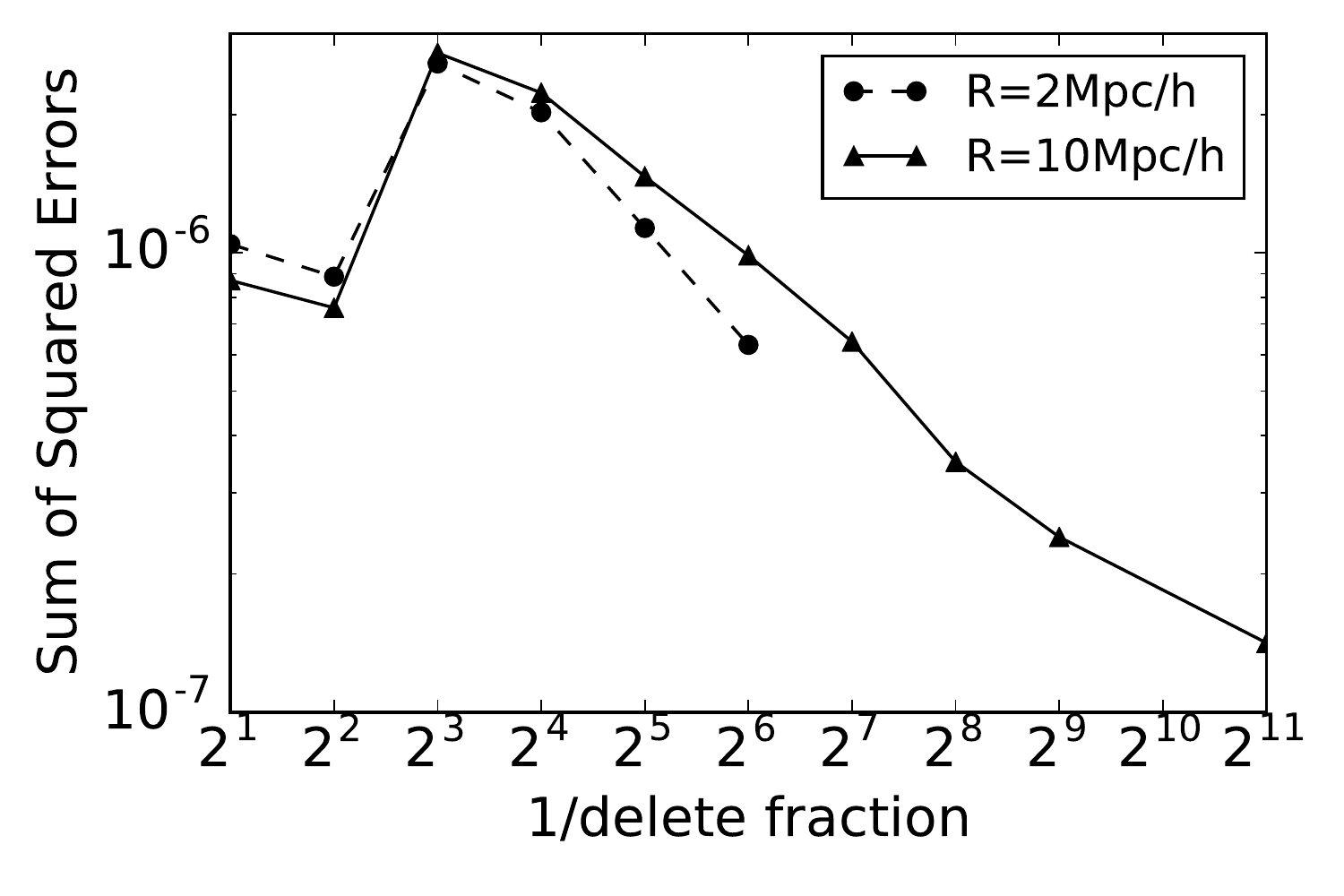}
\caption{Sum of jackknife errors for cell radii $R=2h^{-1}$ Mpc and $R=10h^{-1}$ Mpc at $z=0$ in $\Lambda$CDM as a function of the delete fraction.}
\label{fig:jackknife}
\end{figure}

For two test cases, cells of radii $R=2h^{-1}$ Mpc and $R=10 h^{-1}$ Mpc at $z=0$ in $\Lambda$CDM, the sums of jackknife errors for a given delete fraction are similar (Figure \ref{fig:jackknife}). Except when half or 1/4 of the cells are deleted, the sum of jackknife errors decreases as the delete fraction decreases. The jackknife errors are slightly larger than the uncertainties due to uncertain counts near the boundaries of cells in the resolution study, but are much smaller than the least squares of best-fit analytical models. The jackknife errors for the CiC distributions of other cell sizes, redshifts and cosmologies are expected to be similar and well below the deviations between the analytical models and CiC distributions. The origin of the uncertainties in our CiC measurement is discussed further in Section \ref{sec:discussion}.

\subsection{CiC Distribution}
\begin{figure*}
\plottwo{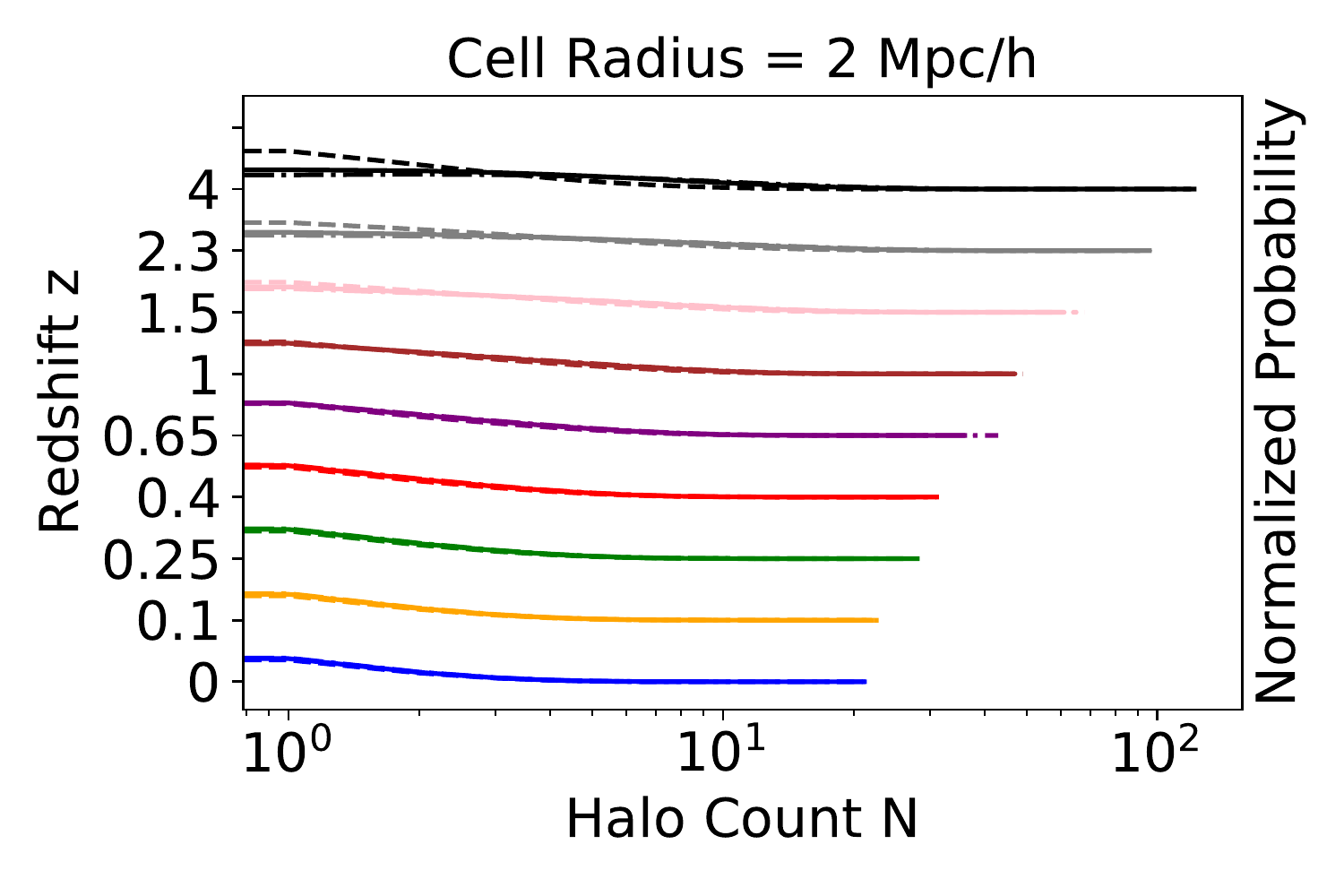}{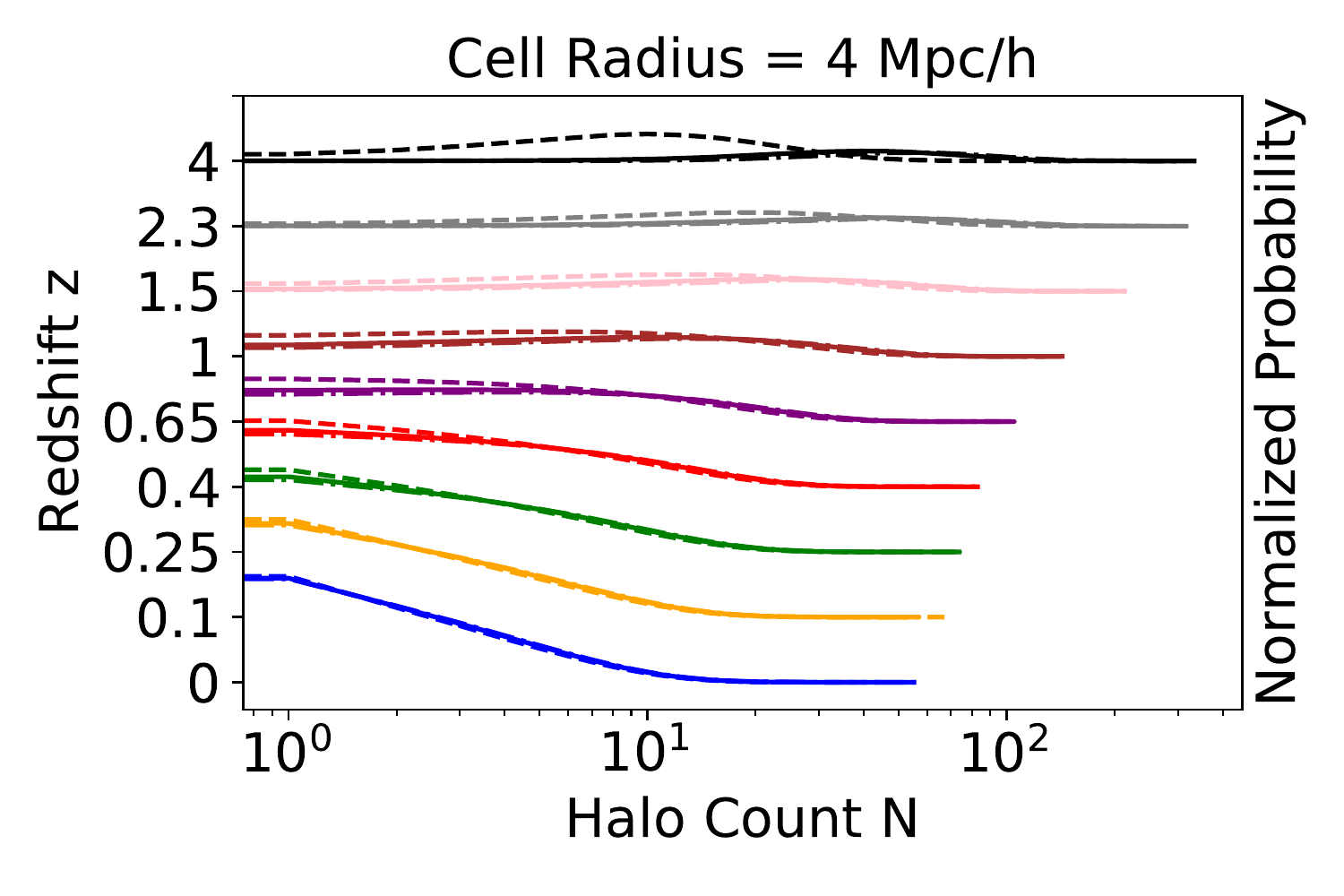}
\plottwo{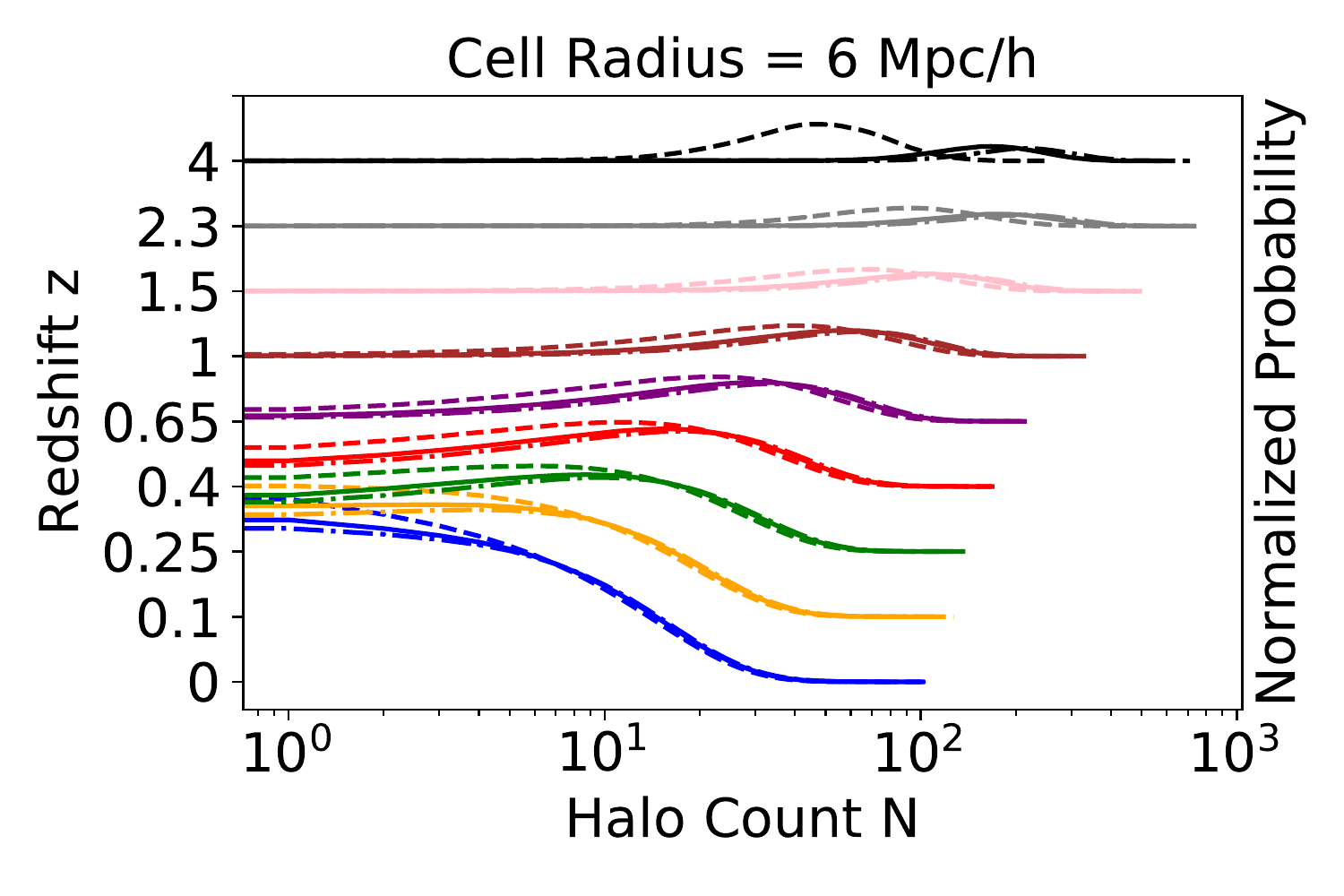}{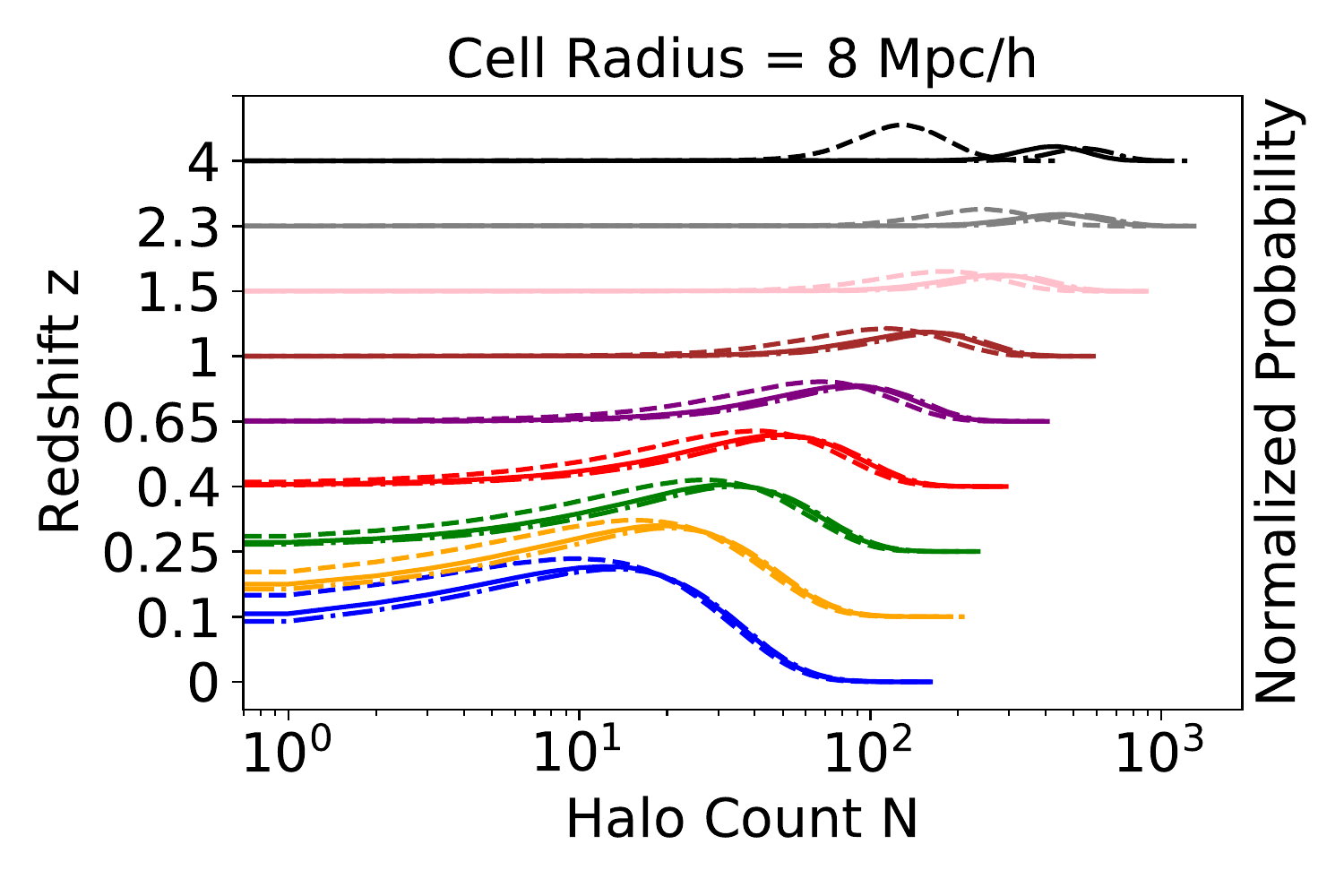}
\plottwo{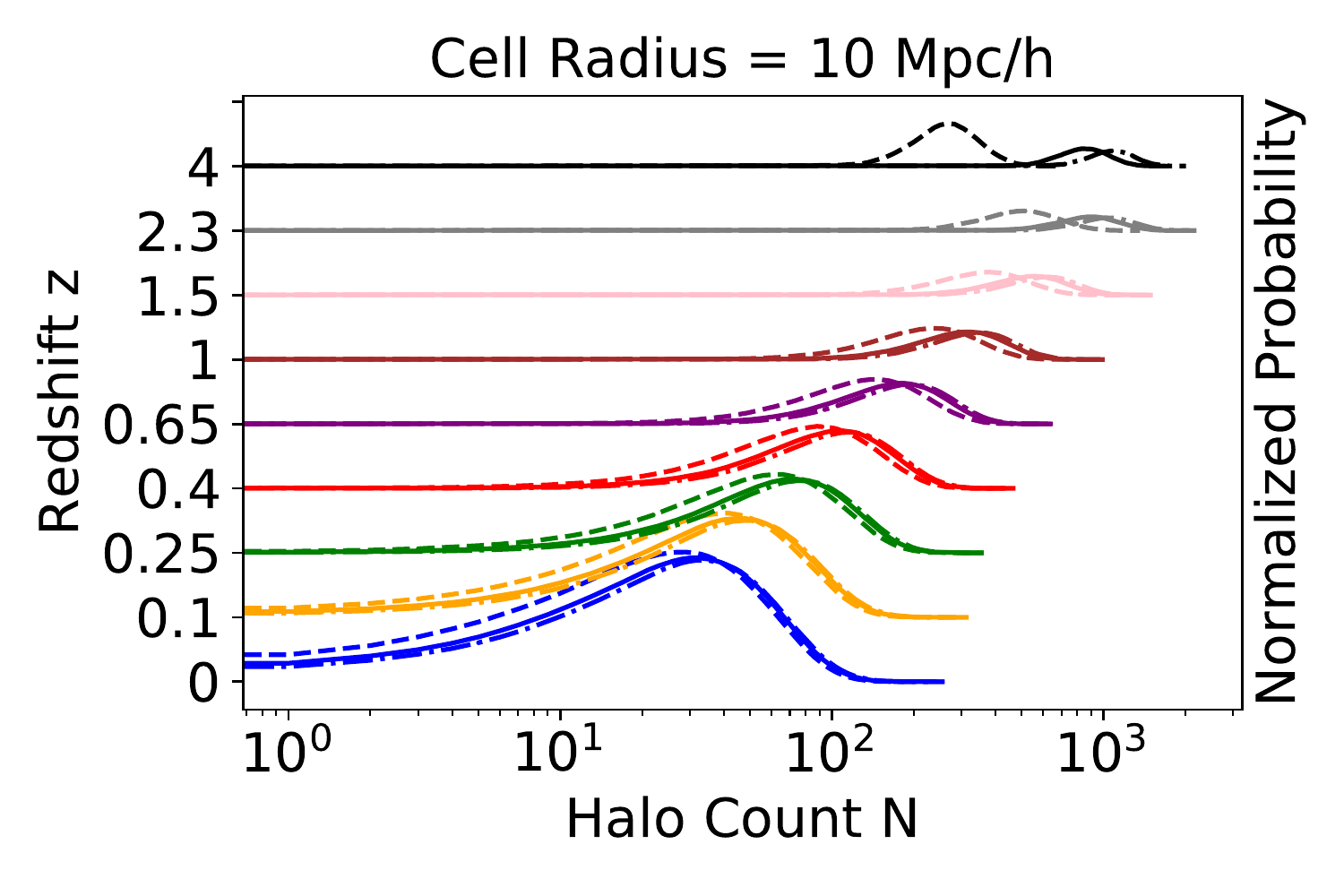}{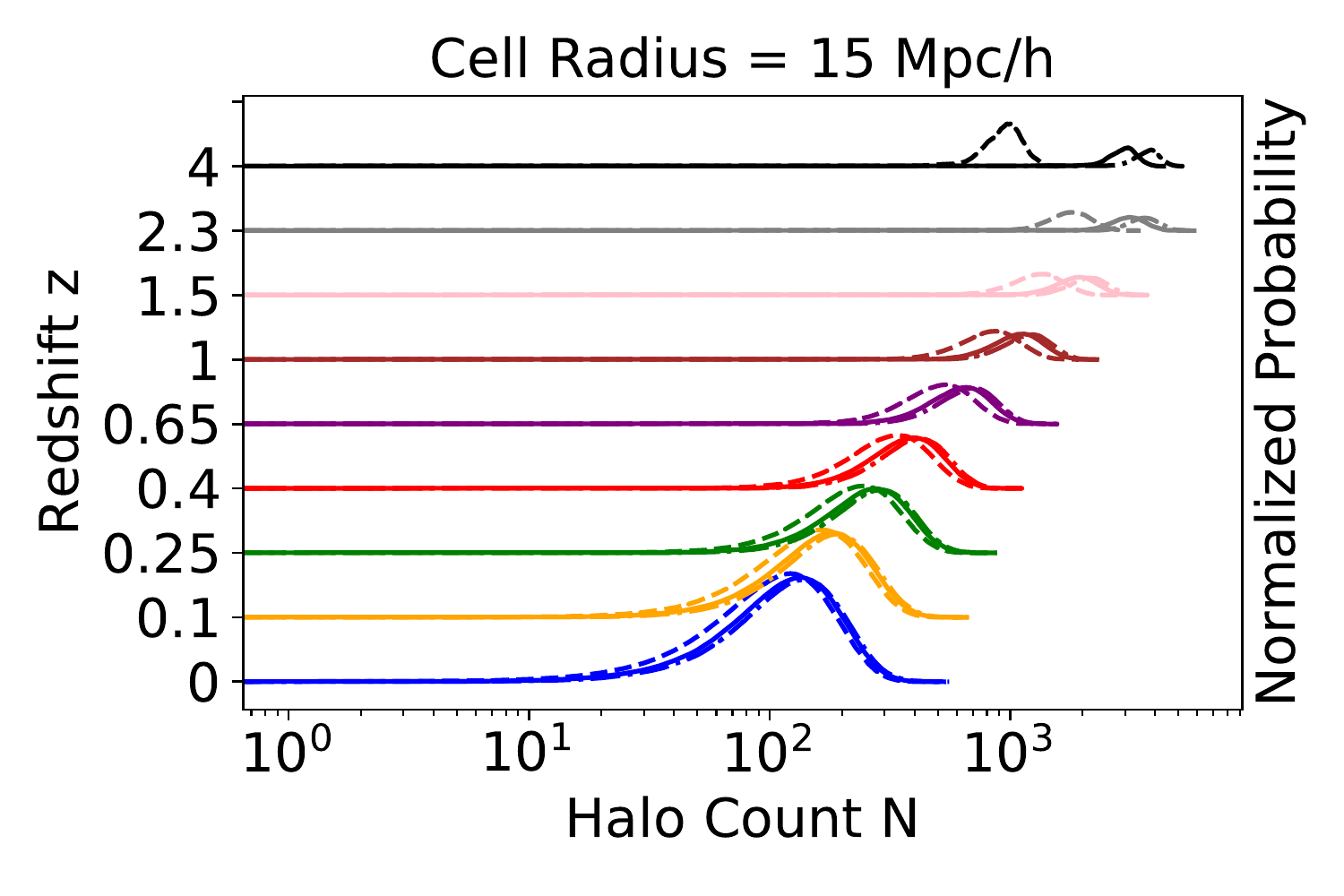}
\plottwo{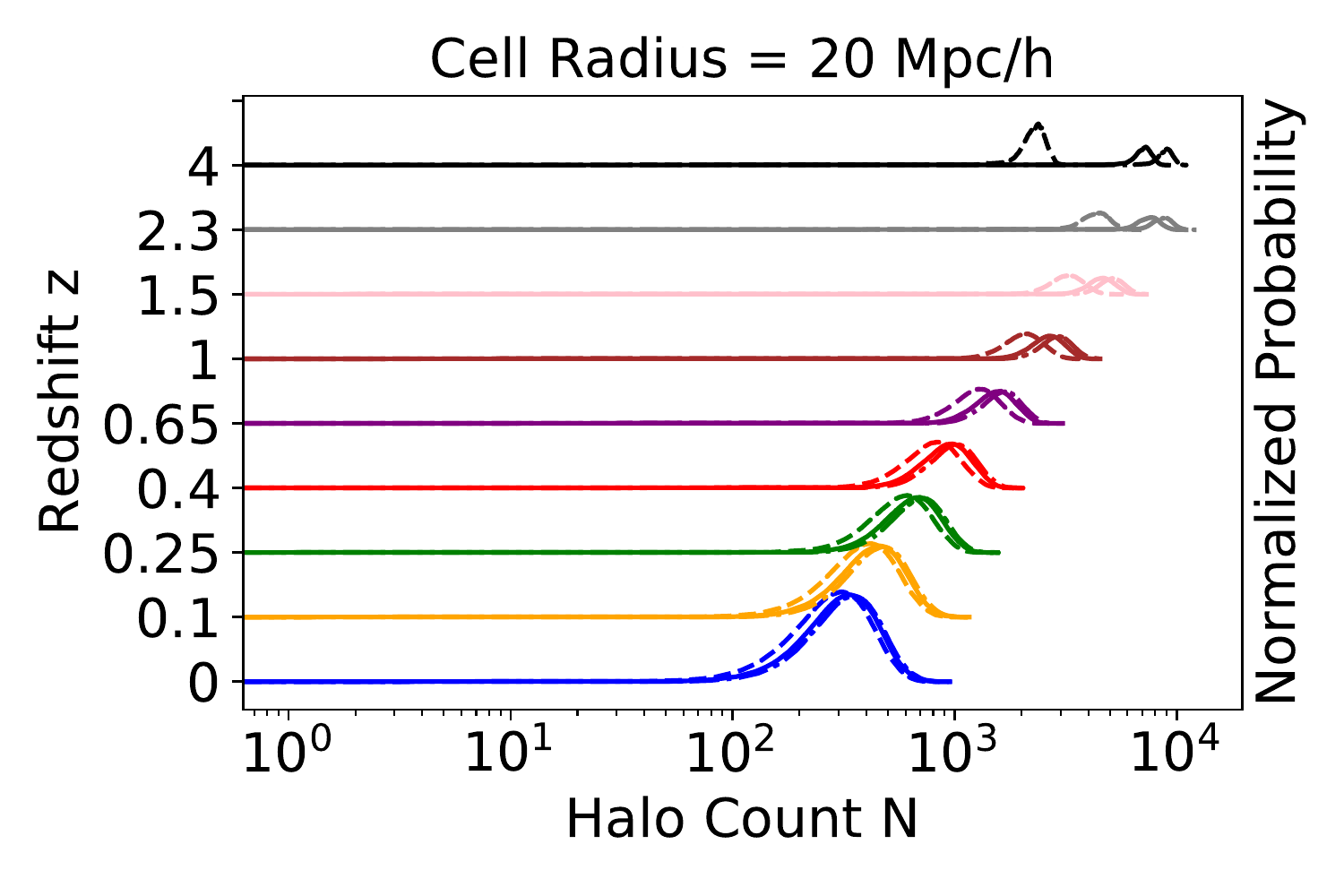}{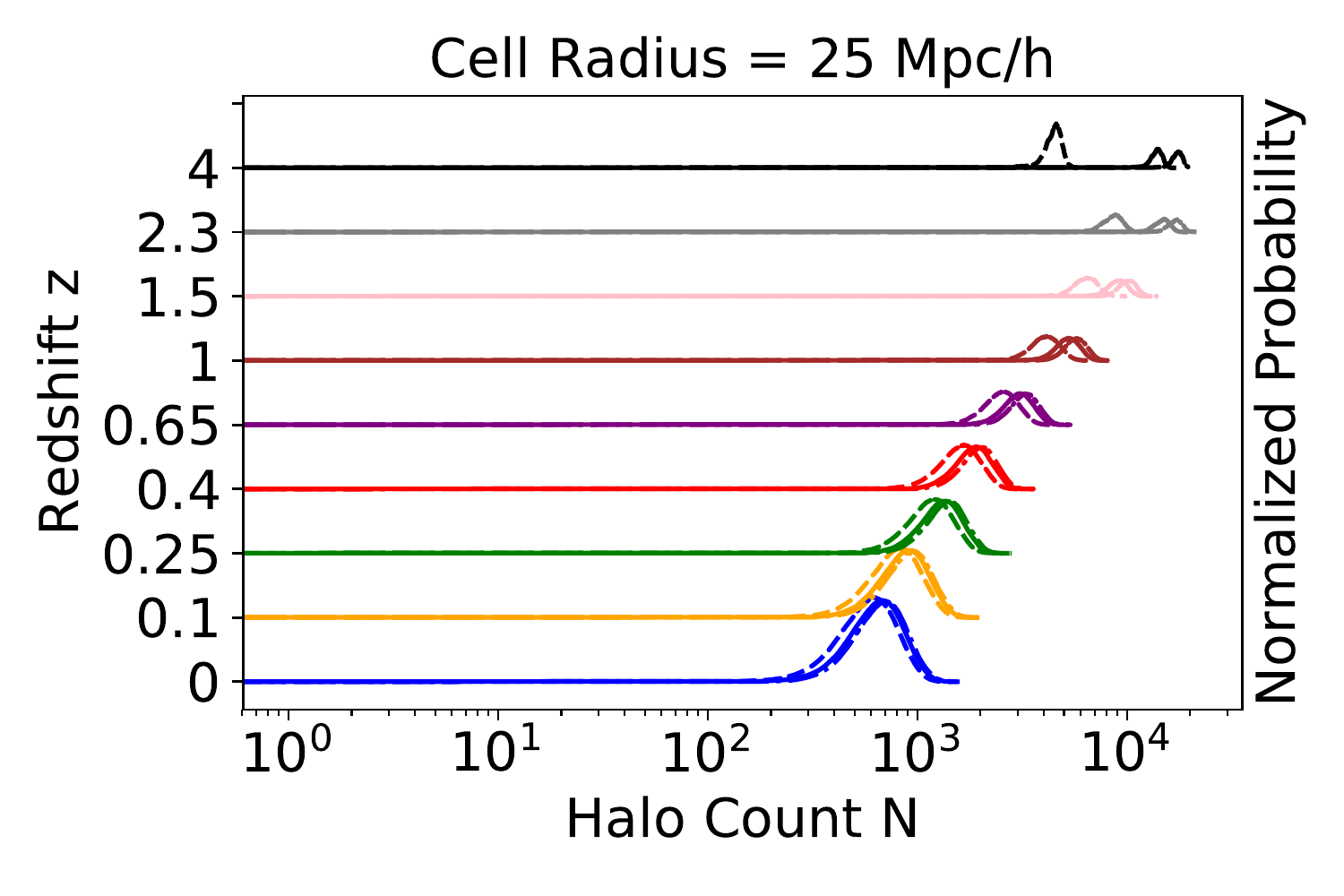}
    \caption{Counts-in-cells distributions $f(N)$ for the $\Lambda$CDM (solid line), RPCDM (dash line) and $w$CDM (dash dot line) at various spherical cell radii. In every panel, the colors of the line series indicate redshift $z=0$ (blue), $z=0.1$ (orange), $z=0.25$ (green), $z=0.4$ (red), $z=0.65$ (purple), $z=1$ (brown), $z=1.5$ (pink), $z=2.3$ (gray) and $z=4$ (black).}
\label{fig:cic_R}
\end{figure*}

\begin{figure*}
\epsscale{0.93}
\plottwo{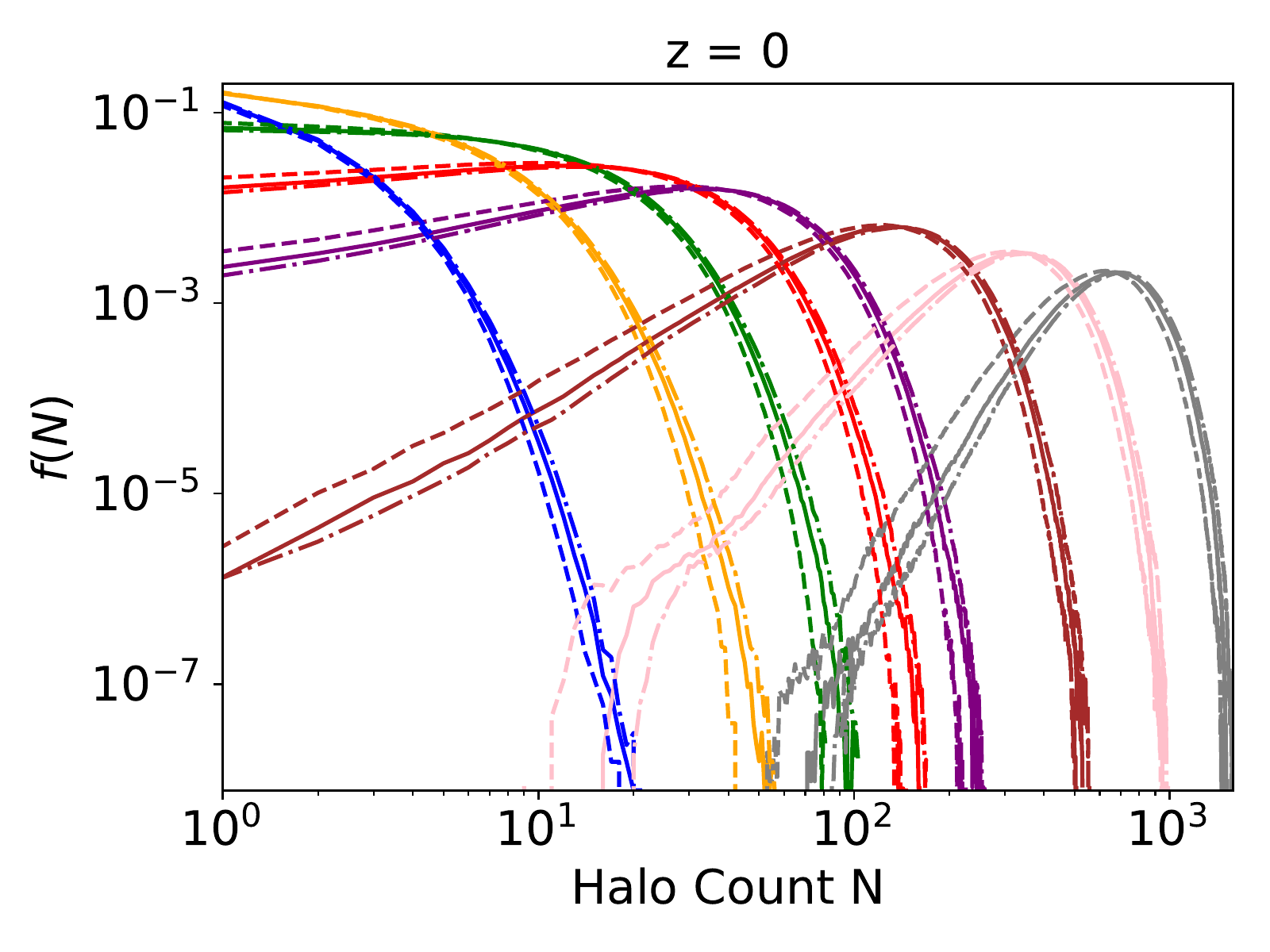}{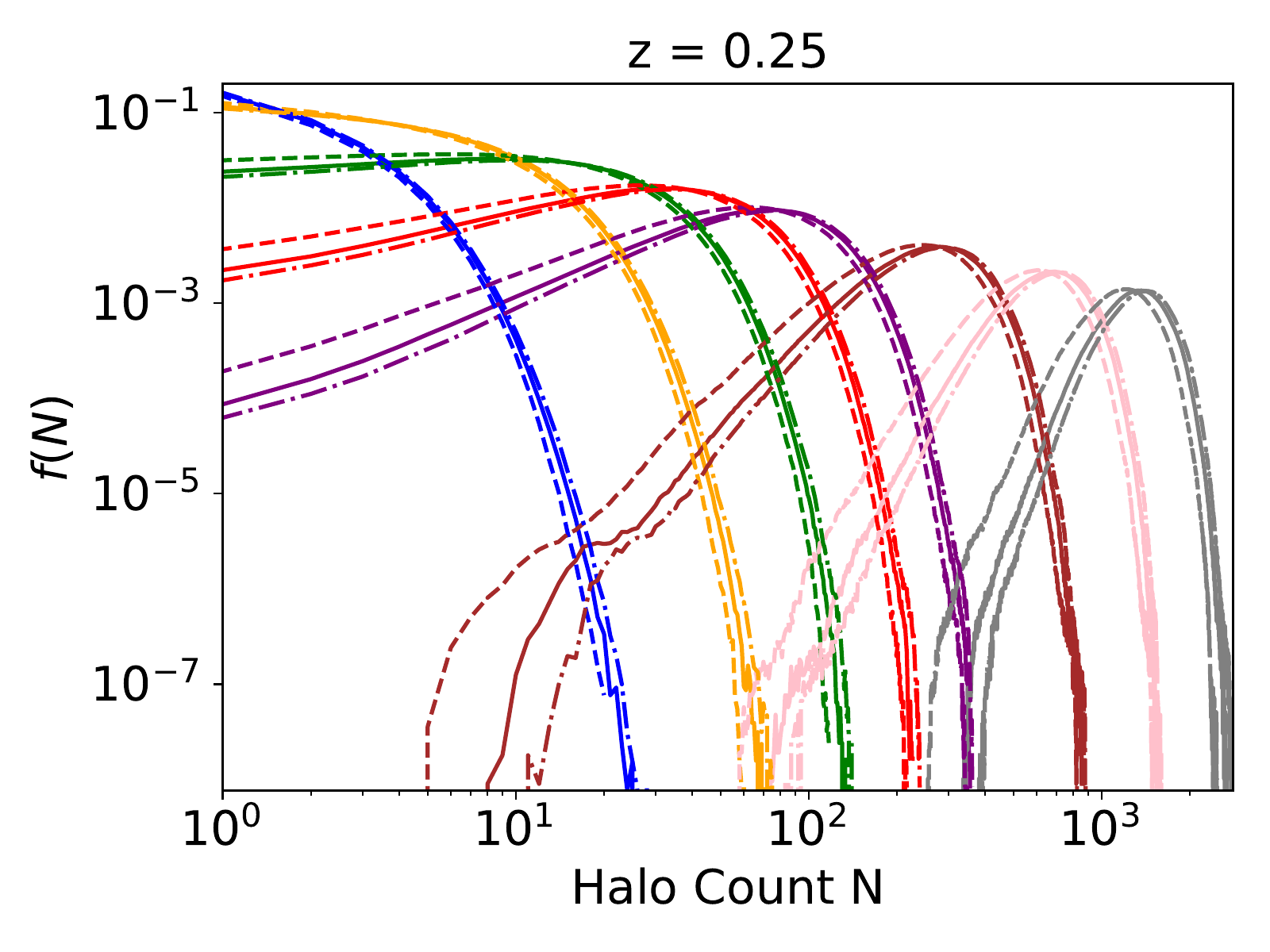}
\plottwo{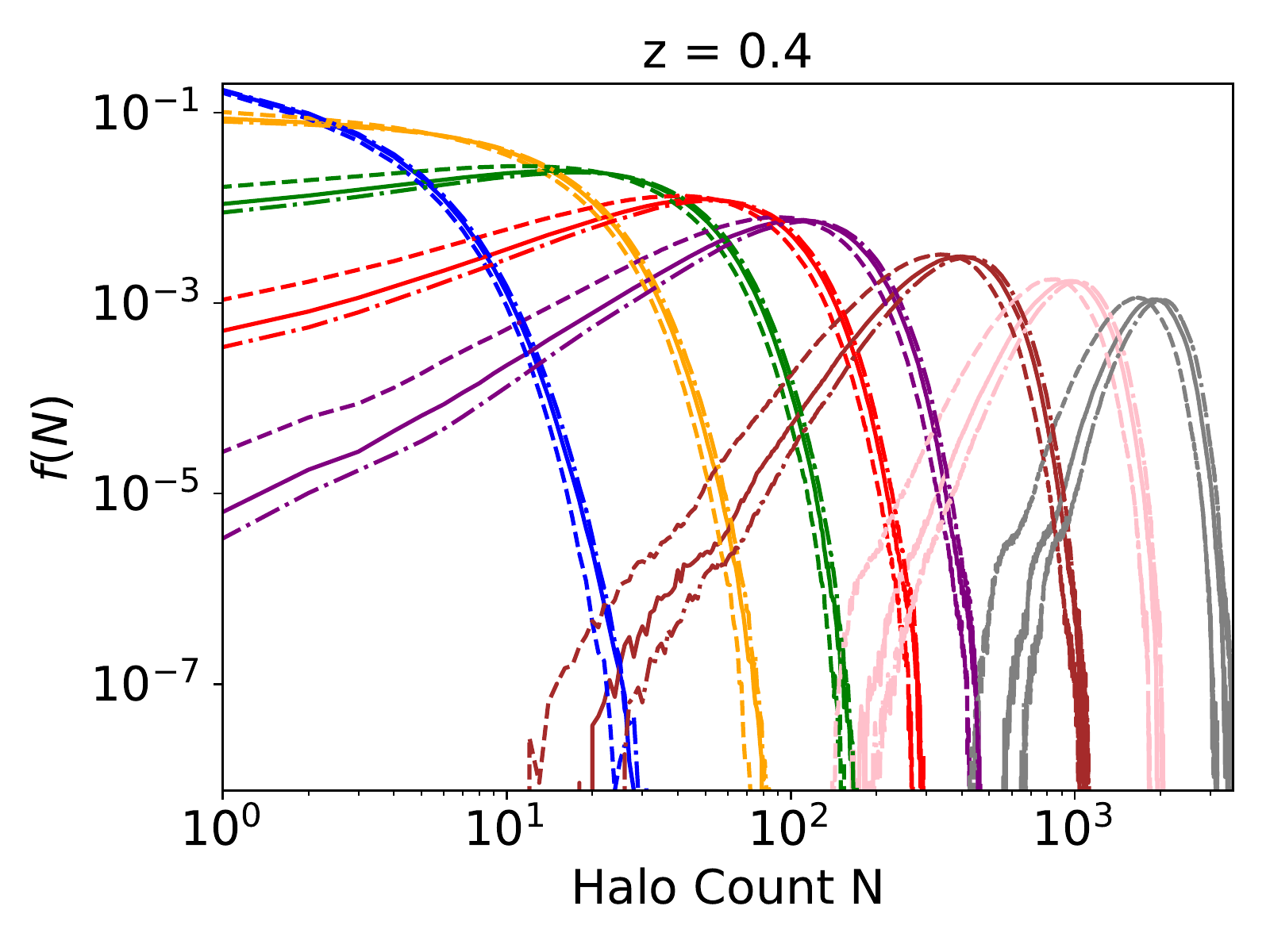}{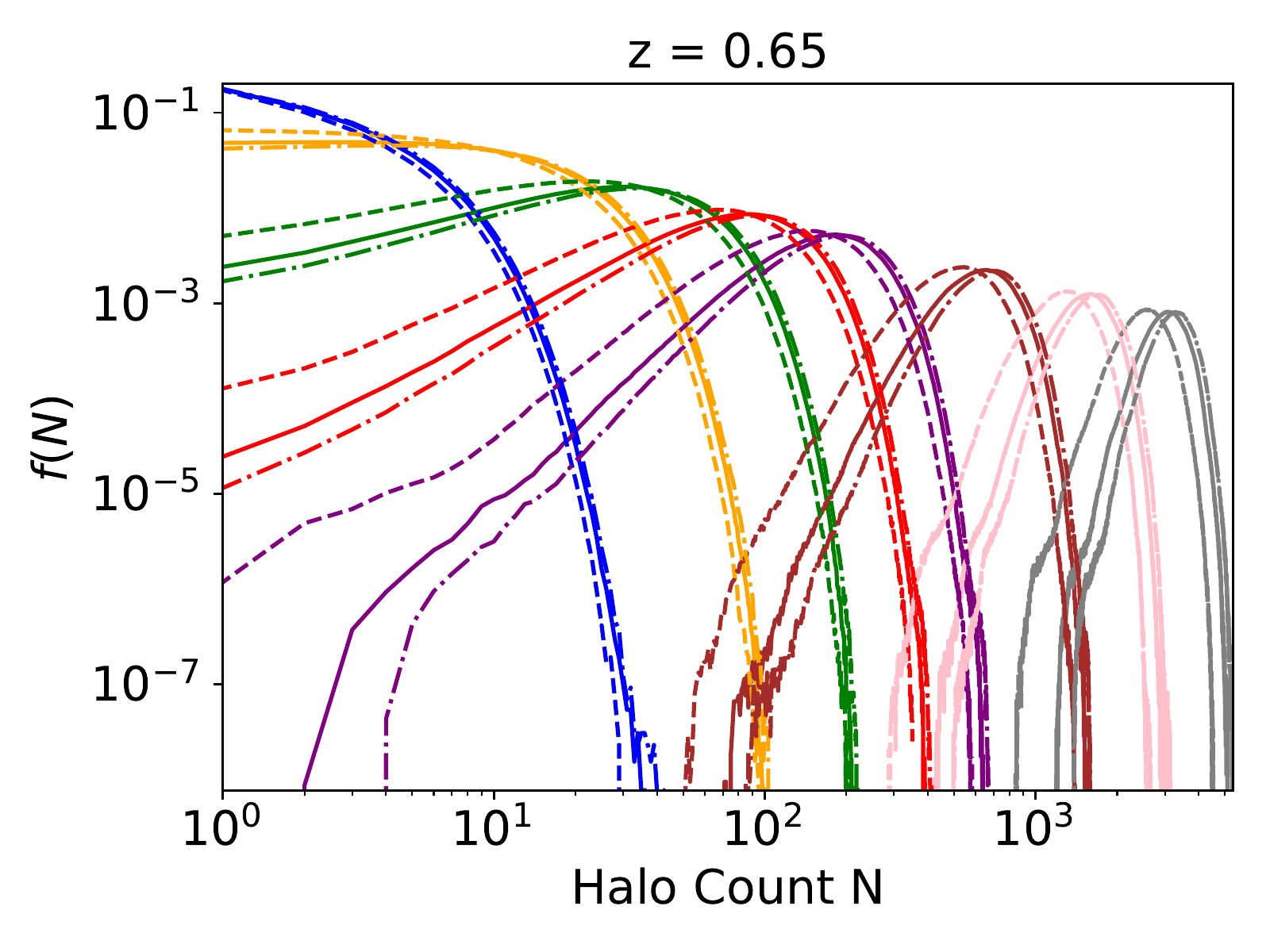}
\plottwo{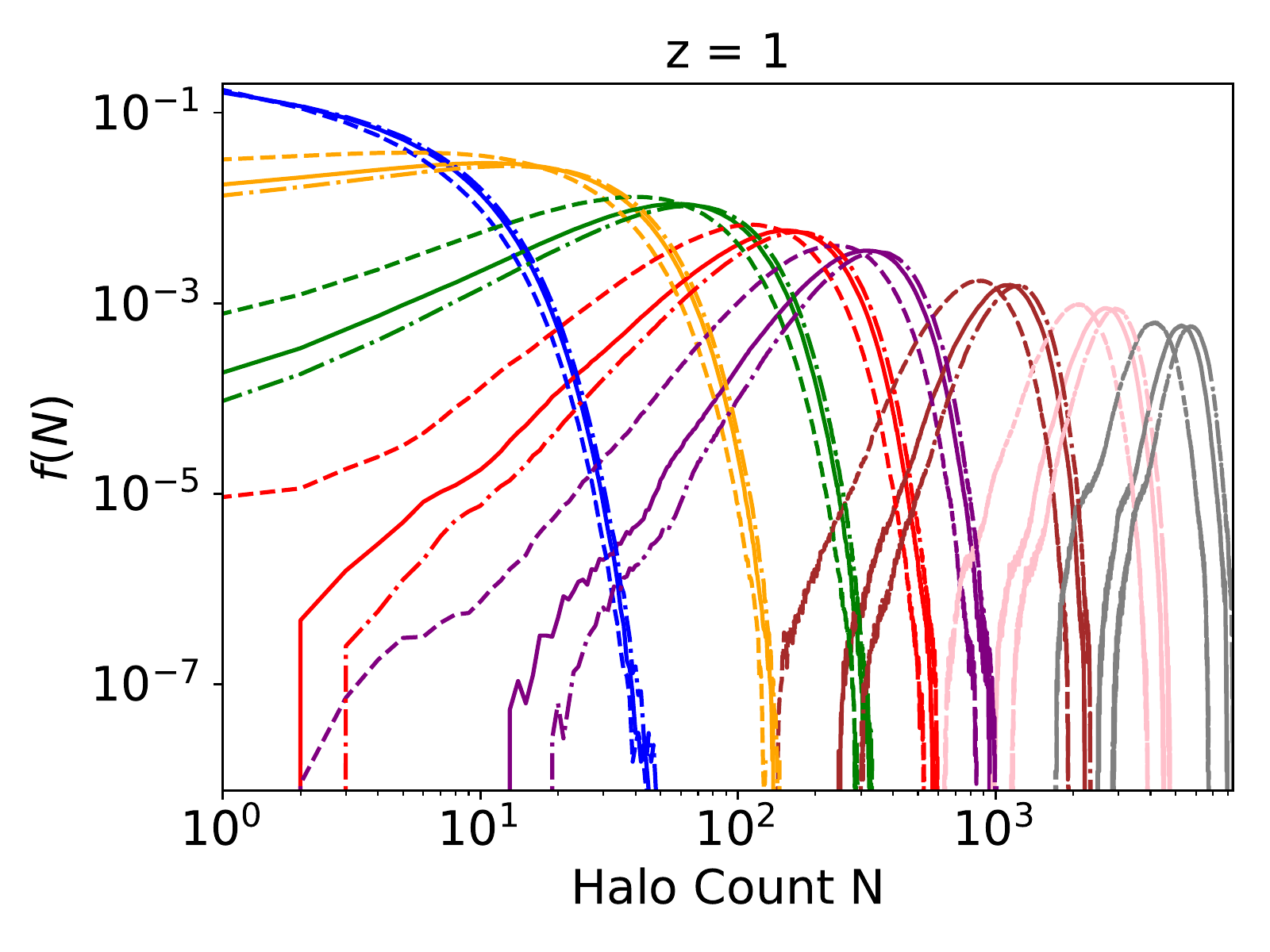}{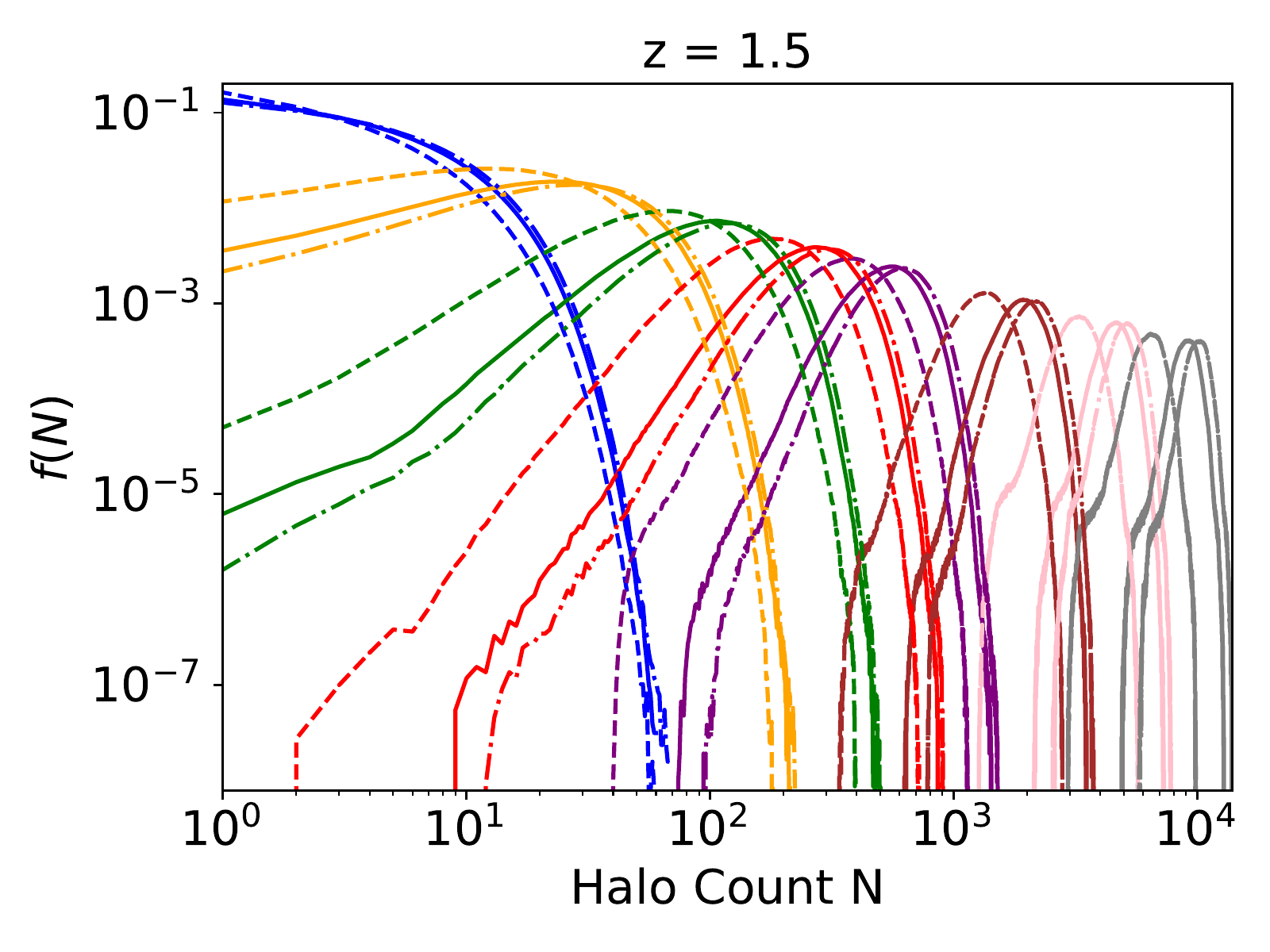}
\plottwo{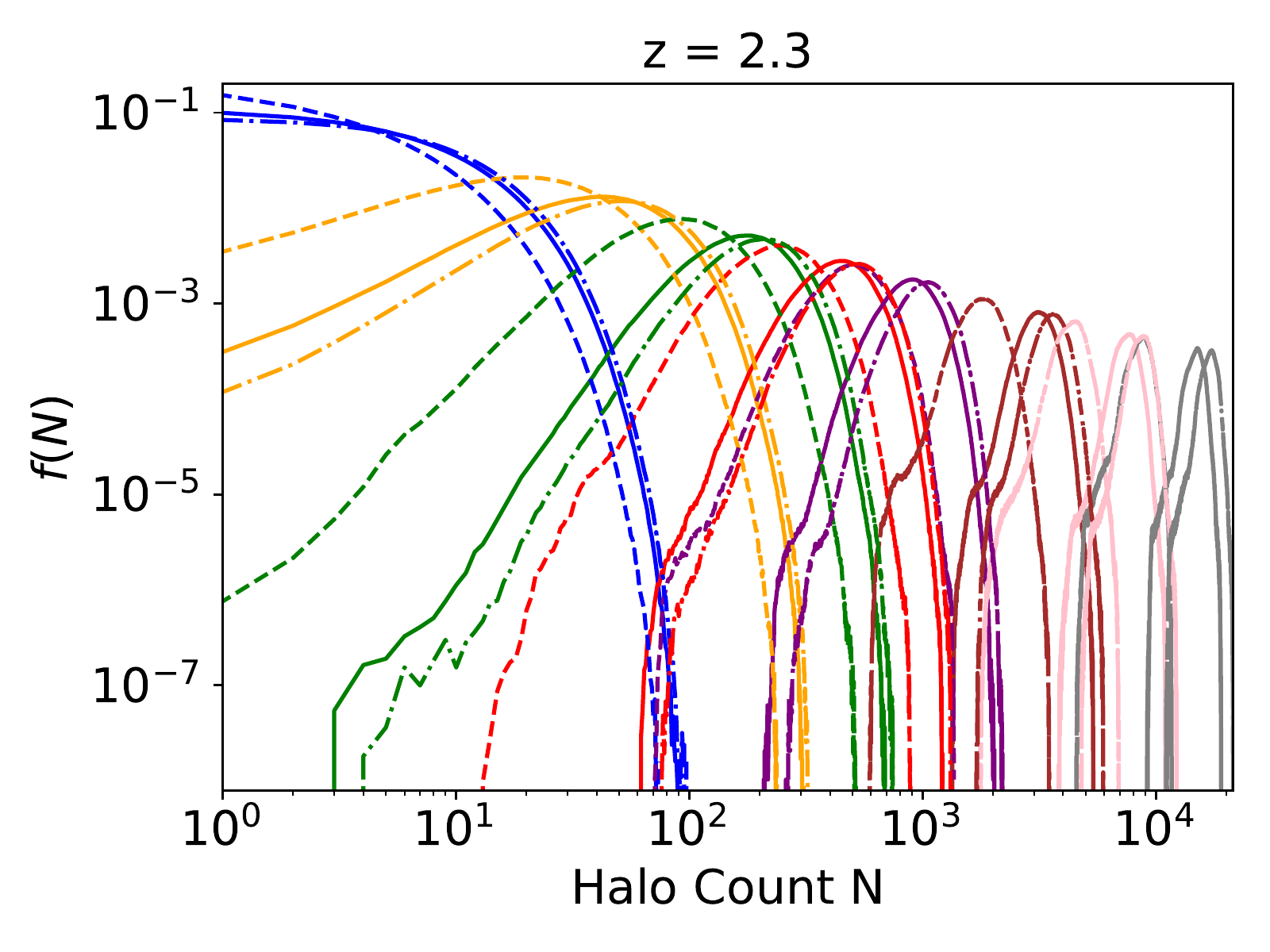}{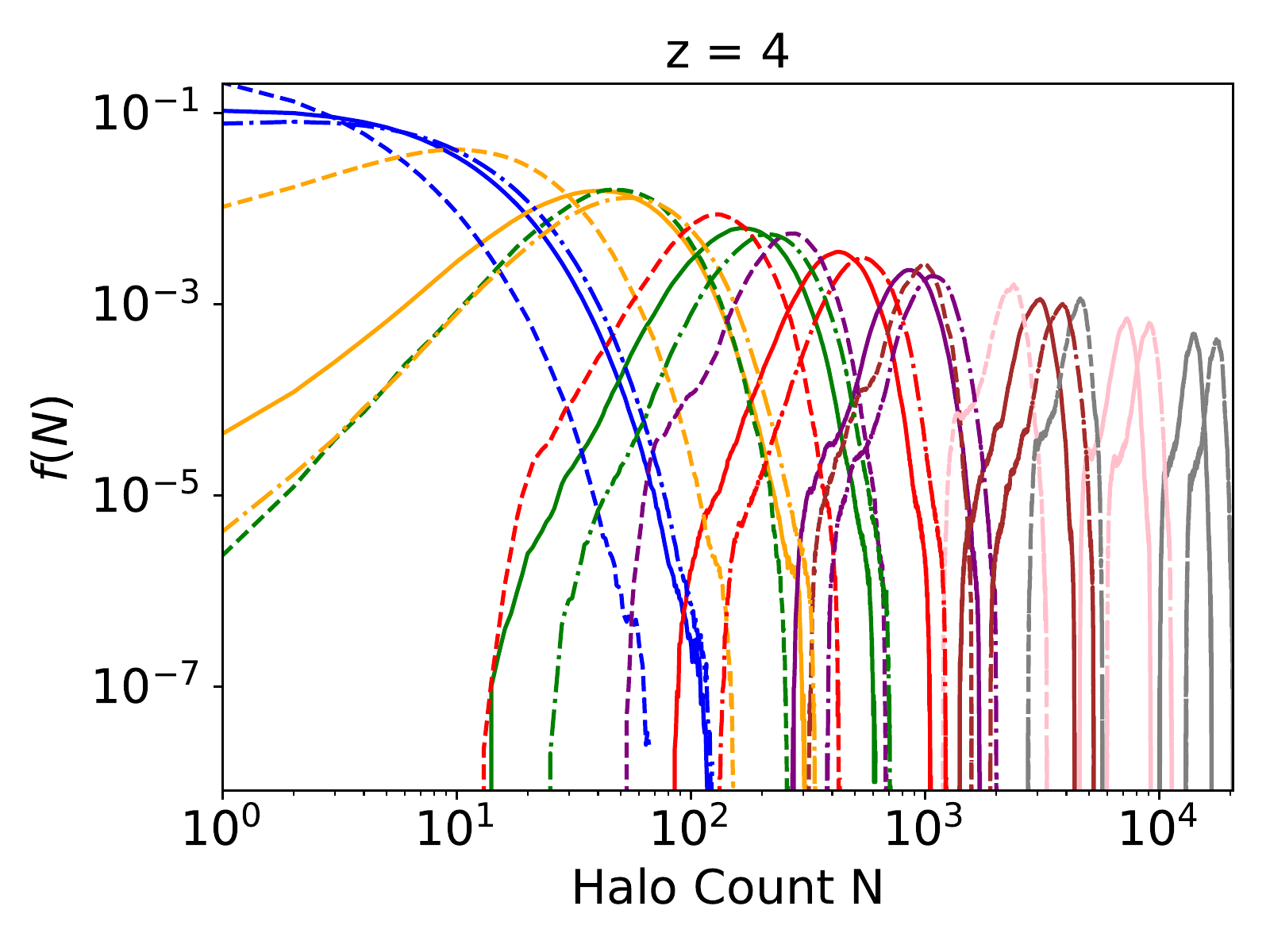}
\caption{Counts-in-cells distributions $f(N)$ for the $\Lambda$CDM (solid line), RPCDM (dash line) and $w$CDM (dash dot line) at various redshifts. In every panel, the colors of the line series indicate spherical cell radii $R=2h^{-1}$Mpc (blue), $4h^{-1}$Mpc (orange), $6h^{-1}$Mpc (green), $8h^{-1}$Mpc (red), $10h^{-1}$Mpc (purple), $15h^{-1}$Mpc (brown), $20h^{-1}$Mpc (pink) and $25h^{-1}$Mpc (gray).}
\label{fig:cic_z}
\end{figure*}

\begin{figure*}
\epsscale{0.85}
\plottwo{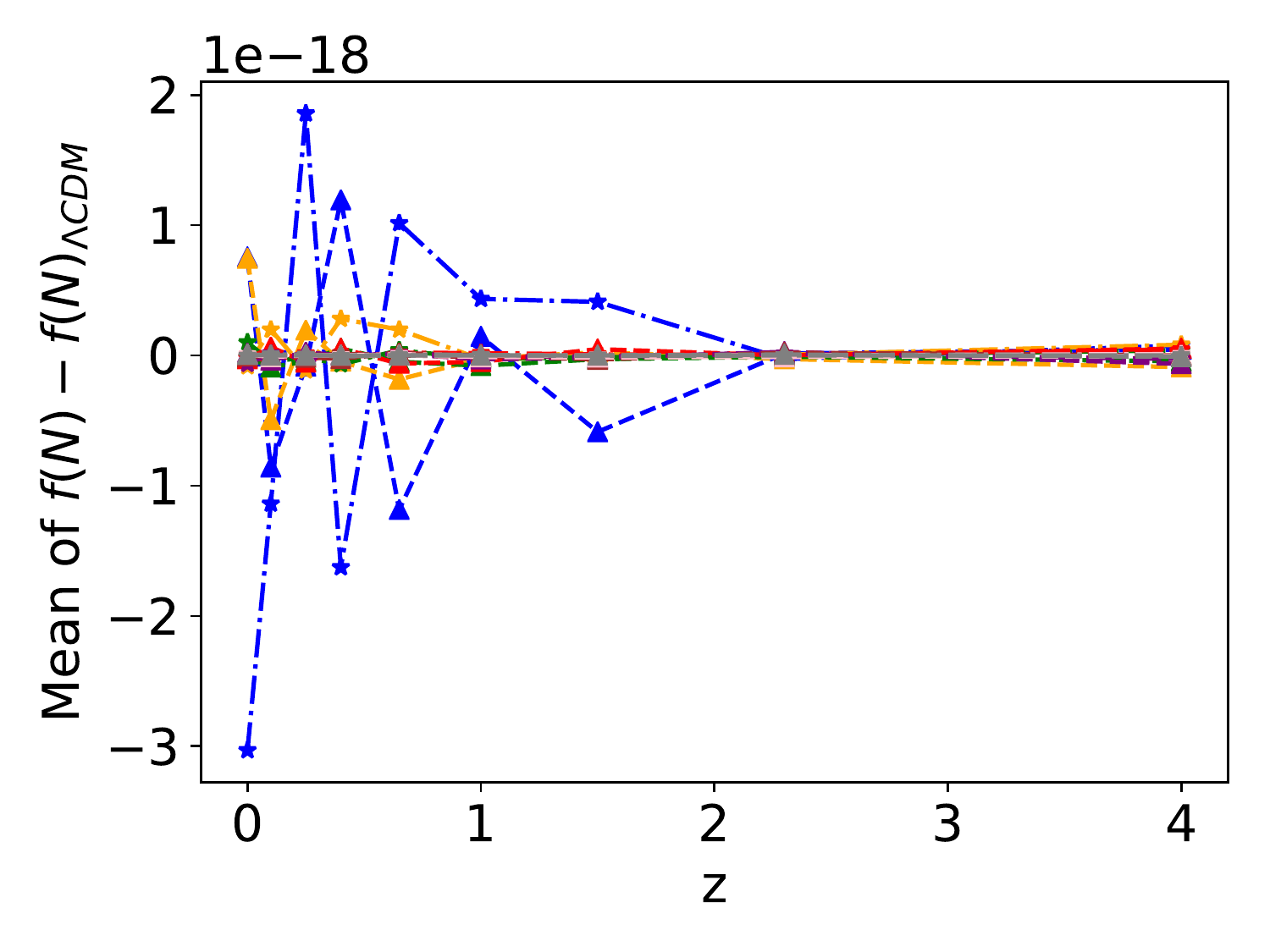}{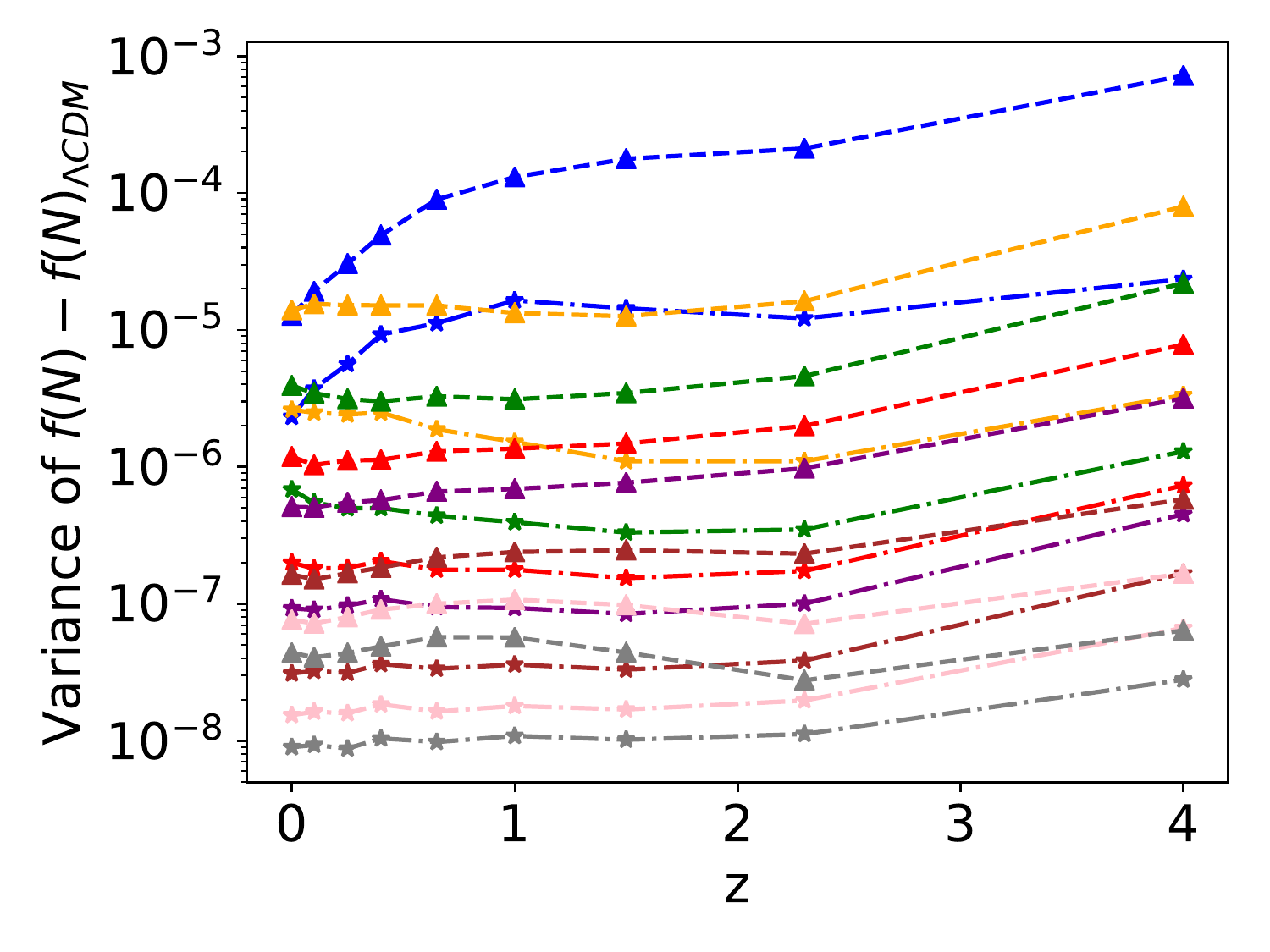}
\plottwo{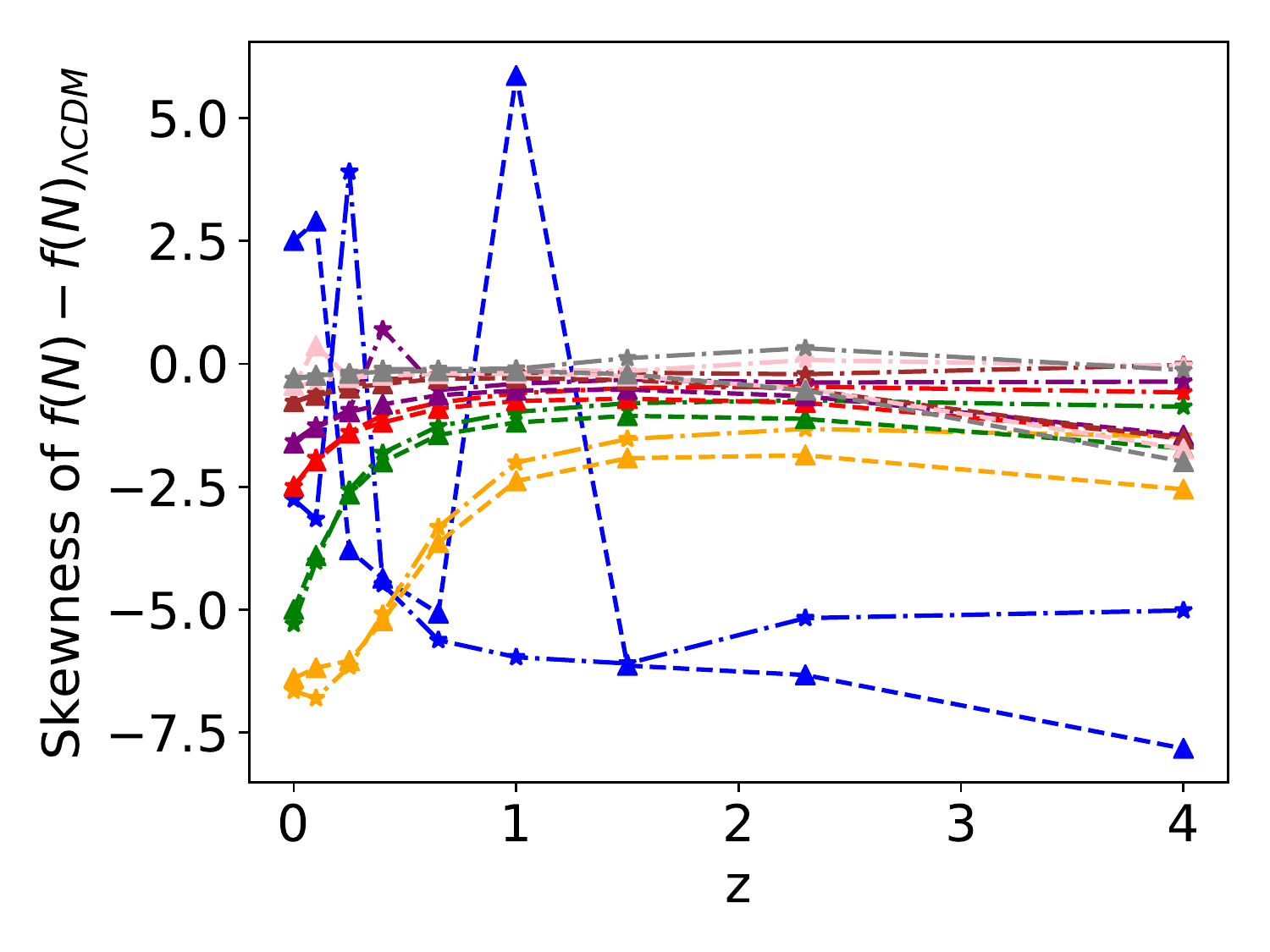}{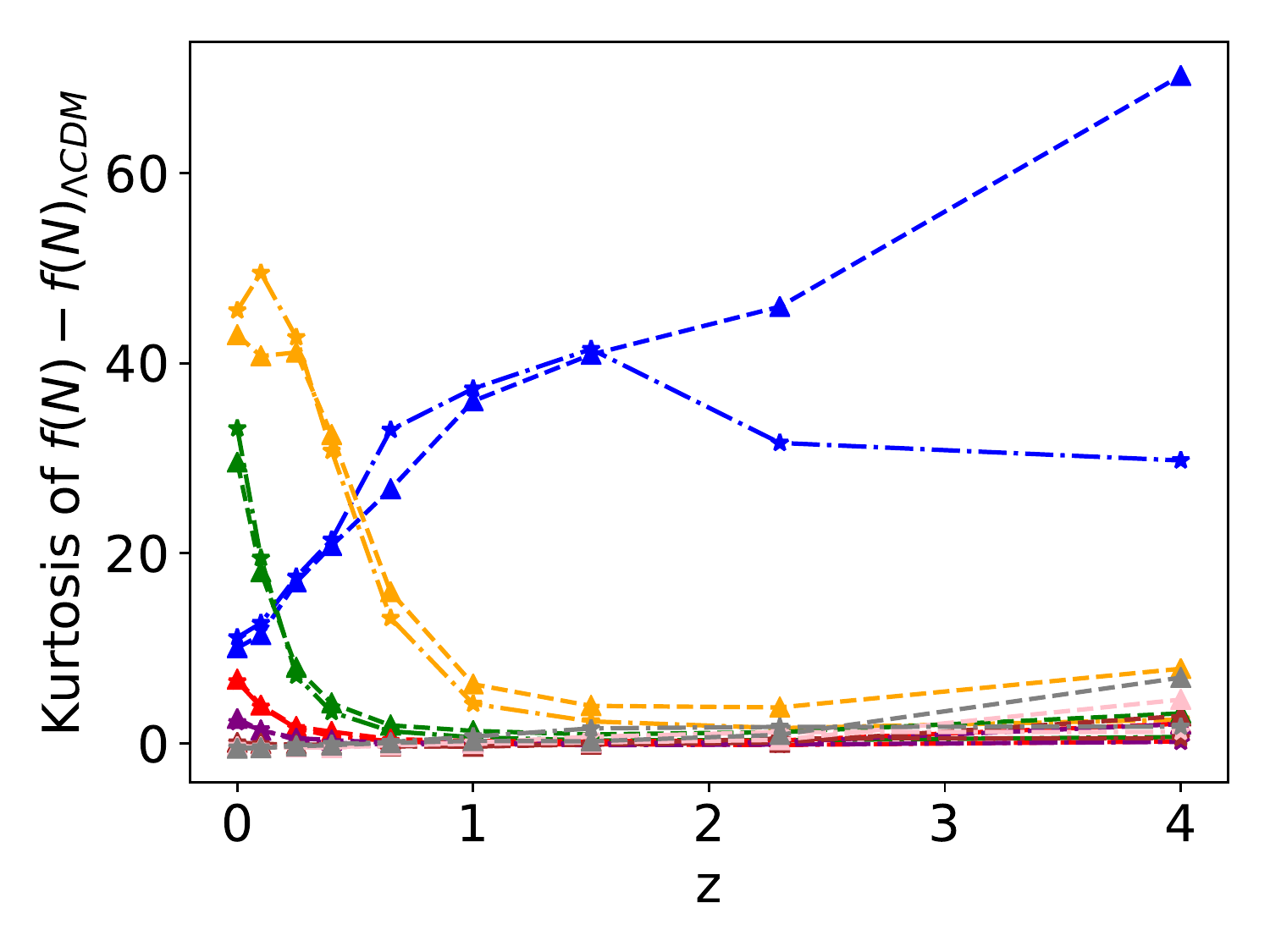}
\caption{Mean, variance, skewness and kurtosis of $\triangle f(N)_{RP-\Lambda}=f(N)_{RPCDM}-f(N)_{\Lambda CDM}$ (dash line with triangles)} and $\triangle f(N)_{w-\Lambda}=f(N)_{wCDM}-f(N)_{\Lambda CDM}$ (dash dot line with stars) as a function of redshift. Colors represent the same cell radii as those in Figure \ref{fig:cic_z}.
\label{fig:residual_moments}
\end{figure*}

\begin{figure*}
\epsscale{0.82}
\plottwo{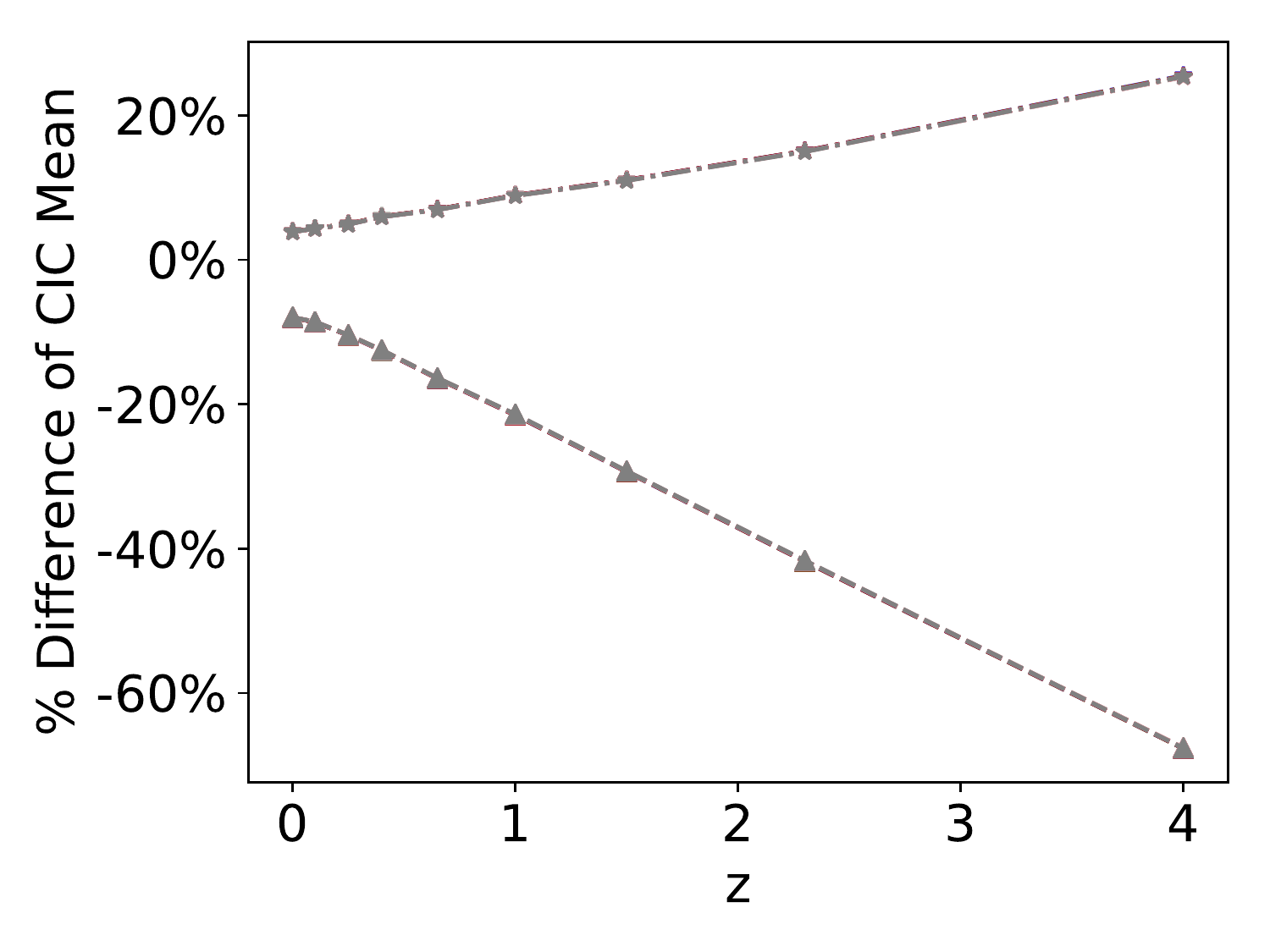}{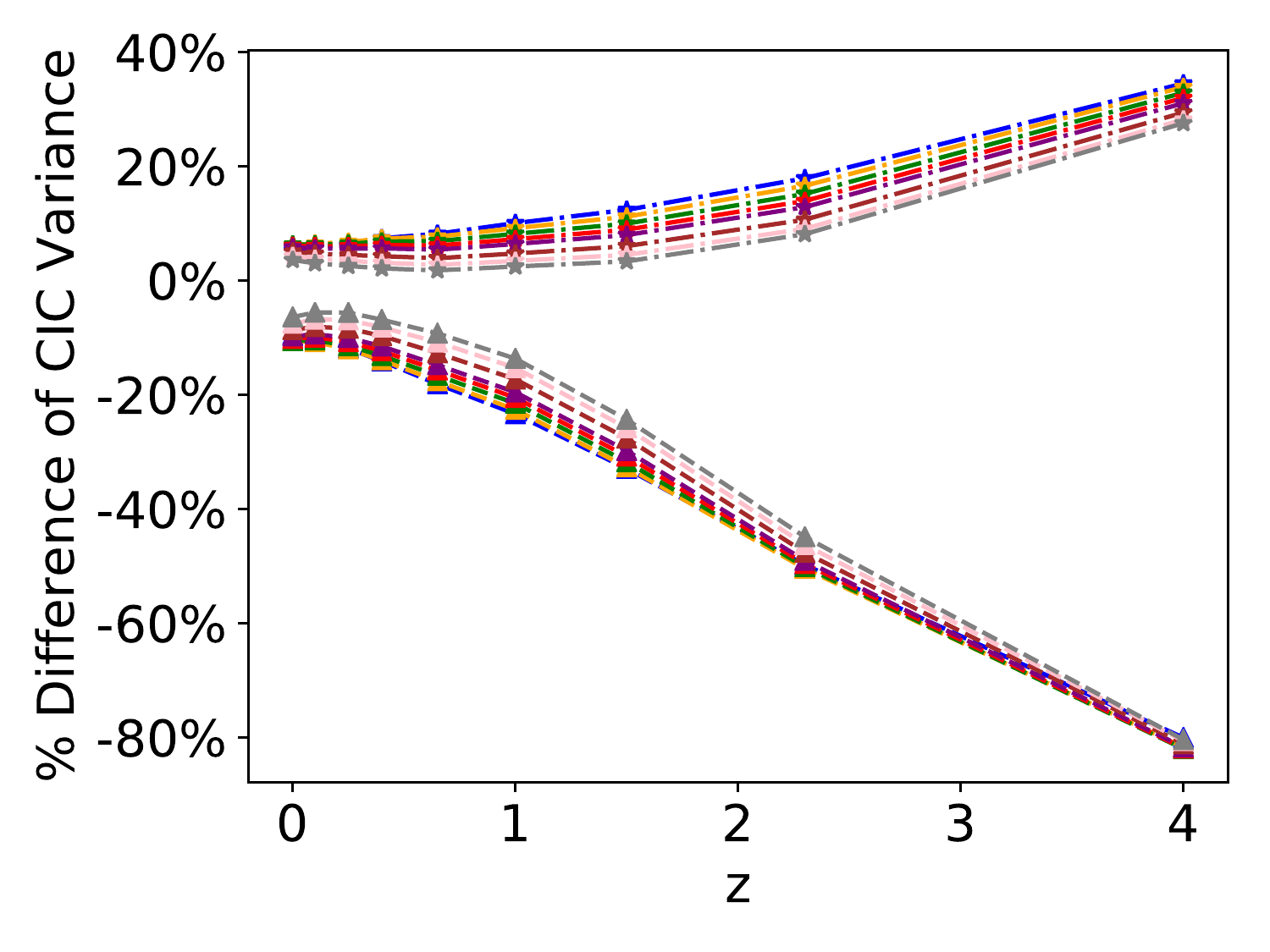}
\plottwo{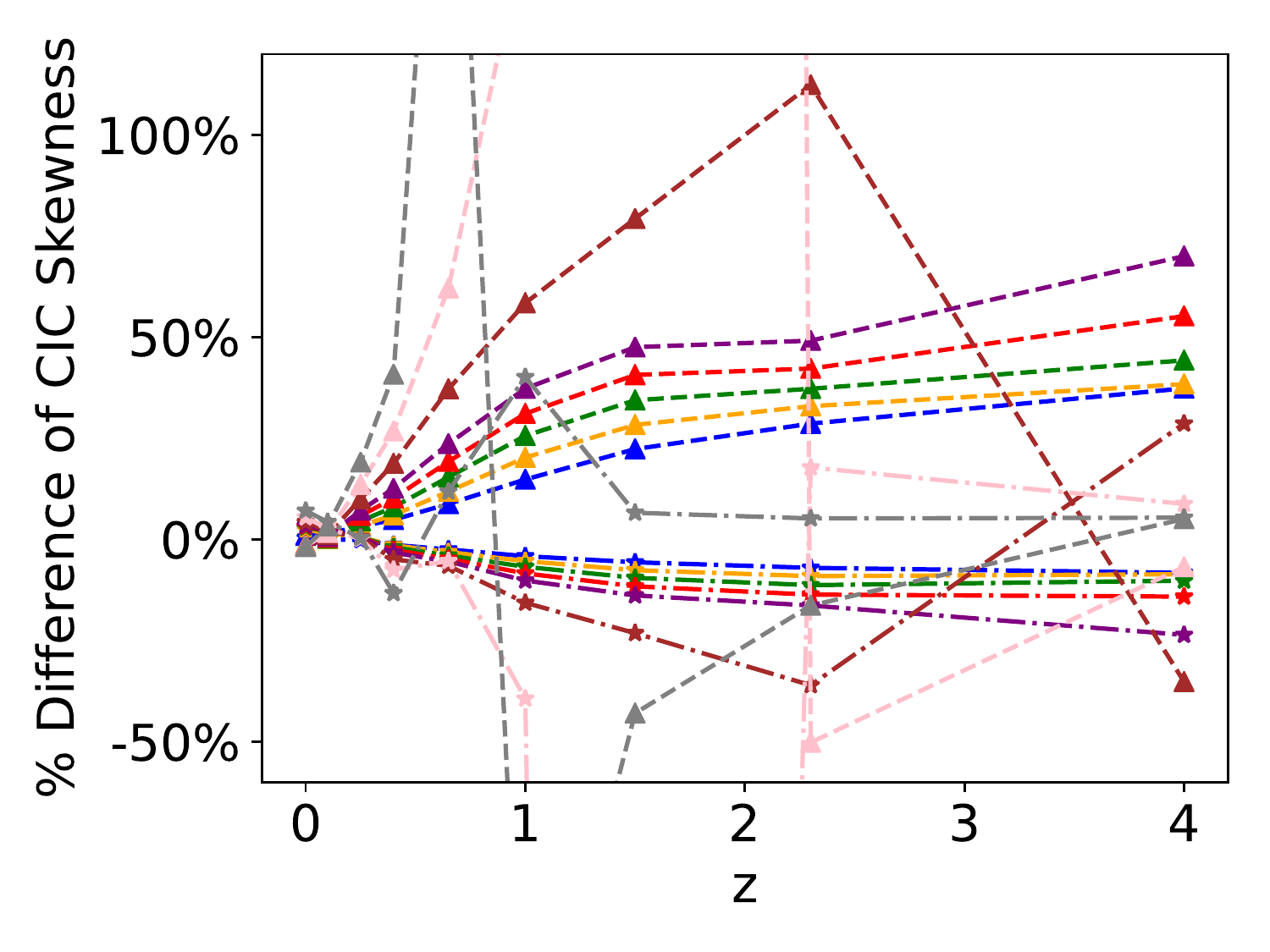}{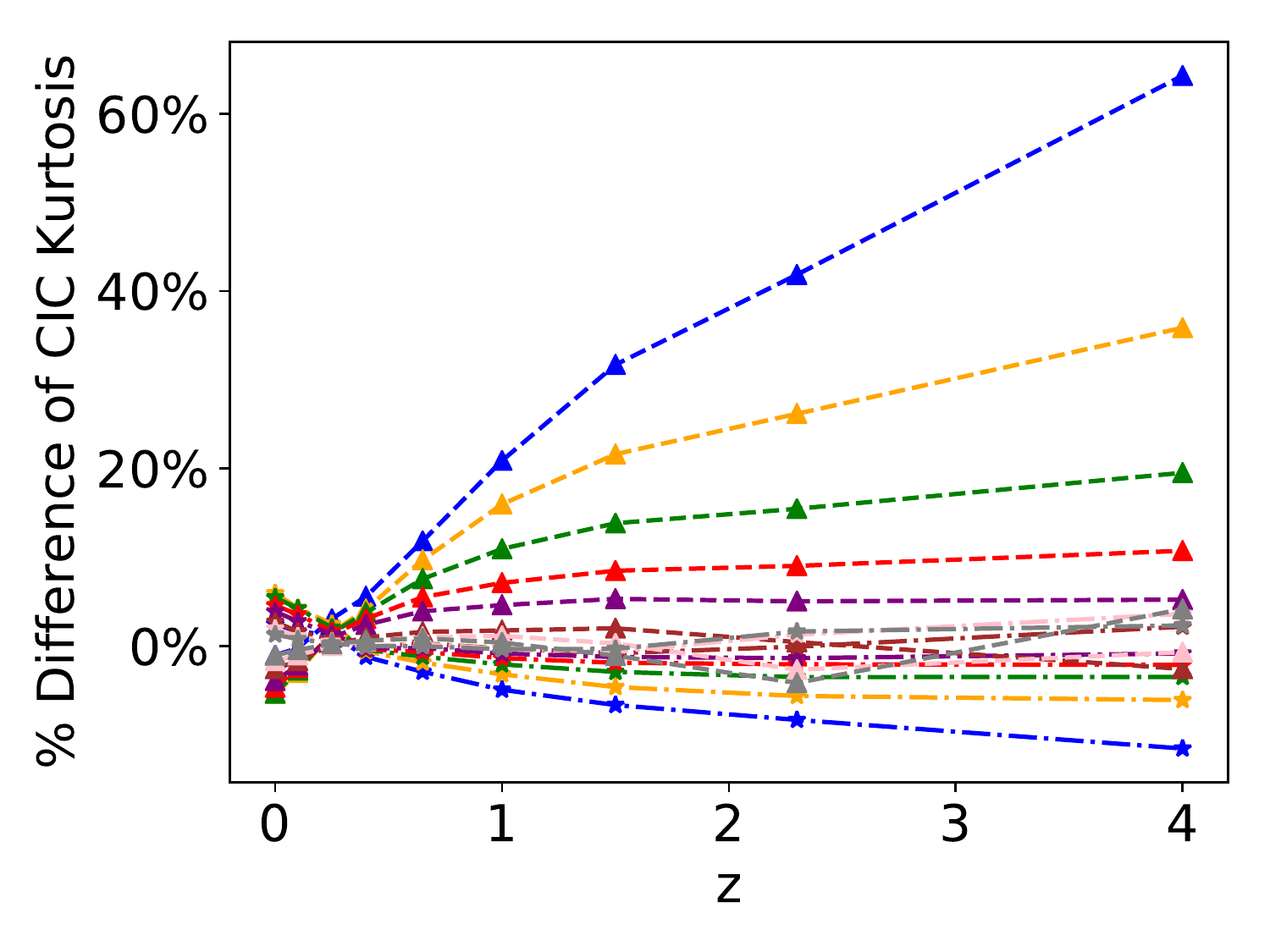}
\caption{Percentage difference in the counts-in-cells $f(N)$ mean, variance, skewness and kurtosis for RPCDM (dash line with triangles) and $w$CDM (dash dot line with stars) compared with $\Lambda$CDM as a function of redshift. In the upper panels, the decreasing lines are the dash lines with triangles and the increasing lines are the dash dot lines with stars; the overlying lines show that the trends are scale-independent. Colors represent the same cell radii as those in Figure \ref{fig:cic_z}.}
\label{fig:percent_moments}
\end{figure*}

The measured CiC distributions $f(N)$ shown in Figure \ref{fig:cic_R} and \ref{fig:cic_z} are smooth unimodal probability distribution functions. In Figure \ref{fig:cic_z}, at a given redshift the peak of the distribution moves to larger halo counts as the cell radius increases, because larger cells contain more halos. On a log scale, $f(N)$ is less smooth in larger cells at higher redshifts. For a fixed cell radius (Figure \ref{fig:cic_R}), the peak of $f(N)$ shifts towards larger $N$ at higher redshifts, because the same physical volume encloses a larger comoving volume. The peak also decreases at higher redshifts, as the probability spreads into a wider range of $N$. The differences between distributions in different dark energy models in smaller cells and lower redshifts are smaller, but grow very significantly as a function of both cell radius and redshift.

In Figure \ref{fig:residual_moments} we examine the lowest-order statistical moments of the residuals between quintessence and $\Lambda$CDM and phantom dark energy and $\Lambda$CDM, denoted by $\triangle f(N)_{RP-\Lambda}=f(N)_{RPCDM}-f(N)_{\Lambda CDM}$ and $\triangle f(N)_{w-\Lambda}=f(N)_{wCDM}-f(N)_{\Lambda CDM}$ respectively. The mean, variance, skewness and kurtosis of the residuals of $f(N)$ between both alternative dark energy models and $\Lambda$CDM are larger in smaller cells. The mean of the residuals $\bar \triangle f(N)_{RP-\Lambda}$ and $\bar \triangle f(N)_{w-\Lambda}$ decrease at higher redshifts. The residual means are computed over all $N$ (from 0 to the largest N with non-zero count), so their values are vanishingly small and do not distinguish well between dark energy models due to this averaging. The variance of of the residuals $\triangle f(N)_{RP-\Lambda}$ is larger than that of $\triangle f(N)_{w-\Lambda}$ for all cell radii and redshifts. Generally, as redshift increases, the variance of the difference tends to first decrease and then increase, although this trend is not seen for all cell radii. The skewness of the residuals $\triangle f(N)_{RP-\Lambda}$ is generally smaller than that of $\triangle f(N)_{w-\Lambda}$. Both values approach zero at higher redshifts and in larger cells except for cell radius $R=2h^{-1}$Mpc. The kurtosis of the differences $\triangle f(N)_{RP-\Lambda}$ and $\triangle f(N)_{w-\Lambda}$ are not strongly distinguished by dark energy cosmological model; the kurtosis of these differences generally decreases for larger cells and higher redshifts. It increases very slightly from $z=2.3$ to $z=4$. Figure \ref{fig:residual_moments} shows that the variance, skewness and kurtosis of the residuals $\triangle f(N)_{RP-\Lambda}$ and $\triangle f(N)_{w-\Lambda}$ on $2h^{-1}$Mpc scale at $z>2.3$ are the largest and the most different among RPCDM, $w$CDM and $\Lambda$CDM. The skewness and kurtosis of the residuals on $4-6h^{-1}$Mpc scales are larger at $z<0.65$. At scales larger than $8h^{-1}$Mpc, the moments of the residuals are not significantly far from zero at $0<z<4$.

\begin{figure*}[hbt!]
\epsscale{0.99}
\plottwo{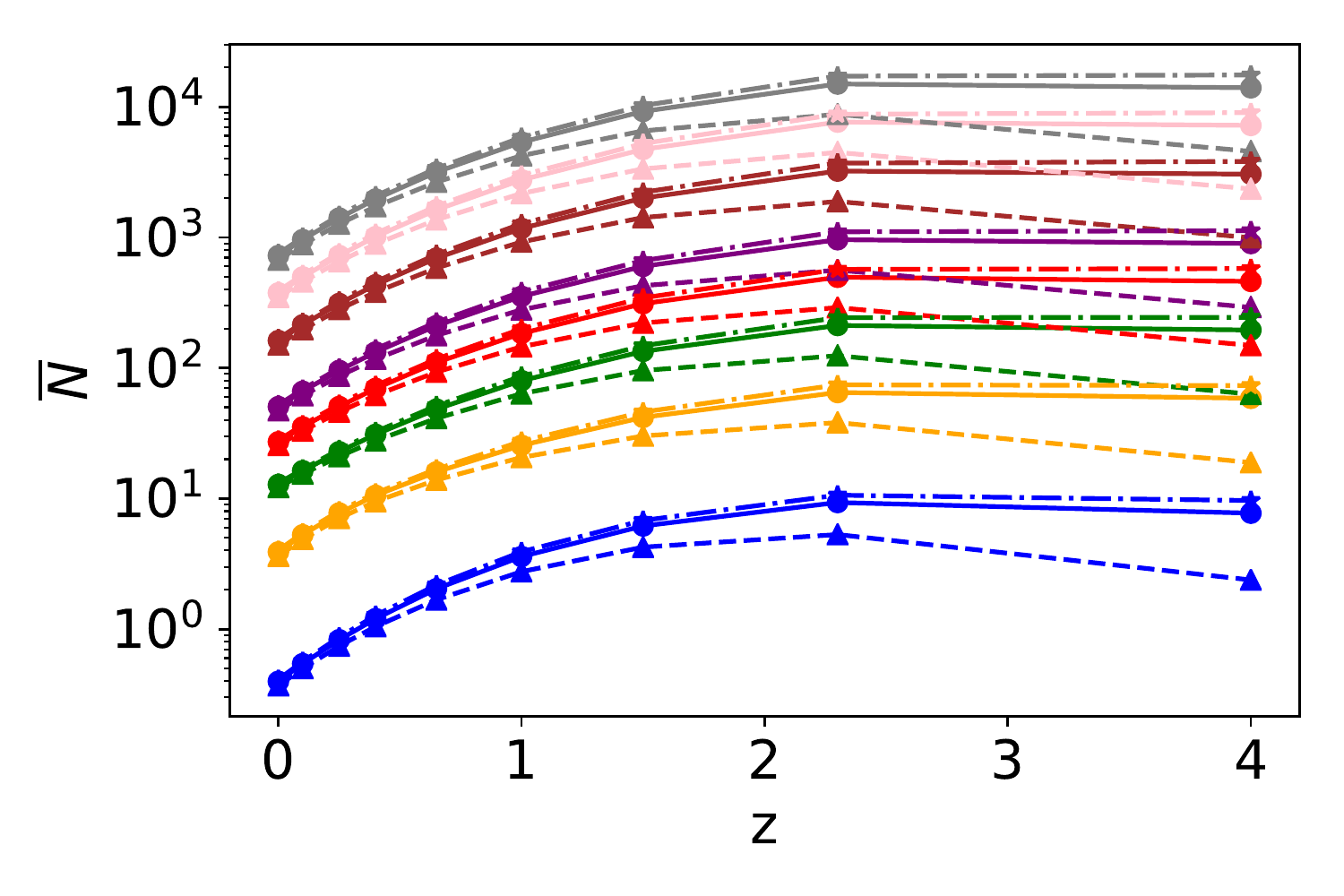}{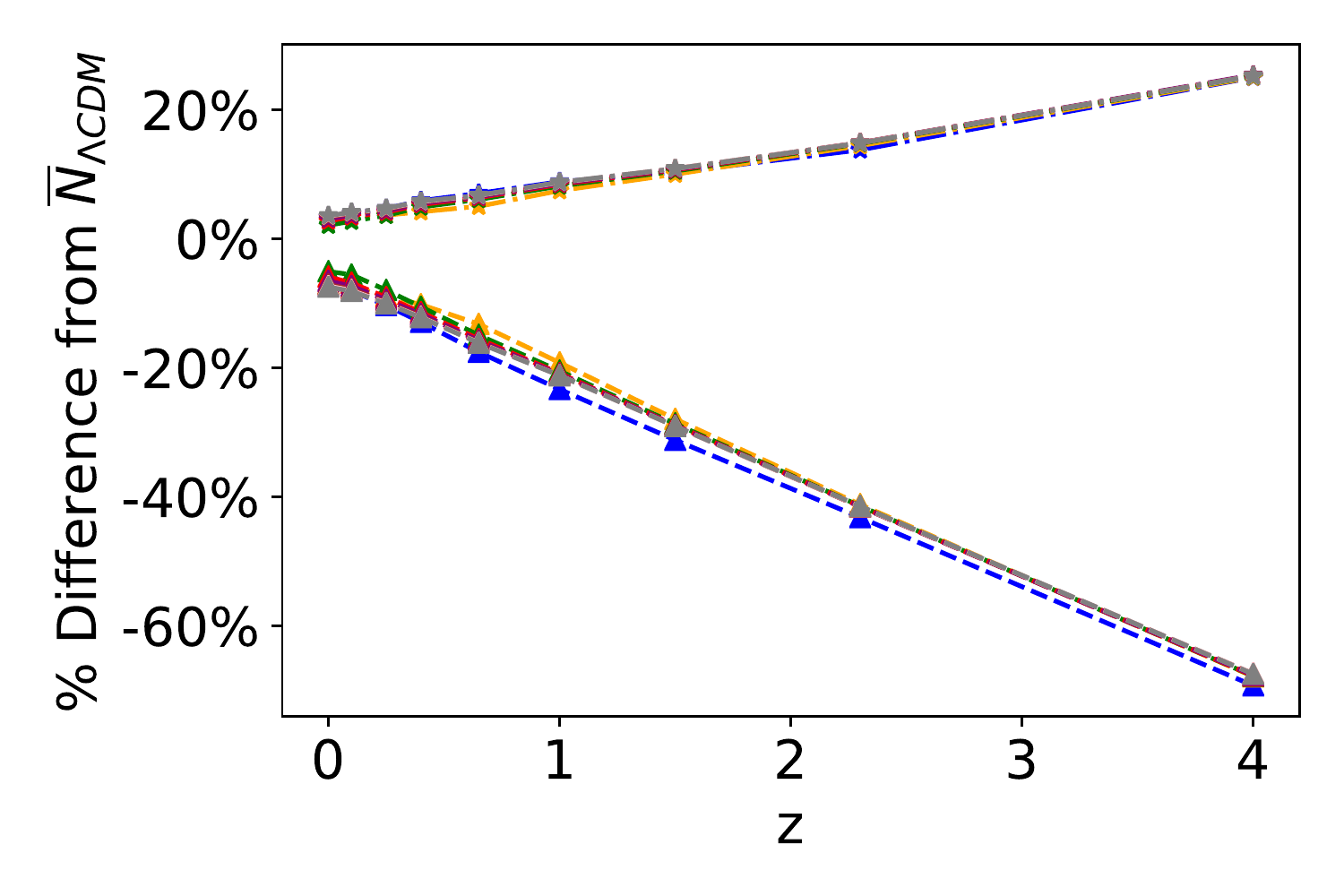}
\plottwo{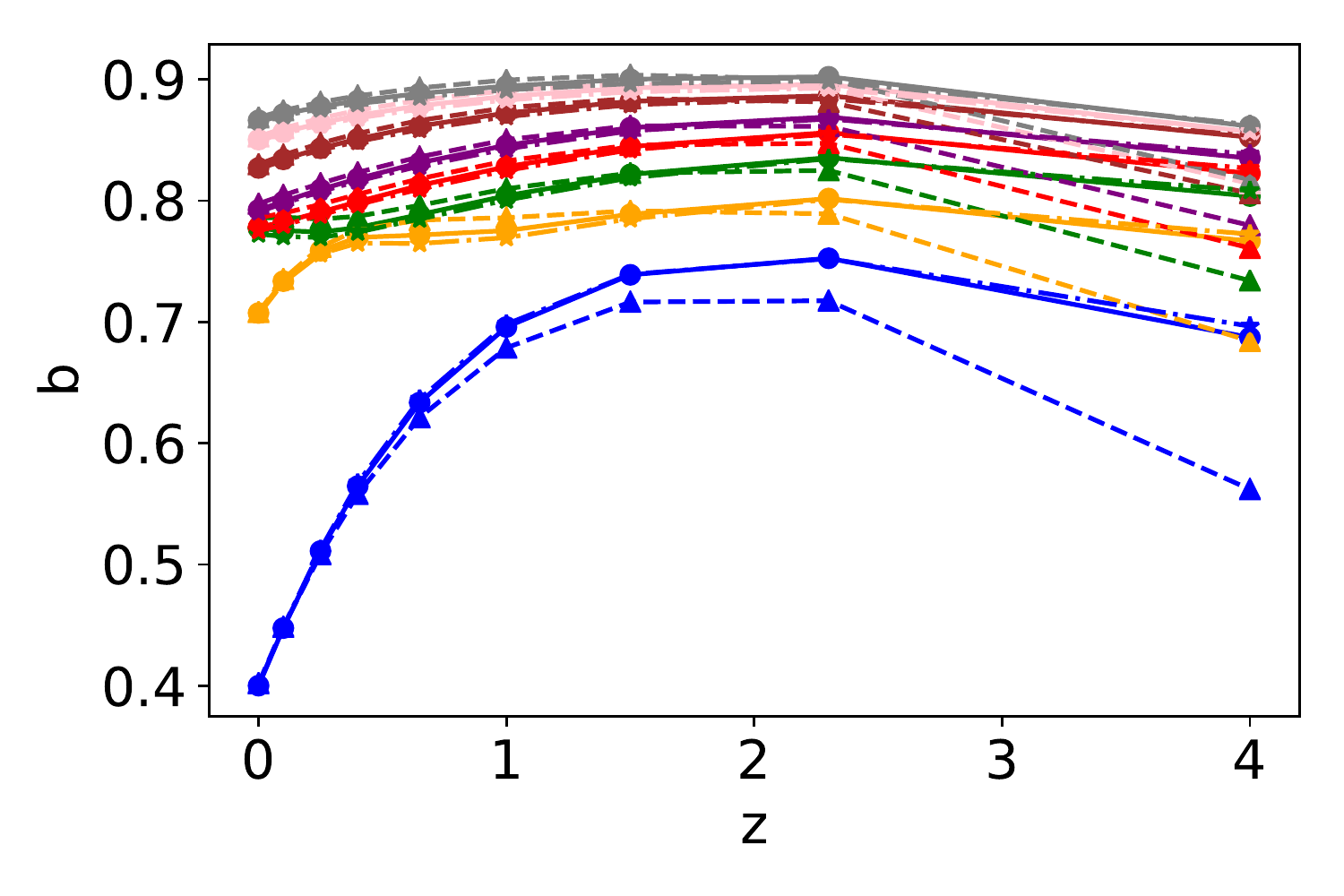}{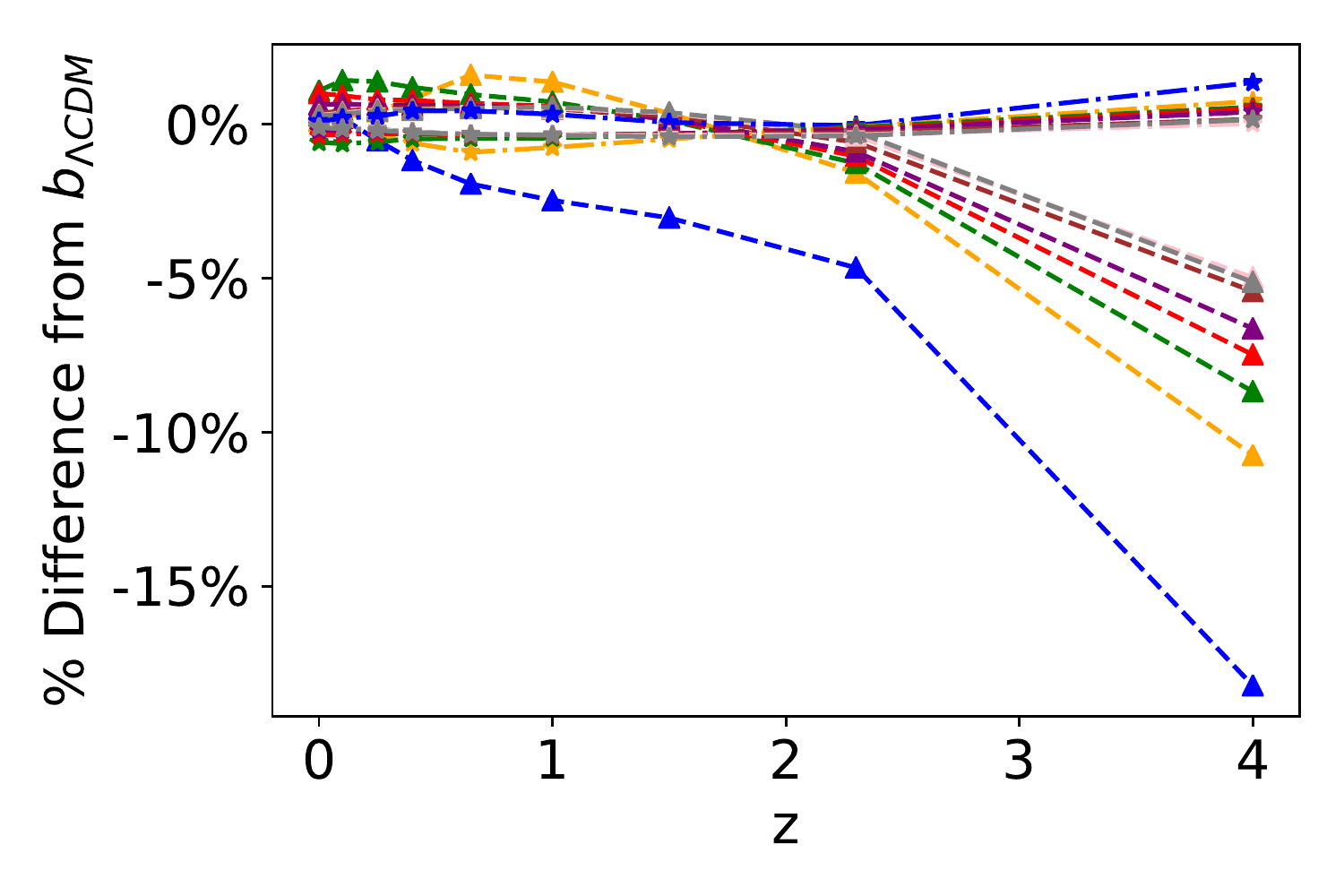}
\caption{The best-fit GQED parameters of RPCDM (dash line with triangles) and $w$CDM (dash dot line with stars) compared with $\Lambda$CDM (solid line with dots). Colors represent the same cell radii as those in Figure \ref{fig:cic_z}.}
\label{fig:gqed}
\end{figure*}

\begin{figure*}[hbt!]
\epsscale{0.99}
\plottwo{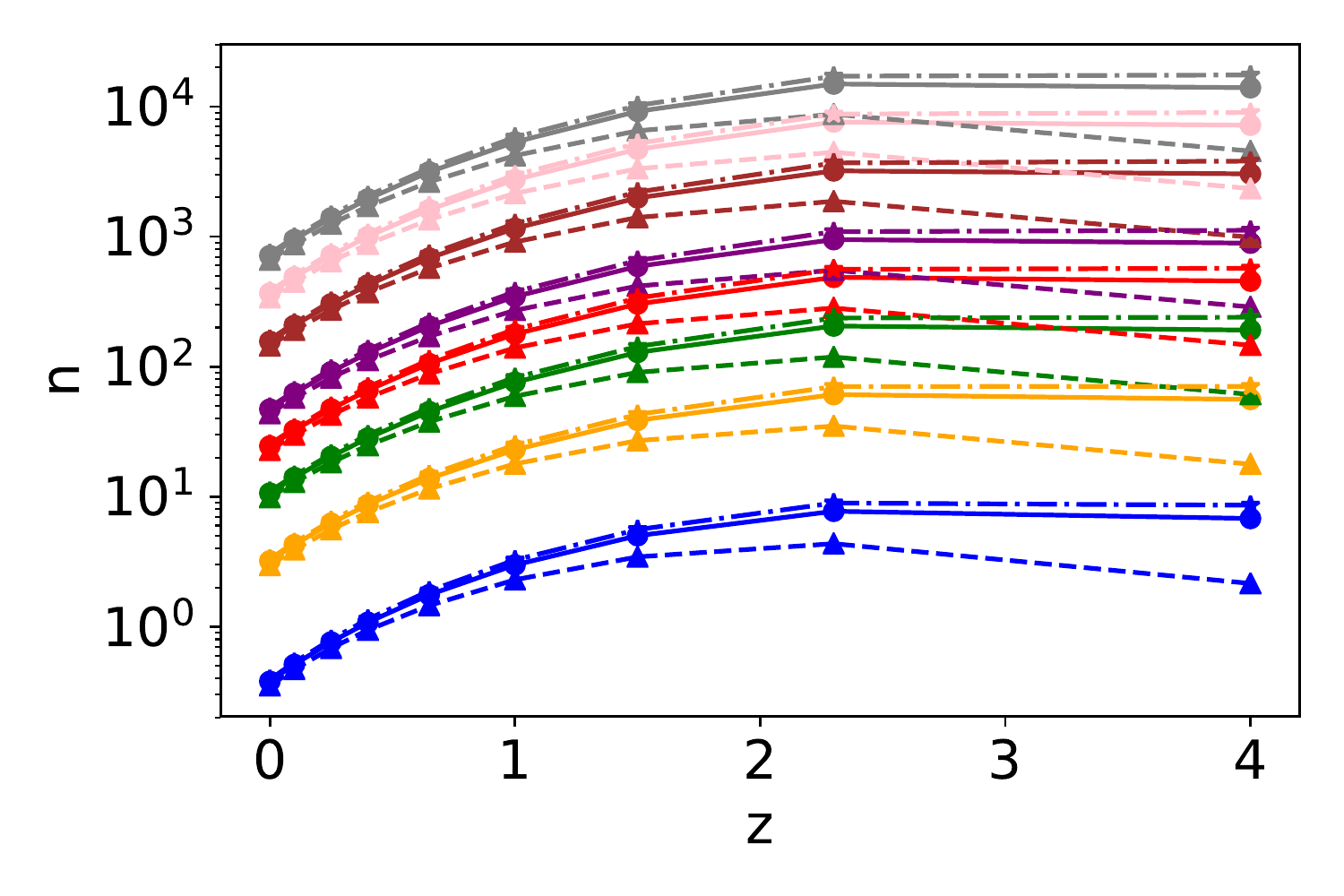}{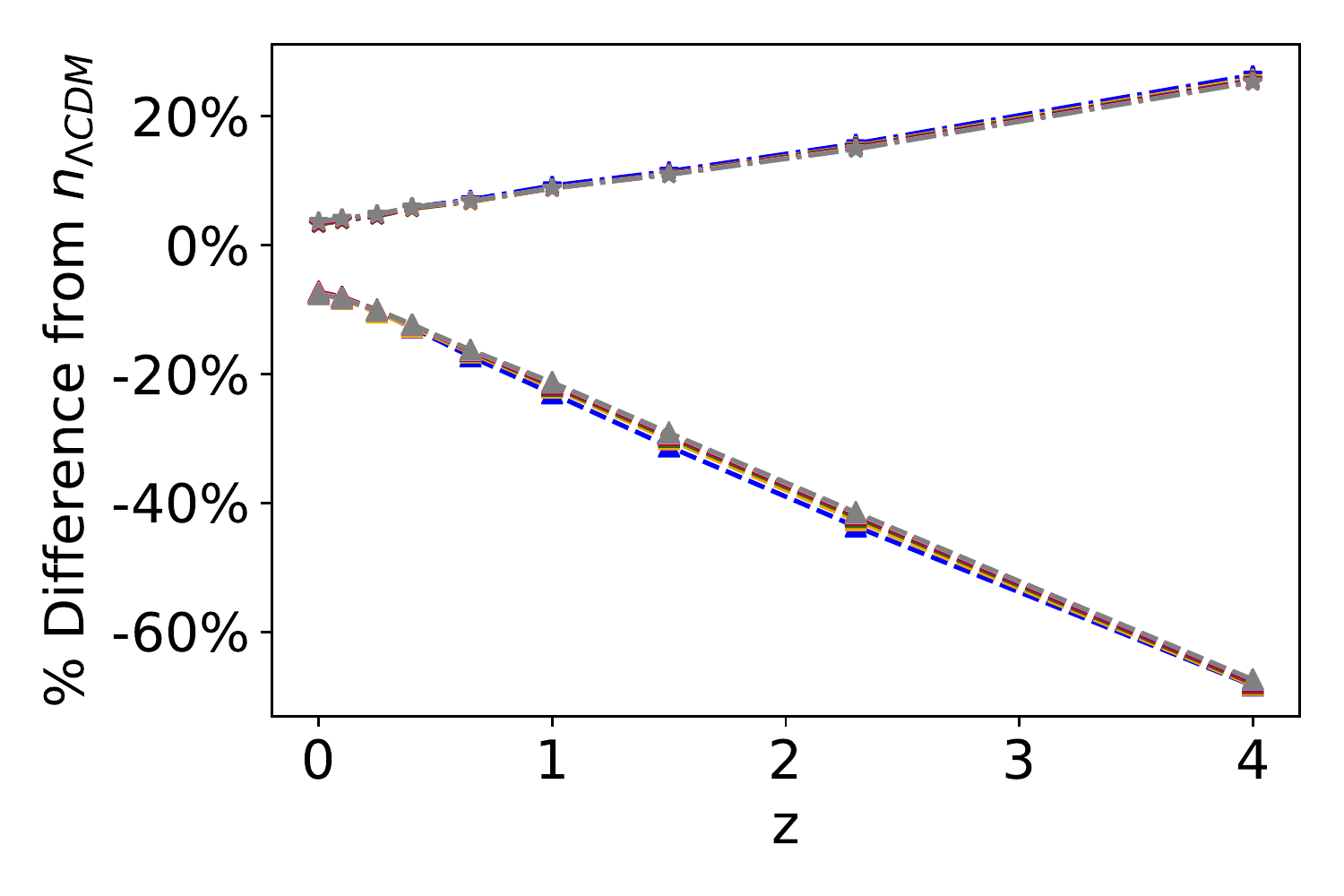}
\plottwo{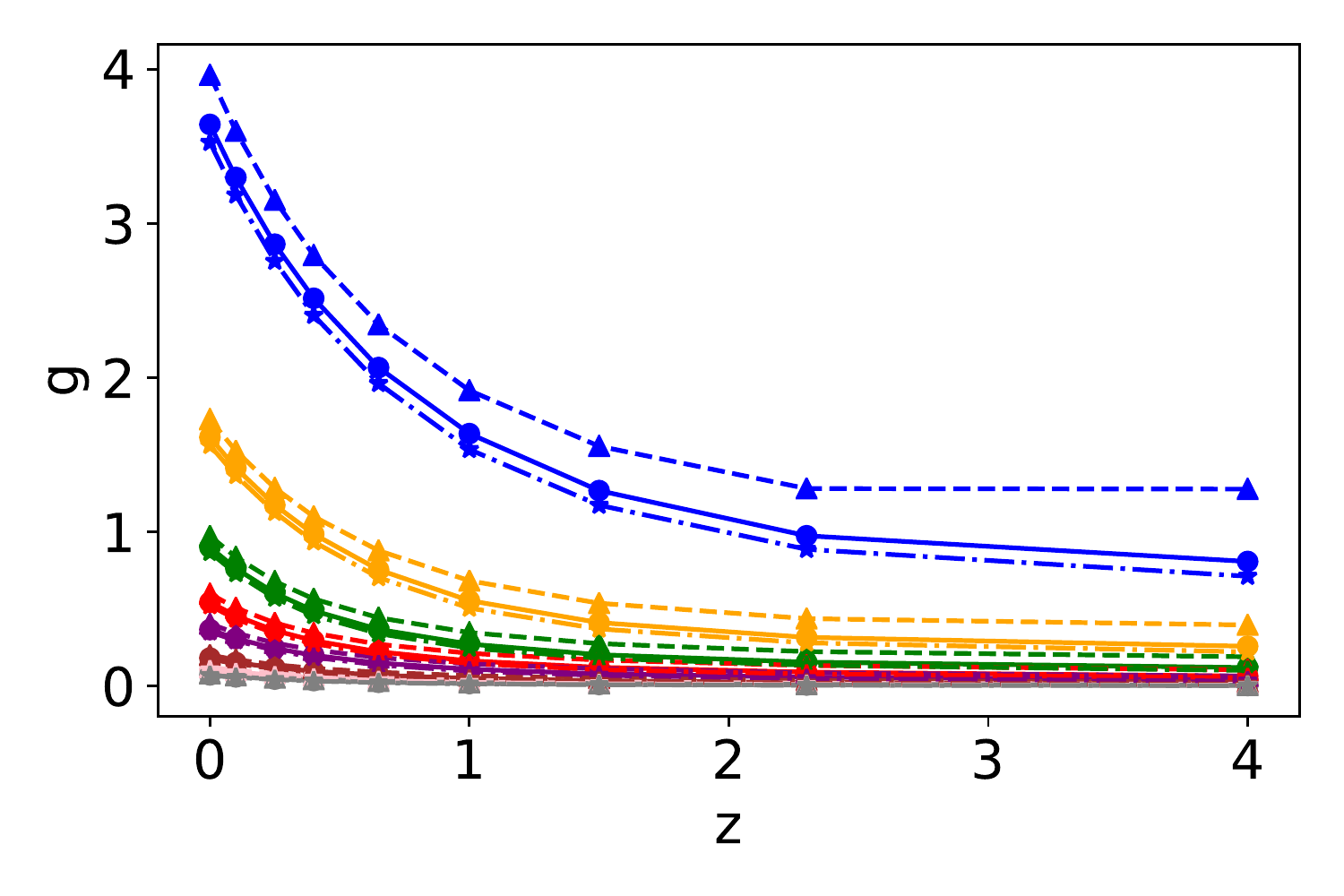}{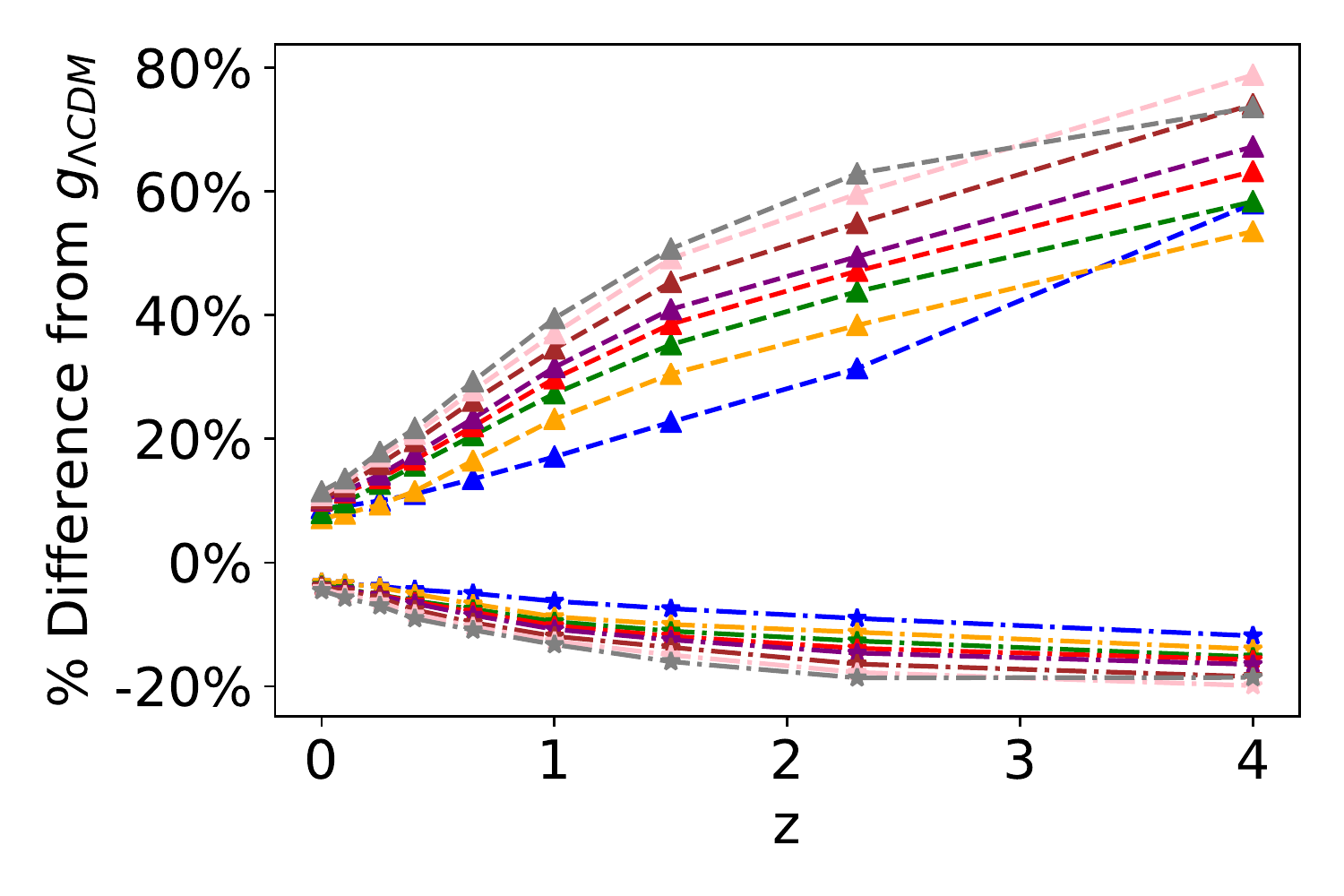}
\caption{The best-fit NBD parameters of RPCDM (dash line with triangles) and $w$CDM (dash dot line with stars) compared with $\Lambda$CDM (solid line with dots). Colors represent the same cell radii as those in Figure \ref{fig:cic_z}.}
\label{fig:nbd}
\end{figure*}

\begin{figure*}[hbt!]
\plottwo{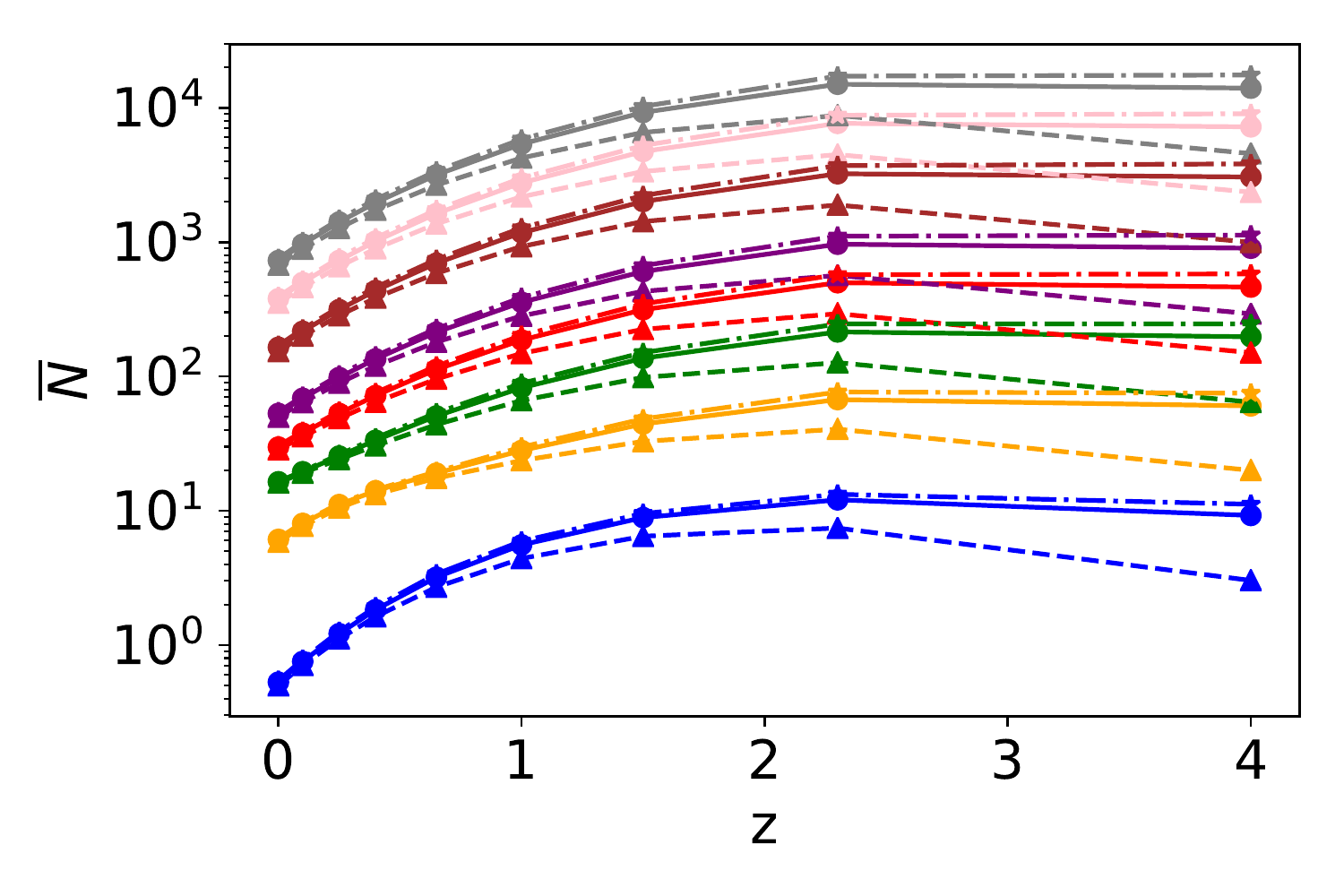}{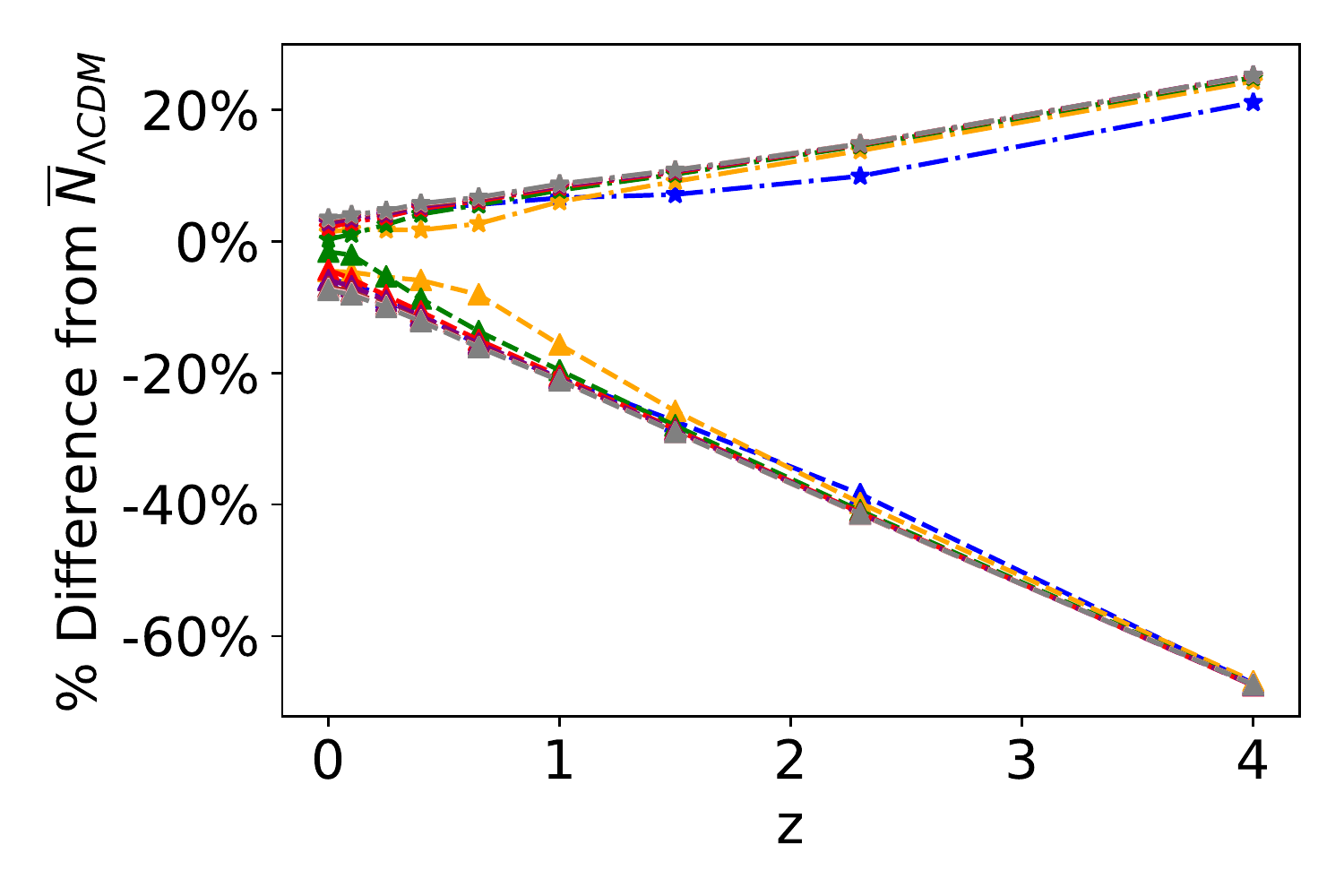}
\plottwo{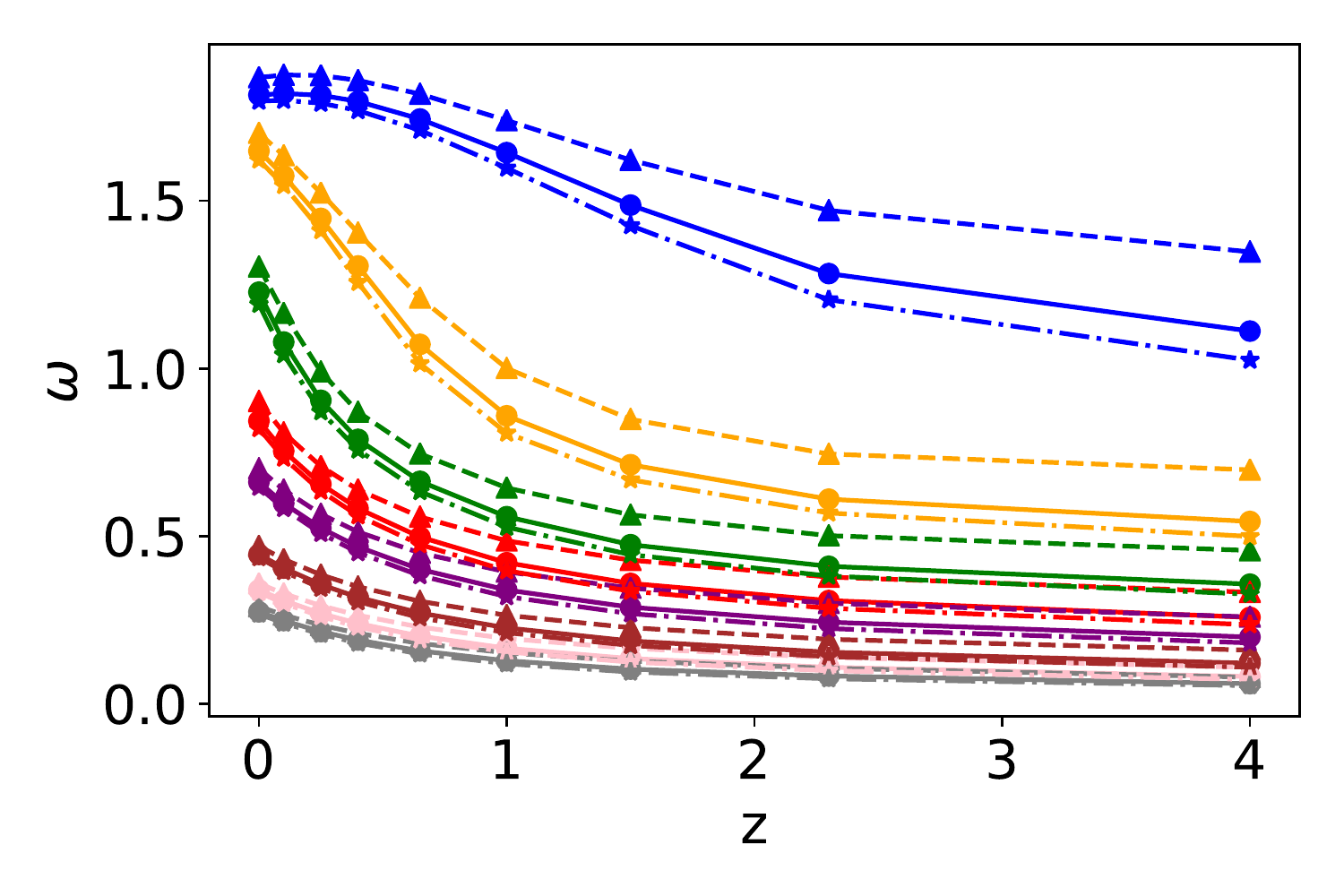}{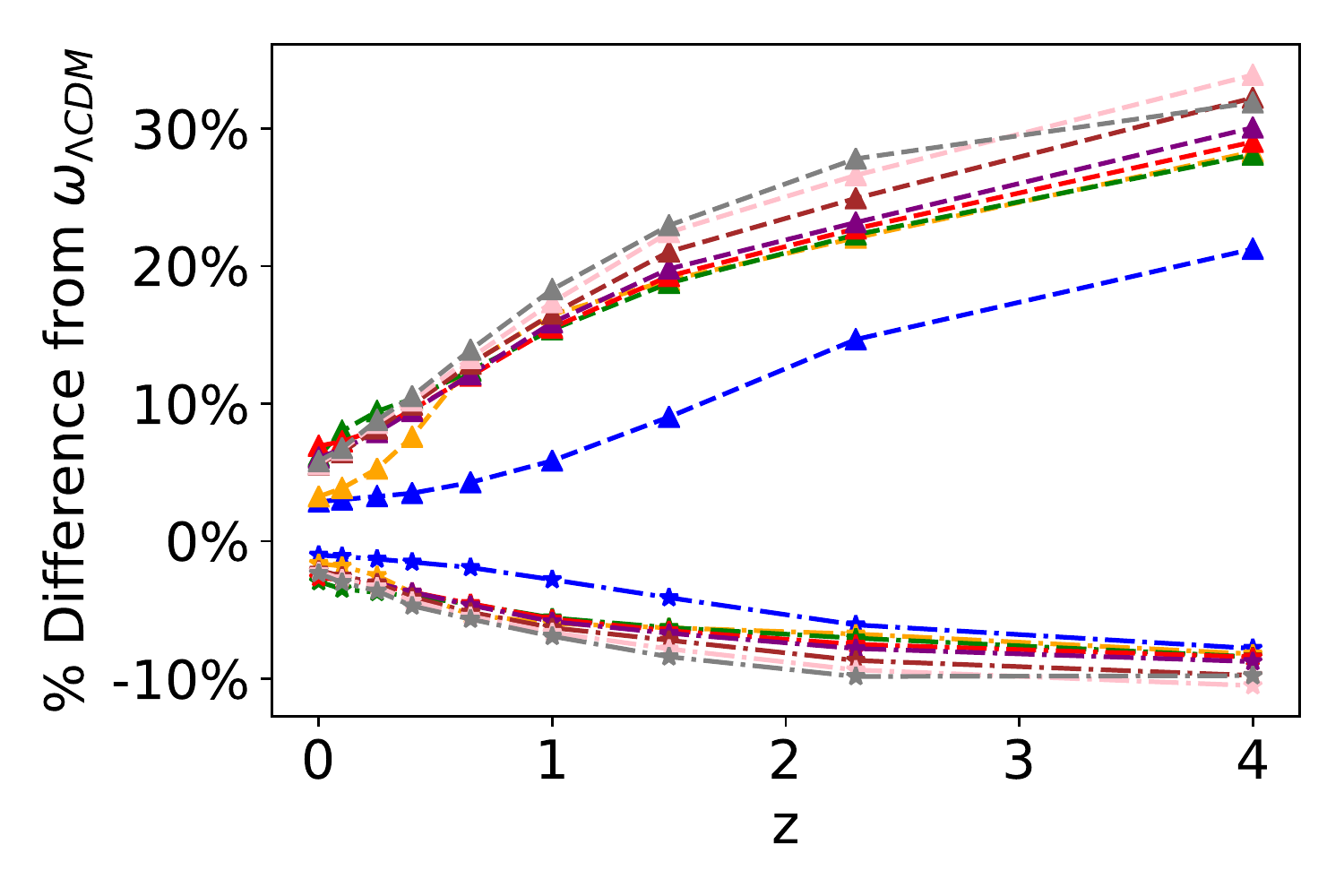}
\caption{The best-fit PLN parameters of RPCDM (dash line with triangles) and $w$CDM (dash dot line with stars) compared with $\Lambda$CDM (solid line with dots). Colors represent the same cell radii as those in Figure \ref{fig:cic_z}.}
\label{fig:pln}
\end{figure*}

\begin{figure*}[hbt!]
\plottwo{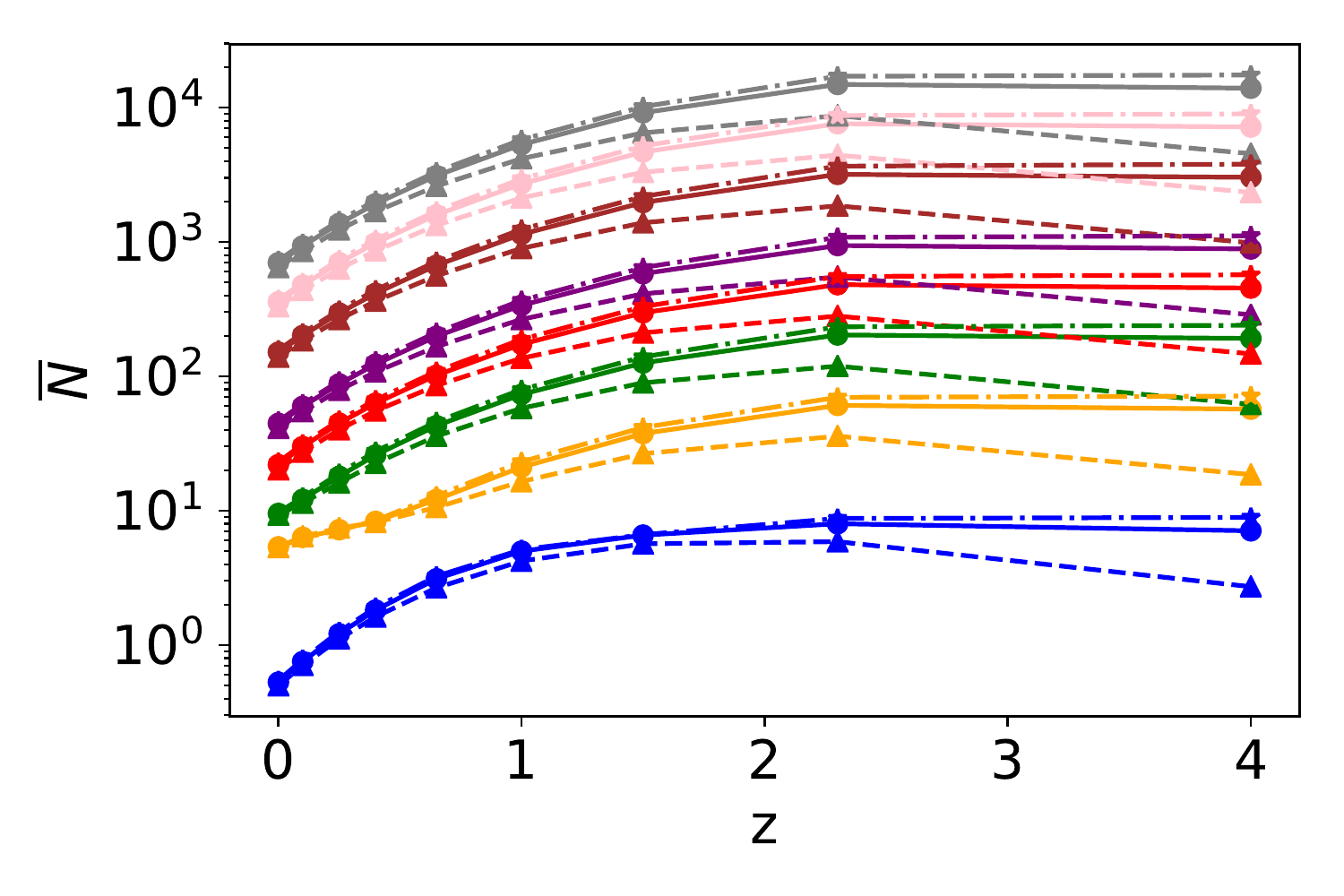}{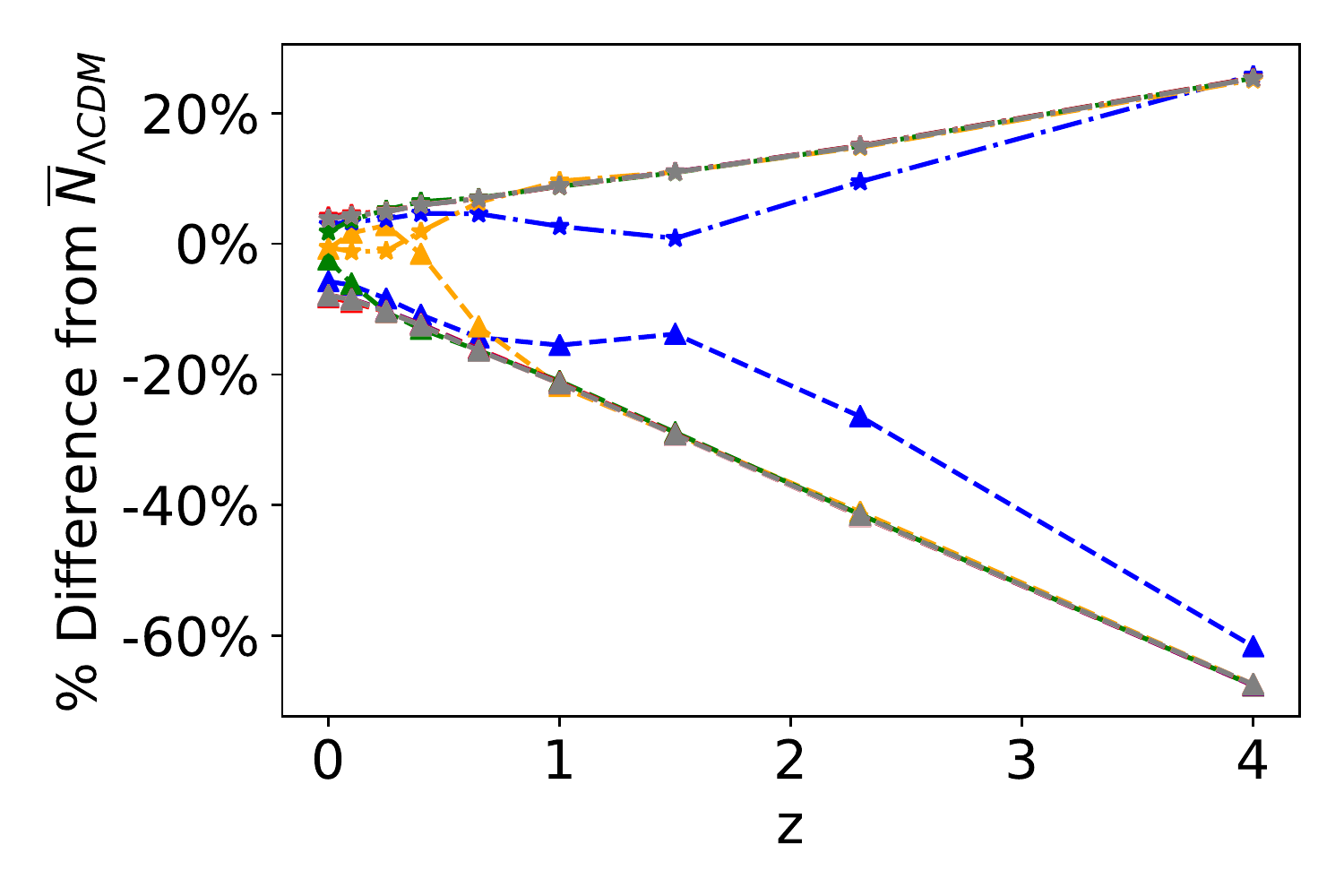}
\plottwo{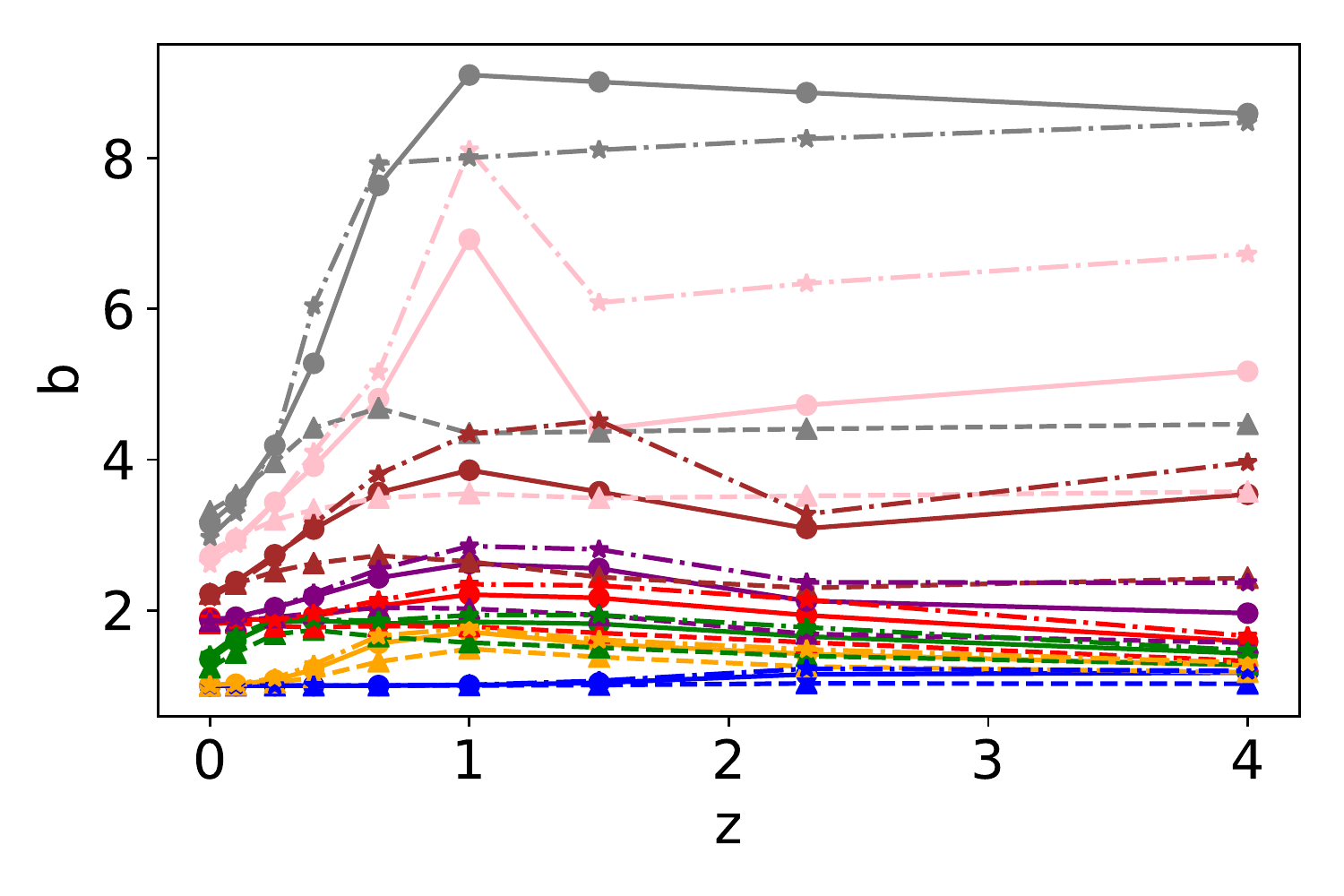}{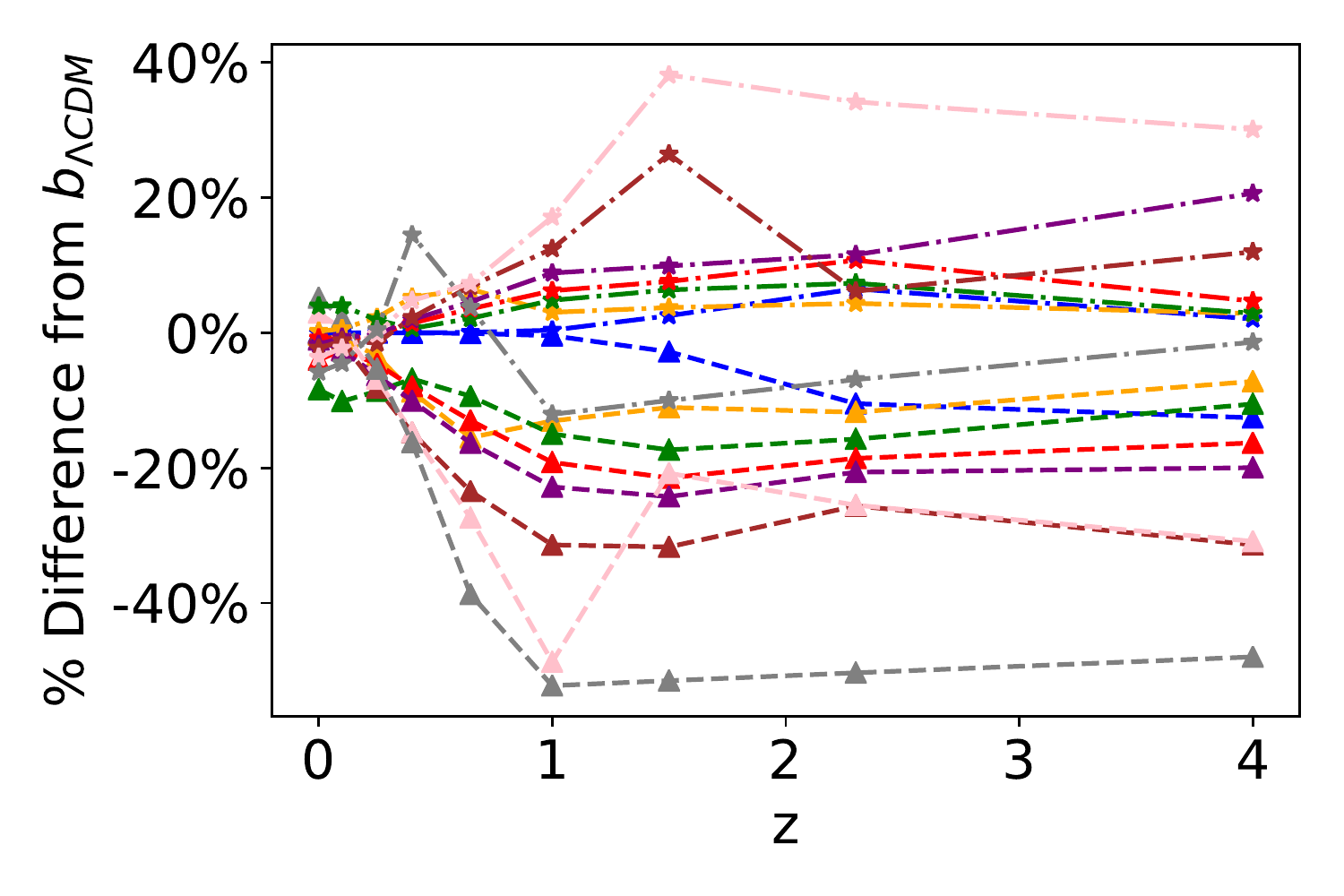}
\plottwo{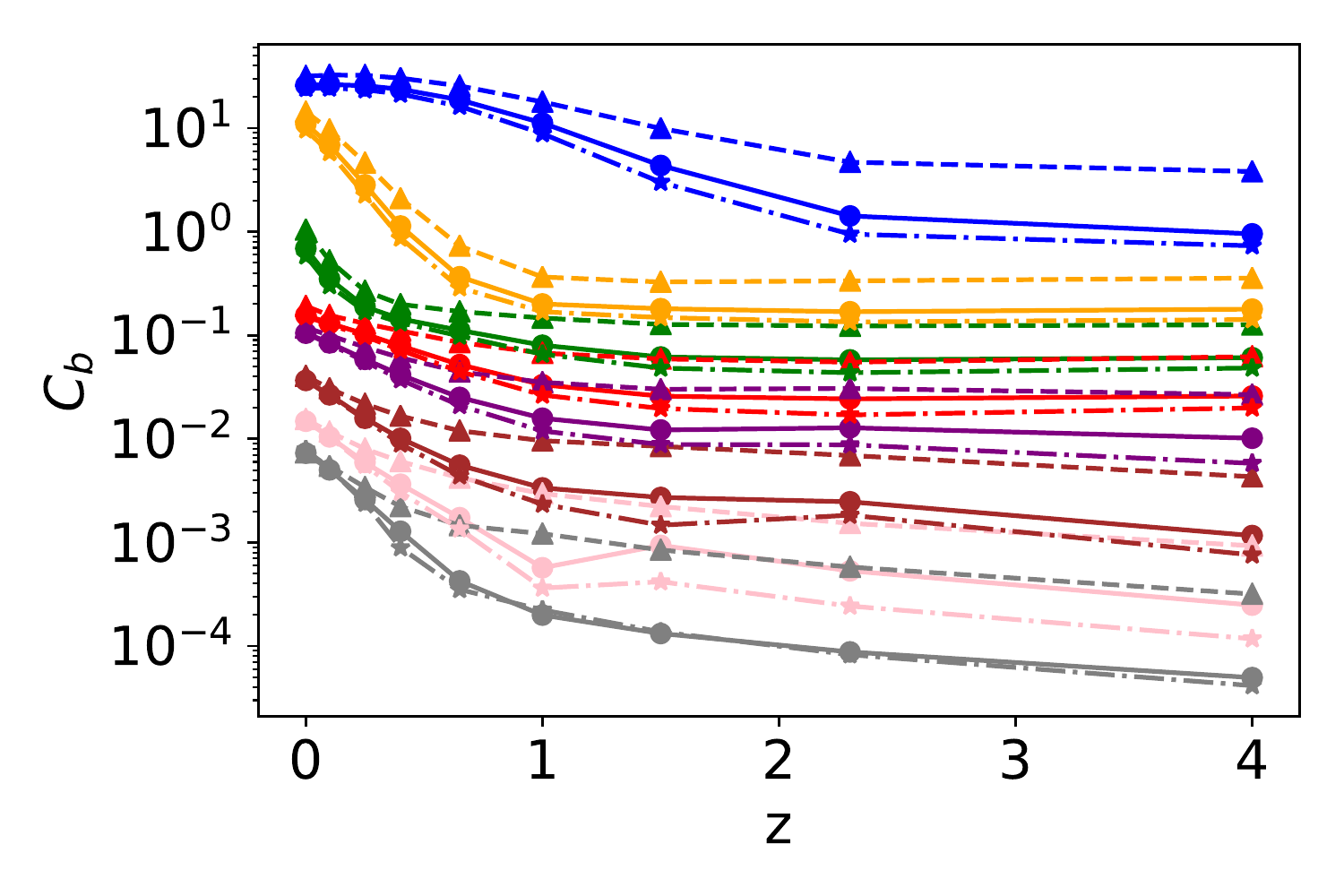}{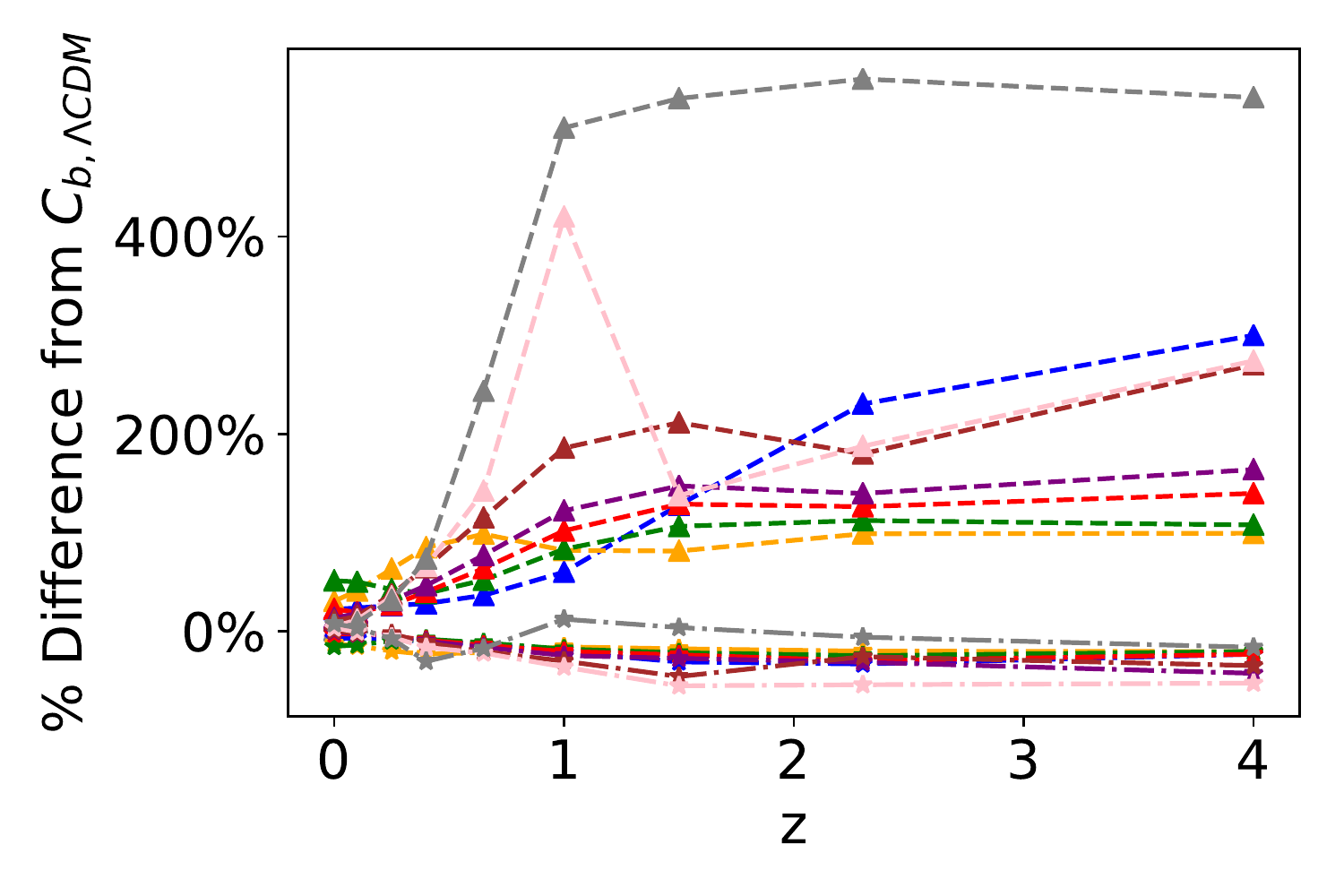}
\caption{The best-fit PLNB parameters of RPCDM (dash line with triangles) and $w$CDM (dash dot line with stars) compared with $\Lambda$CDM (solid line with dots). Colors represent the same cell radii as those in Figure \ref{fig:cic_z}.}
\label{fig:plnb}
\end{figure*}
The percentage differences of the lowest moments of the measured CiC $f(N)$ in other dark energy models compared to the moment values obtained in $\Lambda$CDM are shown in Figure \ref{fig:percent_moments}. Overall, RPCDM clearly shows larger amplitude and opposite signs in percentage differences than $w$CDM when compared with $\Lambda$CDM. The percentage difference of the CiC $f(N)$ mean between RPCDM and $\Lambda$CDM decreases almost linearly as a function of redshift regardless of cell radius from about -8.0\% at $z=0$ to about -67.7\% at $z=4$. The percentage difference of the CiC mean between $w$CDM and $\Lambda$CDM increases almost linearly as a function of redshift regardless of cell radius from about 3.9\% at $z=0$ to about 25.5\% at $z=4$. The colored lines for different cell radii overlap almost exactly in the upper left panel of Figure \ref{fig:percent_moments}. This shows that the percentage difference of the CiC means between RPCDM/$w$CDM and $\Lambda$CDM depends only on redshift, not scale. The trend for variance is similar, but shows slightly larger percentage differences in smaller cells up to a maximum $\sim -80\%$ for RPCDM and about 30\% for $w$CDM at $z=4$. The percentage difference of skewness as a function of redshift shows only smooth trends for cell radii up to 10$h^{-1}$Mpc, but has erratic and non-monotonic variation in larger cells. The percentage difference in kurtosis between RPCDM and $\Lambda$CDM increases up to more than 60\% and that between $w$CDM and $\Lambda$CDM more than -10\%. The percentage difference changes sign and crosses 0 at $z\sim0.25$. Smaller cells show larger percentage differences in kurtosis. Overall, Figure \ref{fig:percent_moments} shows that in the redshift range of $0<z<4$, the mean, variance, skewness and kurtosis of the CiC PDFs are more promising in showing differences among RPCDM, $w$CDM and $\Lambda$CDM at high redshifts than at low redshifts. The percentage differences in mean and variance of the CiC PDFs do not depend strongly on scale between $2-25h^{-1}$Mpc. The percentage difference in skewness is a smooth function of redshift on $2-10h^{-1}$Mpc scales and increases as a function of scale. The percentage differences in kurtosis are more pronounced on small scales between $2-6h^{-1}$Mpc.

\begin{figure*}
\plottwo{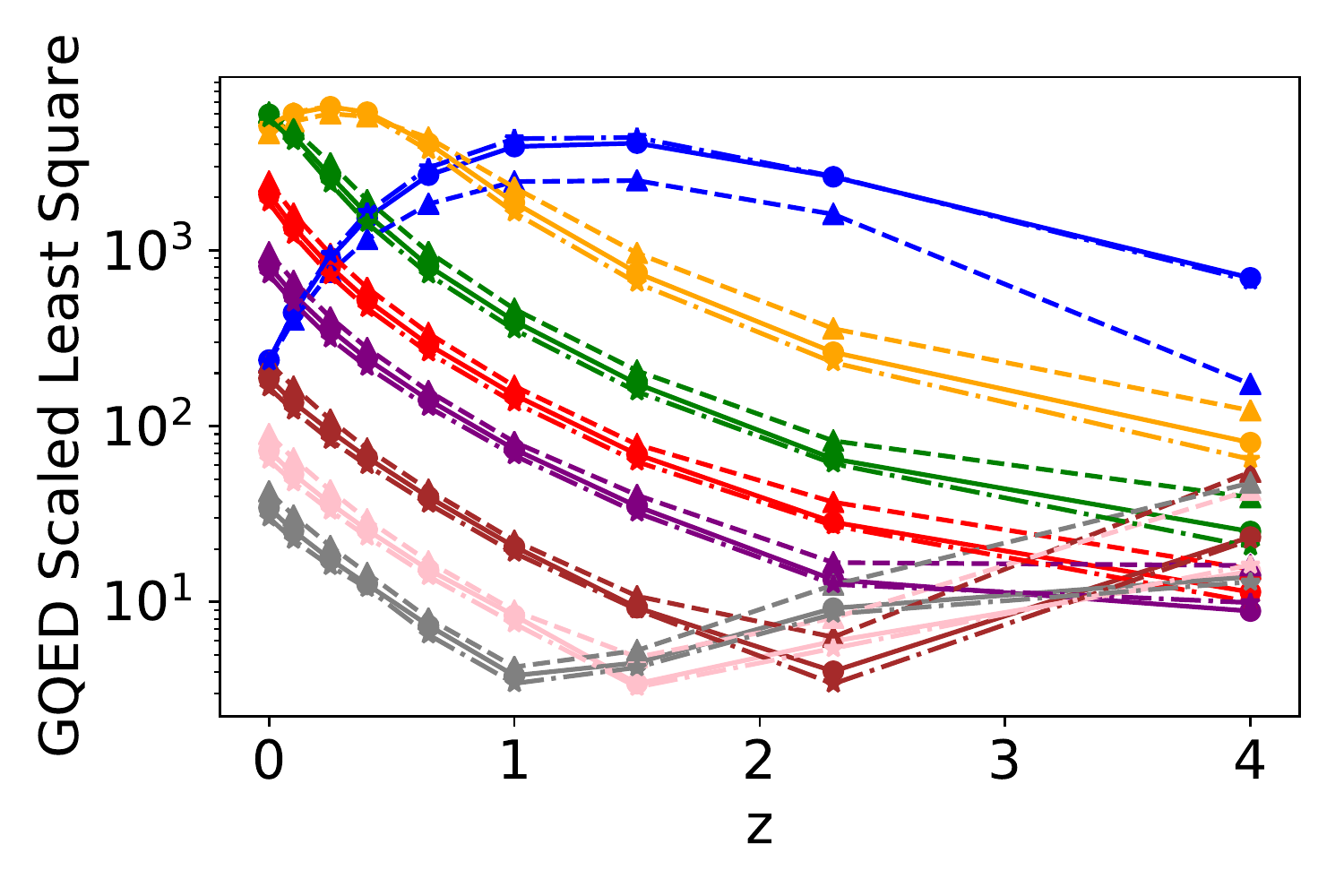}{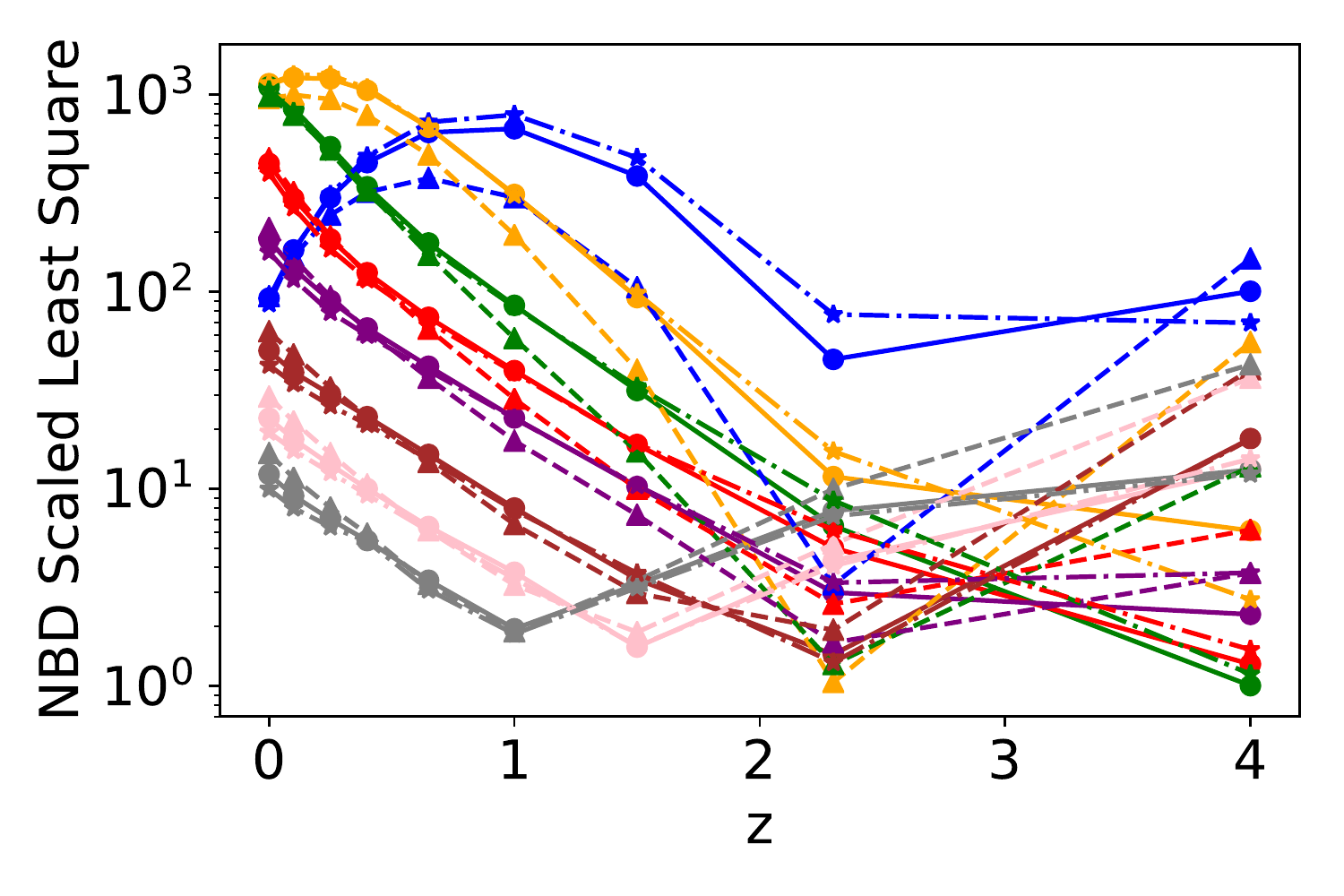}
\plottwo{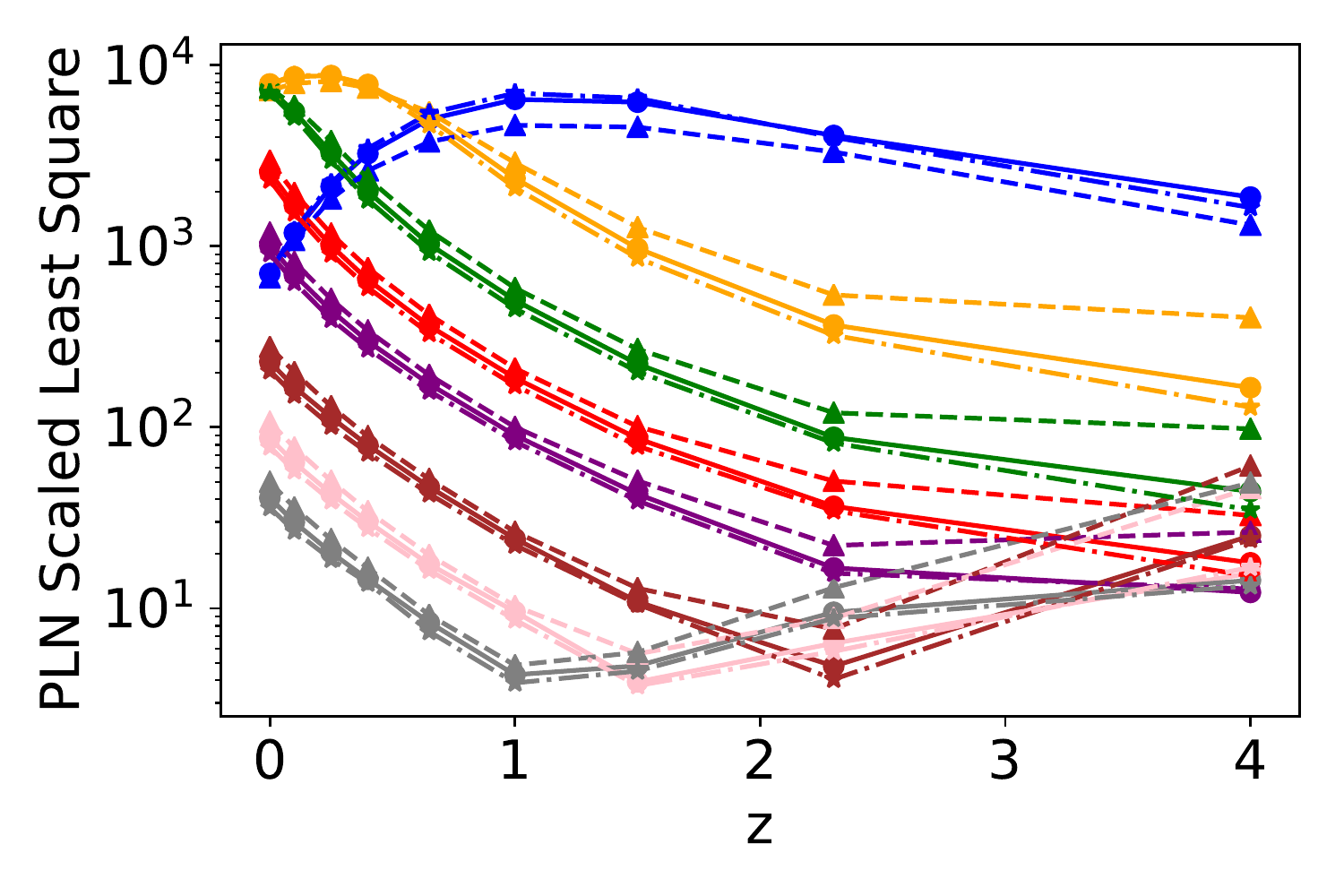}{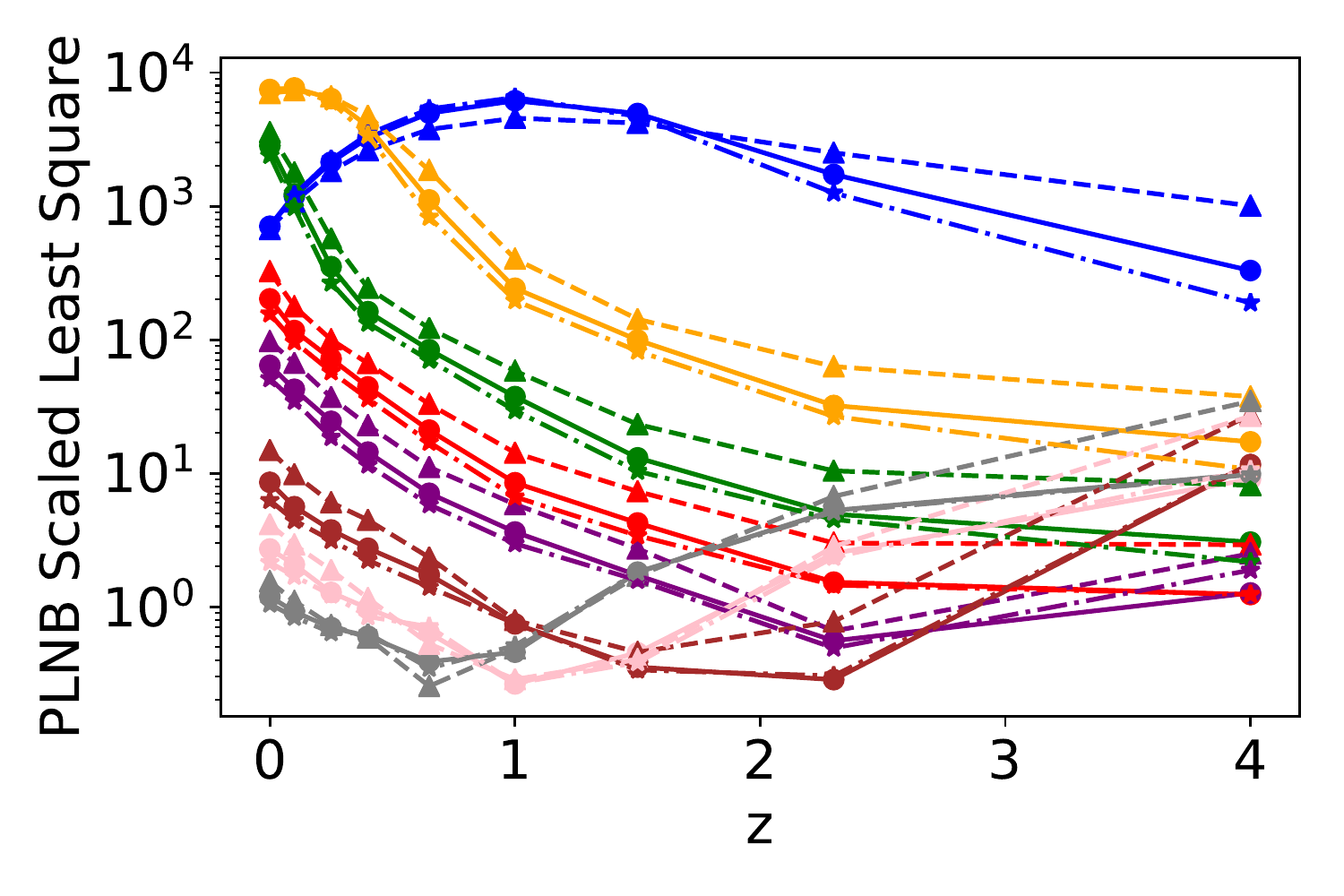}
\caption{Scaled least squares of the best-fit GQED, NBD, PLN and PLNB for the $\Lambda$CDM (solid line with dots), RPCDM (dash line with triangles) and $w$CDM (dash dot line with stars) as a function of redshifts. Colors represent the same cell radii as those in Figure \ref{fig:cic_z}.}
\label{fig:ls}
\end{figure*}

\begin{figure*}
\plottwo{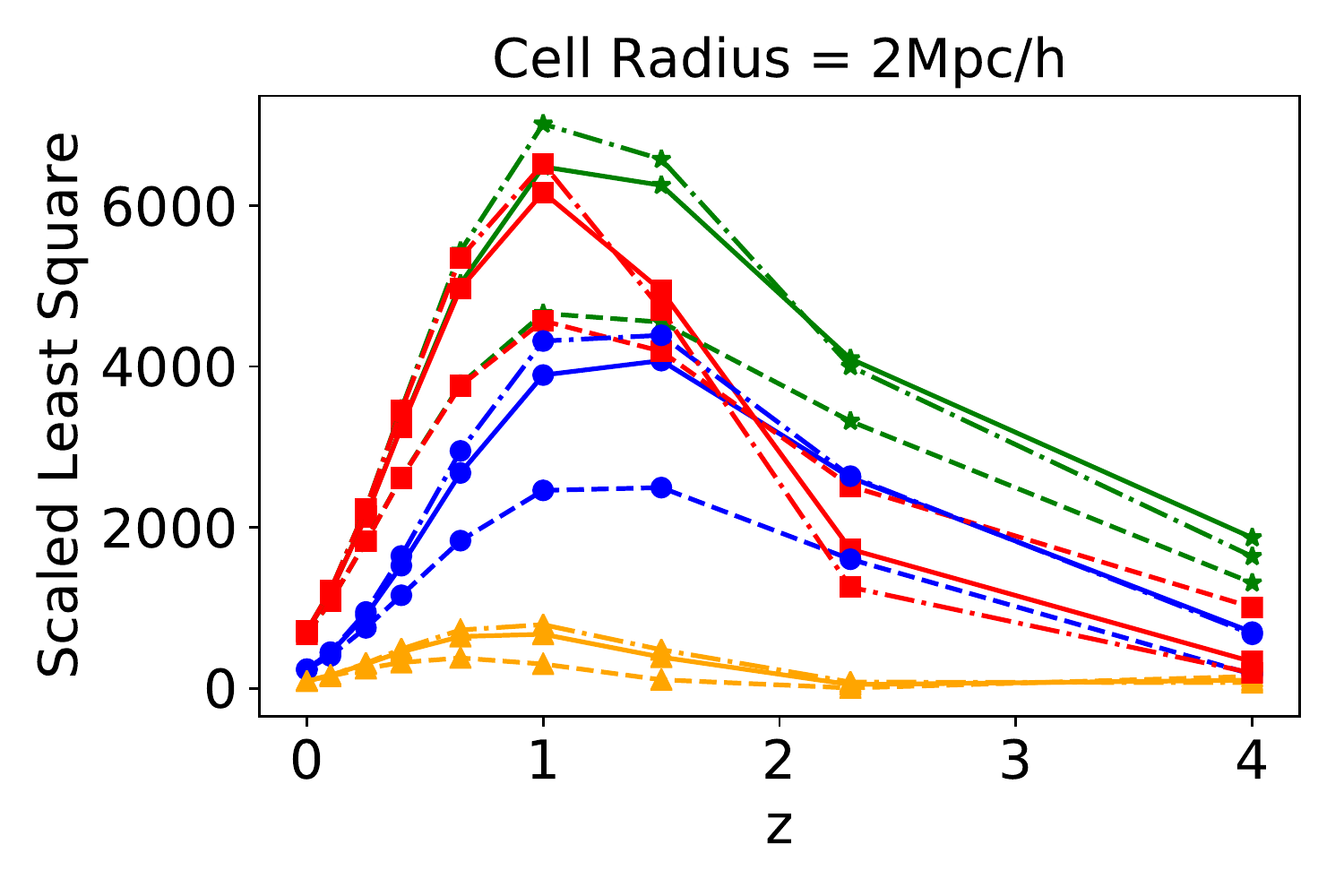}{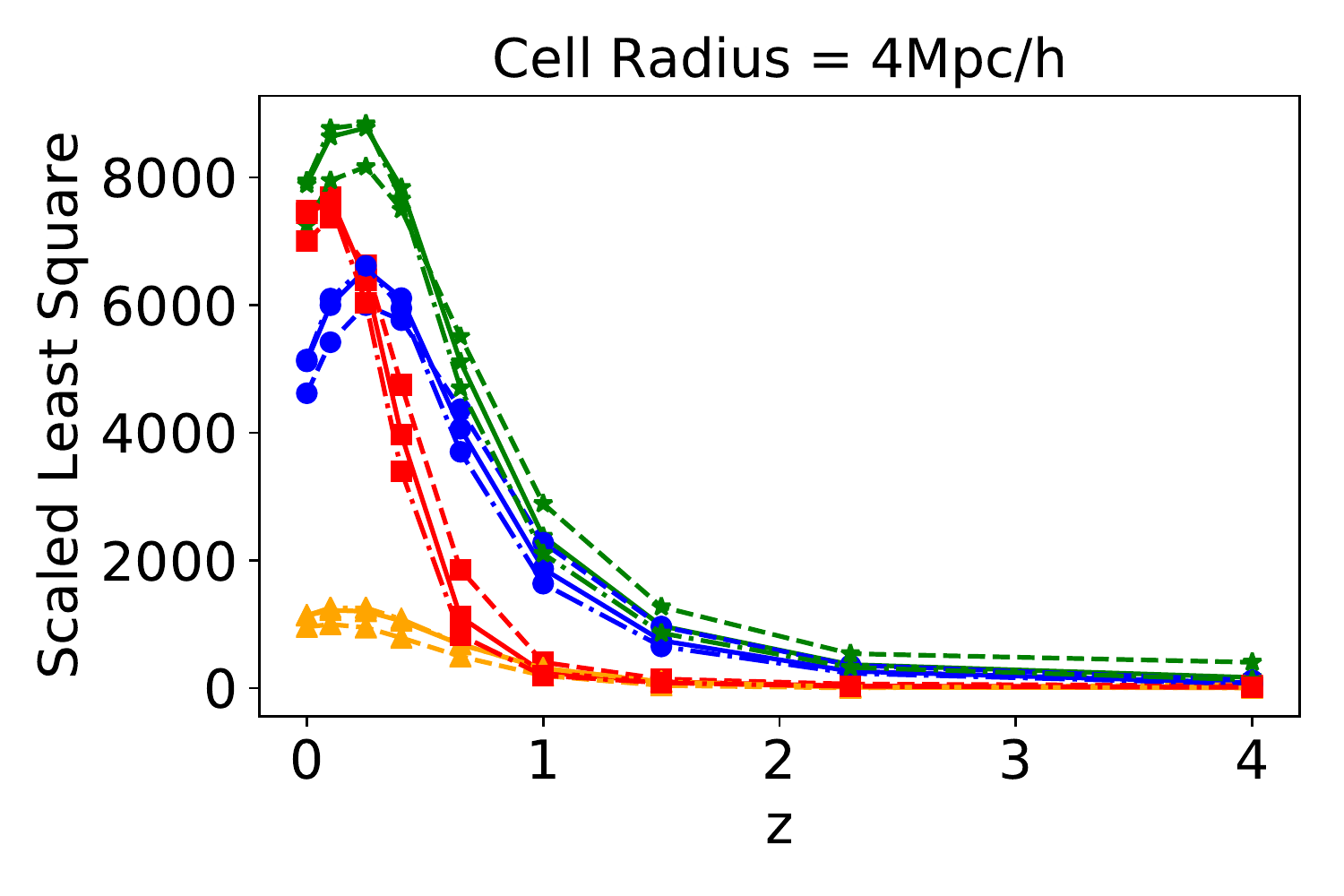}
\plottwo{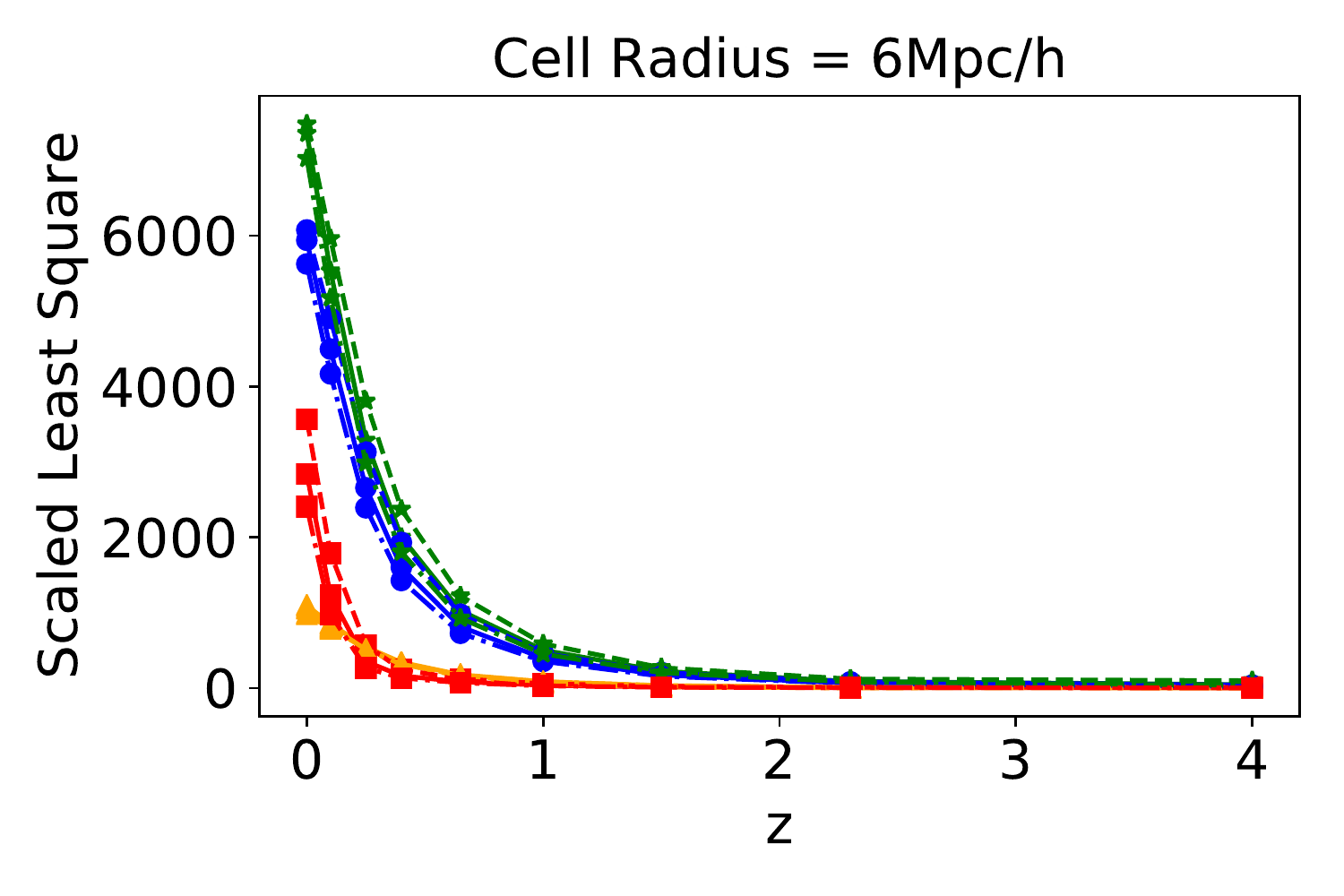}{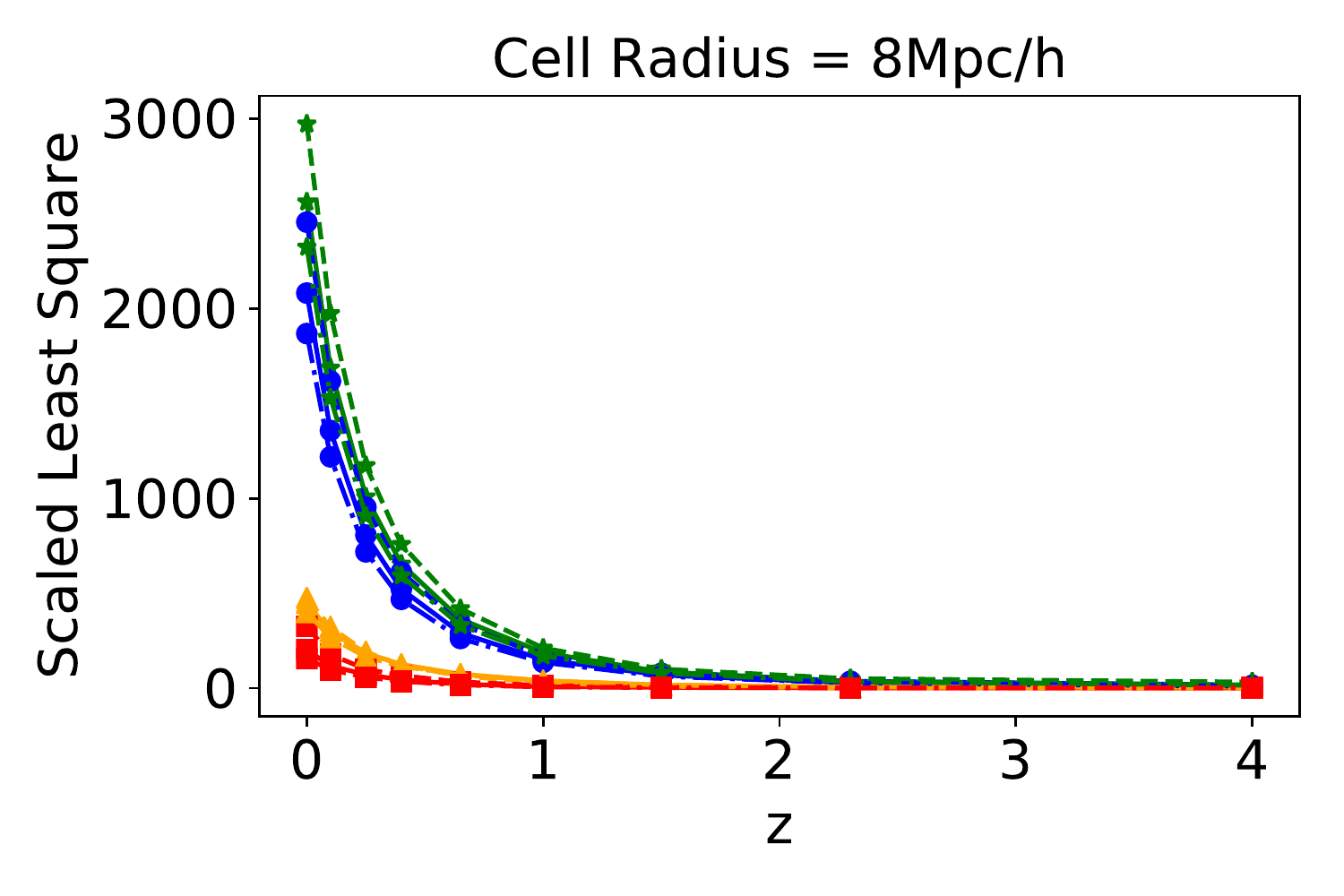}
\plottwo{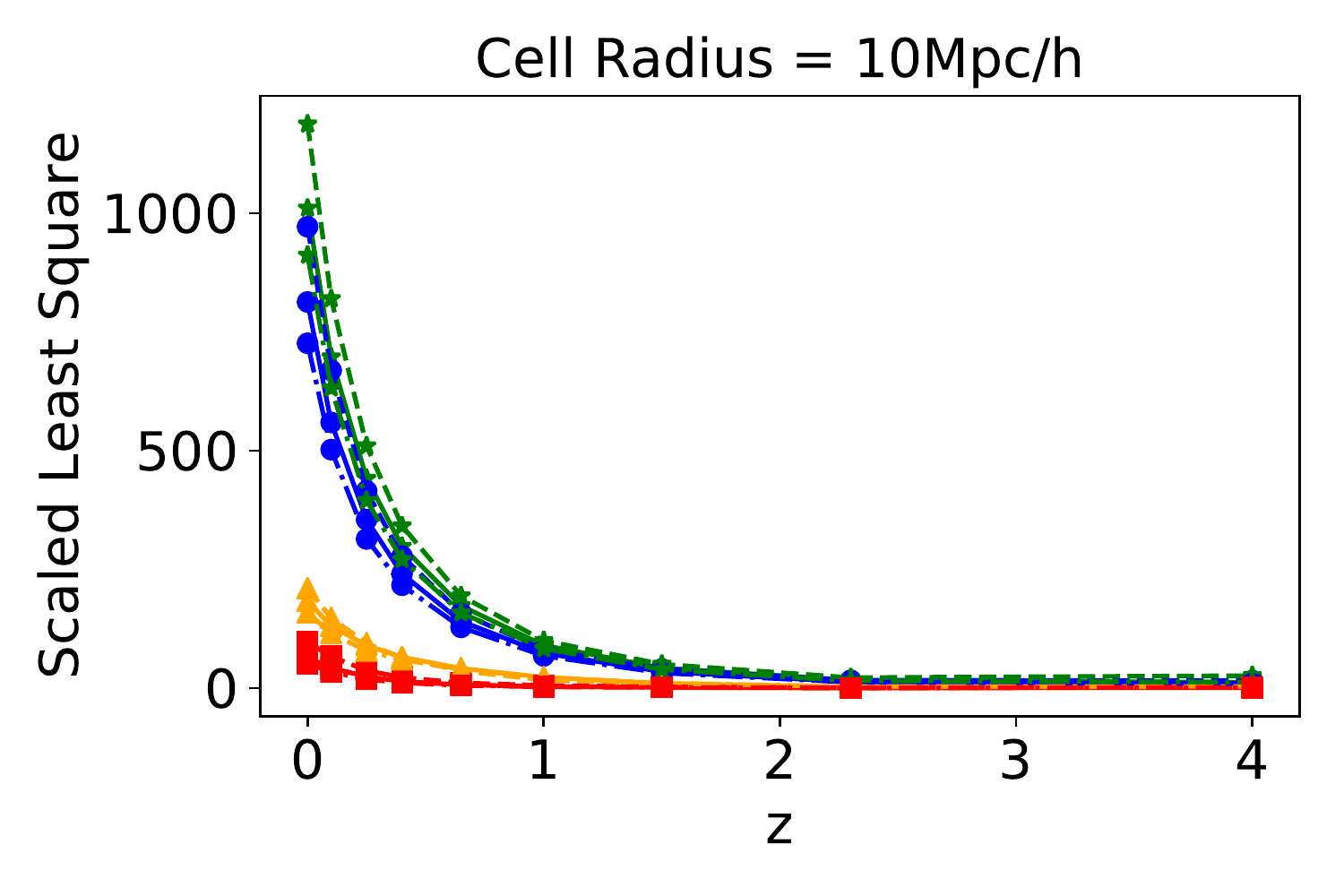}{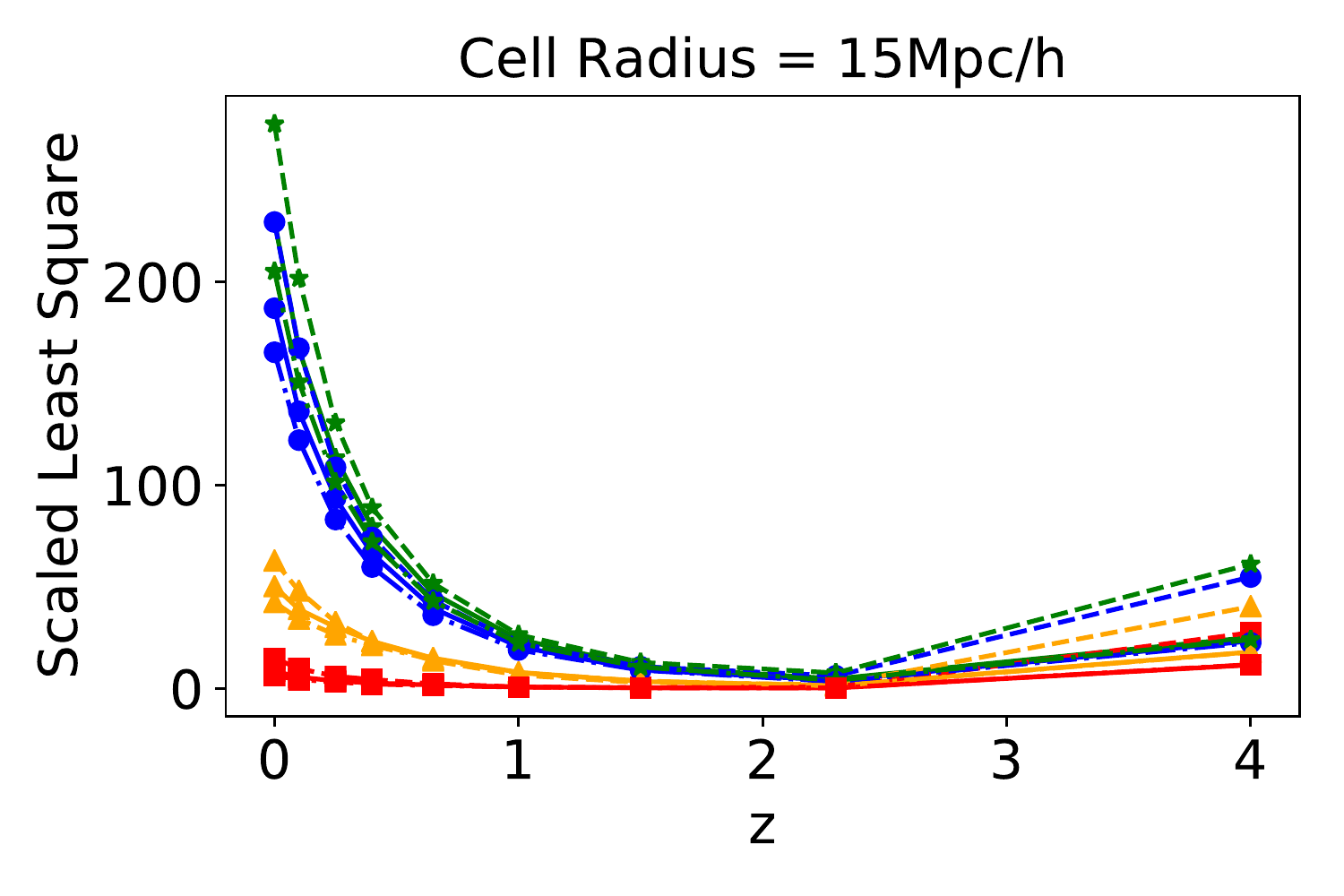}
\plottwo{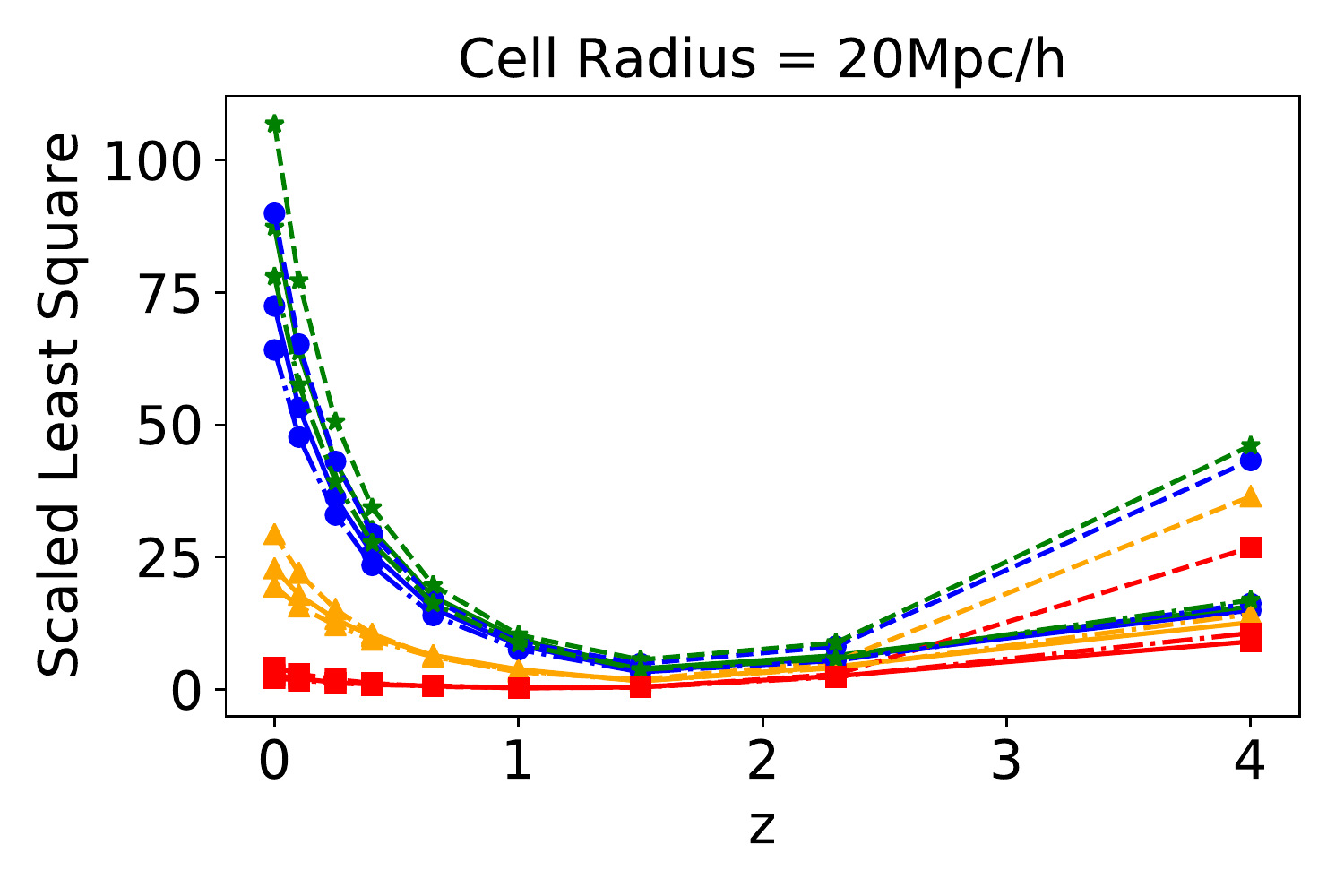}{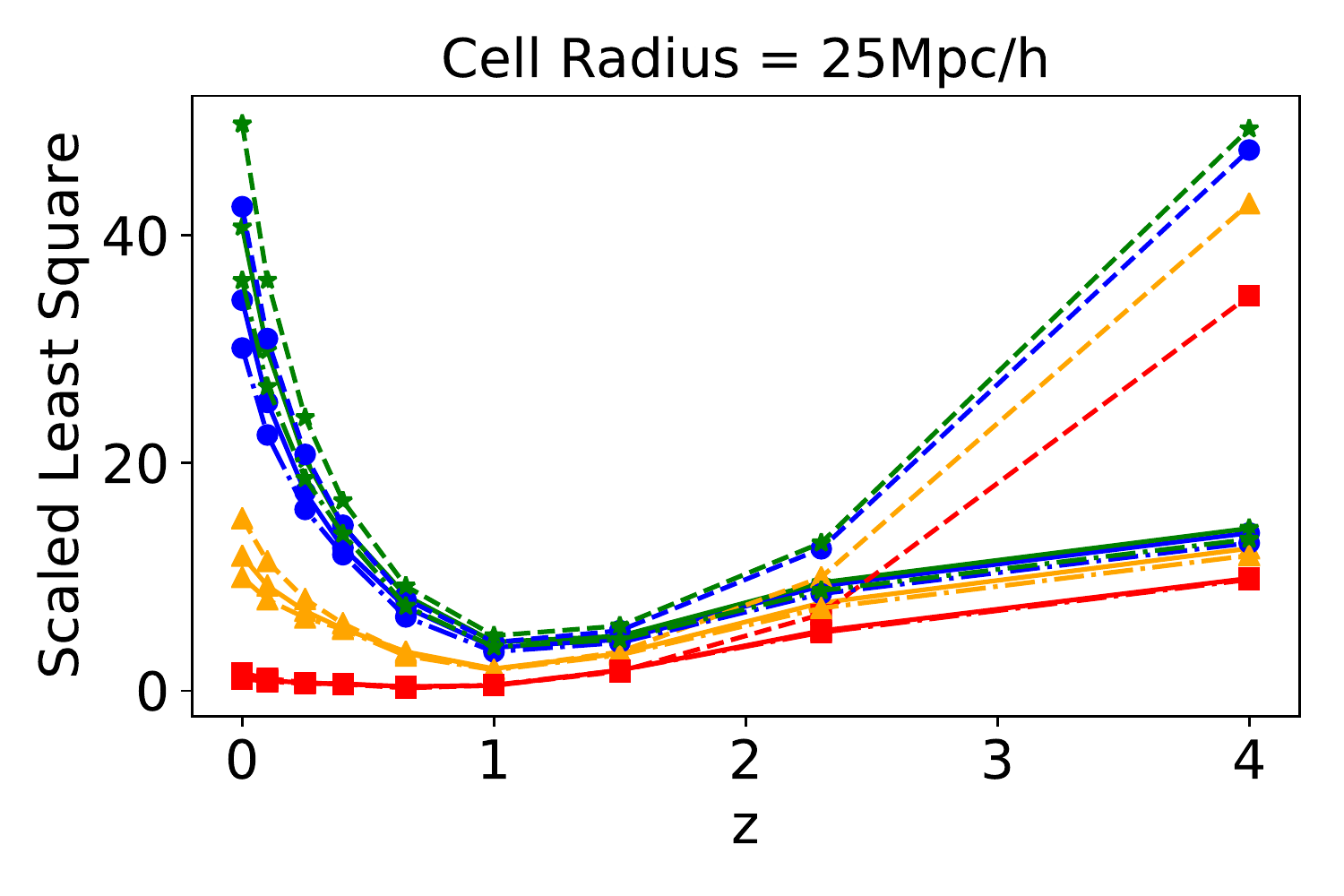}
\caption{Scaled residual least squares for cell radius 2-25 Mpc/h for the $\Lambda$CDM (solid line), RPCDM (dash line) and $w$CDM (dash dot line) cosmologies as a function of redshift. The colors represent the analytical models: GQED (blue dots), NBD (orange triangles), PLN (green stars) and PLNB (red squares).}
\label{fig:ls_cell}
\end{figure*}

\subsection{Modeling the CiC}
The parameters $\bar N$ and $b$ obtained from the best fit of the GQED model (Equation \ref{eq:gqed}) to the measured CiC PDF $f(N)$ are plotted in Figure \ref{fig:gqed} as a function of cell radius and redshift; this Figure shows the fitted parameter values for the RPCDM and wCDM cosmologies and their percentage difference relative to the same fitted parameters for the $\Lambda$CDM cosmology. Comparable plots for best-fit NBD parameters $n$ and $g$ (Equation \ref{eq:nbd}), PLN parameters $\bar N$ and $\omega$ (Equation \ref{eq:pln}), and the PLNB parameters ${\bar N}$, $b$, and $C_b$ (Equation \ref{lnbpdf}) are shown in Figures \ref{fig:nbd}, \ref{fig:pln}, and \ref{fig:plnb} respectively.

Examination of these figures shows that in all four fitted models, the parameters related to the mean halo count (either $\bar N$ or $n$) all show very similar trends in percentage differences with redshift relative to $\Lambda$CDM as the first-moment (mean) of the CiC $f(N)$ computed directly and shown in Figure \ref{fig:percent_moments}. The best-fit $b$ in GQED increases and then decreases as a function of redshift, with larger $b$ in large cells; we remind the reader that the $b$ parameter in the GQED model (Equation \ref{eq:gqed}) is a clustering rather than a bias parameter. The percentage difference of best-fit $b$ between RPCDM and $\Lambda$CDM grows positively but decreases strongly to negative amplitudes as redshift increases. In Figure \ref{fig:gqed} the smaller cells with $R=2h^{-1}$Mpc show more than -18\% difference and the largest cells with $R=25h^{-1}$Mpc show about 5\% difference at $z=4$. The percentage differences in $b$ between RPCDM and $\Lambda$CDM at $z<2.3$ and $w$CDM and $\Lambda$CDM at $z<4$ are all below 2\%. Only RPCDM at very high redshift shows significant percentage difference compared to $\Lambda$CDM. The best-fit $g$ in the NBD is smaller for larger cells and decreases as a function of redshift. The percentage difference of $g$ between RPCDM and $\Lambda$CDM increases from about 8\% to about 50-80\%  and from a few percent negative to more than -10\% between $w$CDM and $\Lambda$CDM. Larger cells generally show larger percentage differences in $g$. The trend for $\omega$ in PLN is very similar with the percentage differences roughly halved. In PLNB, the best-fit bias parameter $b$ generally increases then decreases as a function of redshift. The bias and the percentage differences in bias between alternative dark energy models and $\Lambda$CDM is generally larger in larger cells. The percentage difference in bias in RPCDM and $w$CDM have opposite signs relative to $\Lambda$CDM, ranging in absolute magnitude from zero to 40\%. The best-fit variance parameter $C_{b}$ in PLNB decreases as a function of increasing redshift and cell radius. RPCDM and $\Lambda$CDM show significant percentage difference (more than 100\%) in the variance parameter $C_{b}$, whereas $w$CDM and $\Lambda$CDM show less than 50\% absolute magnitude difference in $C_{b}$. The percentage differences for the bias and the variance parameters in PLNB are greater for large scales up to $25h^{-1}$Mpc at $1<z<4$.

The bias and the variance parameters in the PLNB fit were found to be effectively mathematically degenerate during model fitting. The \textsw{MPFIT} algorithm could not then converge to a global minimum for large cells at high redshifts. To confirm the degeneracy, we fixed $\bar{N}$ and evaluated the sums of squares for different bias and variance parameters on a grid. A two-dimensional grid search for $b$ and $C_{b}$ to locate the global minimum showed that the smallest sums of squares fell on a monotonic curve relating these two parameters. An arbitrarily small variance and a large bias (and vice versa) can therefore yield the same sum of squares. As a result, when the fitted bias was larger than 10 or reached an upper bound of 10, we took this as evidence of strong degeneracy and we used radial basis functions to perform linear and cubic interpolation over existing converged solutions to obtain the bias and variance parameters instead. The residuals of the linearly interpolated parameters were found to be smaller and to vary more uniformly. The linearly interpolated variance could be negative in this approach however, which is not physical, so we adopted the linearly interpolated bias $b$ and fixed its value when fitting the mean count $\overline{N}$ and variance $C_b$. The interpolated valued of $b$ and the fitted $C_b$ are included with other successful three-parameter fits in Figure \ref{fig:plnb}. For the parameter range where degeneracy is not dominant we found the Levenberg-Marquardt algorithm to be robust to choices of initial parameter values. The cases of PLNB with linearly interpolated bias $b$ and fitted $C_b$ and $\bar{N}$ give the smallest least squares among all four analytical models considered.

Figures \ref{fig:ls} and \ref{fig:ls_cell} show the scaled least squares (Equation \ref{eres-lsqdist}) for all four models as a function of redshift, cell radius and dark energy model. The best-fit model for the halo CiC distribution is the NBD model for $R<6h^{-1}$Mpc and the PLNB model for $R>6h^{-1}$Mpc. The choice of dark energy model does not appear to influence significantly the choice of the best-fit analytical model. For $R=6h^{-1}$Mpc, the best-fit model transitions from NBD to PLNB for $z>0.25$. The quality of fits are generally the best at high redshifts for small cells $R<10h^{-1}$Mpc and at medium redshifts for large cells $R>15h^{-1}$Mpc. This improves monotonically as a function of redshift for $R>6h^{-1}$Mpc. For $R<6h^{-1}$Mpc, the scaled least squares have a peak as a function of redshift. In the cell radius range $2\leqslant R\leqslant 25h^{-1}$Mpc, the PLN model always has the lowest residual least squares compared to the other three models. For $R>8h^{-1}$Mpc, the ranking of the best-fit model is always PLNB, NBD, GQED, and PLN. The GQED ranks second for small cells $R<4h^{-1}$Mpc at low redshifts after NBD as the best-fit. The CiC distribution $f(N)$ is most accurately modeled by the GQED, NBD and PLN for cells $R=25h^{-1}$Mpc at $z=1$, $R=20h^{-1}$Mpc at $z=1.5$ and $R=15h^{-1}$Mpc at $z=2.3$. It is most accurately modeled by PLNB for cells $R=25h^{-1}$Mpc at $z=0.65$, $R=20h^{-1}$Mpc at $z=1$ and $R=15h^{-1}$Mpc at $z=2.3$. In Figure \ref{fig:ls_cell}, for cell $R=2h^{-1}$Mpc, RPCDM is best fit by the four models, but for larger cells RPCDM produces worse fits than $\Lambda$CDM and $w$CDM. Only cell radius $R=2h^{-1}$Mpc shows significant differences in the quality of fits between various dark energy models. The differences in the quality of fits between various dark energy models are smaller than the differences in the quality of fits between the analytical models. The best-fits of models for small cells $R=2h^{-1}$Mpc are at low and high redshifts. The best-fits of models for cells $R=4-10h^{-1}$Mpc are at high redshifts. The best-fits of models for large cells $R=15-25h^{-1}$Mpc are at medium redshift ($1<z<2.3$). These results are discussed further in Section \ref{sec:discussion}.

\section{Discussion}\label{sec:discussion}
\subsection{Gravitational Clustering Timescale}
Consider the Friedmann equation for $\Lambda$CDM \citep{peacock99}
\begin{equation}
    H^{2} = \frac{8\pi G}{3} \rho + \frac{\Lambda c^{2}}{3}
\end{equation}
where $H$ is the Hubble parameter, $\rho$ is the energy density and $\Lambda$ is the cosmological constant. The timescale for the dark matter halos to merge in an expanding universe can be approximated by the gravitational timescale
\begin{equation}
    \tau_{grav}^{-2} \equiv \frac{8\pi G}{3} \rho = H^2-\frac{\Lambda c^2}{3} = H^2(1-\Omega_{\Lambda})
\end{equation}
where $\Omega_{\Lambda}$ is the density parameter for the cosmological constant. Because $\Omega_{\Lambda}>0$, the gravitational timescale $\tau_{grav}$ should always be longer than the universe's expansion timescale $\tau_{Hubble} = 1/H$. In the thermodynamic description of galaxy clustering that gives rise to the GQED \citep{saslaw00}, the assumption for quasi-equilibrium evolution holds if the macroscopic timescale, $\tau_{macro}$, exceeds the microscopic timescale $\tau_{micro}\approx \frac{L}{v}$, where $L$ and $v$ are characteristic microscopic lengths and velocities respectively. Here
\begin{equation}
    \tau_{macro}^{-1} = 2H(1-b)
\end{equation}
where $H$ is the Hubble parameter and $b$ is the clustering parameter in Equation \ref{GQED}. The microscopic timescale $\tau_{micro} \ll \tau_{Hubble}$ in the nonlinear regime and $\tau_{micro} \approx \tau_{Hubble}$ in the linear regime \citep{saslaw00}. The CiC distribution of dark matter halos is a result of halo formation, clustering, merging and the expansion of the universe. Merging halos reduces the number of halos and decreases the two-point correlation function of halos. The effect of merging is more prominent when the halo number density is higher at larger redshifts. The expansion of the universe also reduces the number density of halos and reduces halo correlation functions, but its effect is only dominant at low redshifts after the expansion accelerates. The value of $b$ in the GQED model increases as halos cluster gravitationally, but also decreases because of the competing effects of halo merging and the expansion of the universe. Assuming $\Omega_{\Lambda}=1-\Omega_{m}=1-0.2573=0.7427$, then $\tau_{macro} > \tau_{grav}$ if $b>0.746$ and $\tau_{macro} > \tau_{Hubble}$ if $b>0.5$. In Figure \ref{fig:gqed}, we see $b>0.746$ for cell radii $R>6h^{-1}$Mpc. This means at scales larger than $6h^{-1}$Mpc, the degree of gravitational clustering for a many-body system in $\Lambda$CDM evolves slower than the expansion of the universe and slower than halo formation through merging. In contrast, $b<0.746$ for cell radius $R=2h^{-1}$Mpc at $0<z<4$. At this non-linear scale, local gravitational collapse and mergers happen faster than the global evolution of gravitational clustering. When a halo forms through merging, the number of halos changes for the whole thermodynamic system, so the system changes from one system in quasi-equilibrium to a new system with a different number of halos. Gravitational clustering takes longer to respond to abrupt changes due to mergers, so at small scales the deviation from equilibrium is larger. As a result, the GQED tends to fit less accurately at smaller scales than at large scales in Figure \ref{fig:ls}.

\subsection{The Uncertainties of CiC Measurement}
The CiC distribution measured by our oversampling CiC algorithm has small uncertainties from uncertain counts near cell boundaries. The small jackknife errors indicate that the CiC distribution function measurements are quite precise. The overlapping cells sample the same volume in the simulation box many times and produce very smooth CiC distributions. However, the overlapping cells are inherently correlated, so the deleted jackknife subsamples are not independent and may underestimate the errors. Also, excluding incomplete cells near the edges of the simulation box may introduce additional errors, because the simulation box is then no longer uniformly sampled. This edge effect is larger at high redshifts when a significant fraction ($\sim40\%$) of cells are excluded. The fact that the resultant CiCs at high redshifts are still quite smooth means that the simulation box is large enough. By using the multi-precision library for model evaluation, the uncertainty due to evaluating complex functions should be negligible. The fitting parameters obtained are robust against varied initial guesses as long as the parameters are not degenerate. Only the PLNB model has degenerate parameters in certain parameter ranges and requires appropriate numerical treatment as described above.

\subsection{The Best-Fit Model of CiC}
Among the four analytical models we fitted to the measured CiC PDF $f(N)$, there is no model that fits the CiC best universally in all ranges of cell radii and redshifts. The CiC is best described by the NBD model for cells with $R<6h^{-1}$Mpc and by the PLNB model for cells $R>8h^{-1}$Mpc. Based on comparisons of the residual least squares (Figure \ref{fig:ls_cell}), these individual dark energy models do not favor any particular analytical model of CiC. We find that the CiC $f(N)$ is best modeled for all four analytical models with large cells at medium redshifts. Figure \ref{fig:ls_cell} indicates that various dark energy models show significant differences only at a scale of $R=2h^{-1}$Mpc.

All these various models have ranges of the counts-in-cells distribution function where they more or less agree with the simulations and with each other, as judged by their least squares differences. Yet as \cite{tufte83} discusses, statistical agreement, even (or especially) with many parameters, does not convey the full import of data. Therefore we would be cautious in claiming that small differences in distribution functions are meaningful.

To help overcome this problem it is useful to relate these statistics to the fundamental physical properties of halo clustering. If the statistics need to be modified, so do the physical properties. This is easier to understand in the case of the GQED since its physical origin is much less obscure than for the NBD, PLN and PLNB distributions. This may be because in the GQED only one force, gravity, dominates the evolution.

\subsection{Effects of Dark Energy on CiC}
An important factor concerning the differences shown by different dark energy models lies in the cosmological parameters chosen for this set of simulations (Table \ref{table:DEUSSALL1}). Bouillot et al. chose degenerate parameters the mean cosmic mater density, $\Omega_{m}$, the dark energy equation of state, $w$, and the root-mean-square fluctuation amplitude of the density contrast at 8 $h^{-1}$ Mpc, $\sigma_{8}$ within the marginalized $\sim$ 1 $\sigma$ contour that fits the cosmic microwave background anisotropy power spectra and the luminosity distances measurements to supernova Type Ia \citep{bouillot15}. The Bouillot et al. simulation aims to break the degeneracy between these cosmological parameters by testing non-linear clustering. The accelerated expansion in $\Lambda$CDM is stronger than in RPCDM, so to fit the observations as well as in the $\Lambda$CDM, the smaller acceleration in RPCDM is compensated by a larger dark energy density than is required. This is reflected in the smaller matter density, $\Omega_{m}$, in RPCDM \citep{alimi10}. The opposite is true for $w$CDM. As a result, the CiC mean, as well as the mean number count in the fitted analytical models, all show fewer halos in RPCDM and more halos in $w$CDM throughout their cosmic history.

Figure \ref{fig:cic_z} shows that the CiC distributions measured with the same physical cell radius R have an increasing average halo count and an increasing most probable halo count from $z=0$ to $z=4$ for all three cosmologies. This is mainly because the matter density at a higher redshift is higher and a spherical cell with the same physical volume encloses a larger comoving volume at a higher redshift. In the same comoving volume, more bound halos should have formed at a lower redshift, but this increase in the number of halos is clearly offset by the expansion of the universe at $z\leqslant4$. At sufficiently high $z$, such as $z>4$, large halos above our minimum mass criterion are very few, hence reversing the trend of increasing halo count with increasing redshift. The differences in halo counts between RPCDM or $w$CDM and $\Lambda$CDM are larger at higher redshifts.

Apart from the effect of various dark energy models on the number density of halos through changing the expansion history, a more subtle effect on the degree of clustering as a function of redshift was also shown in the halo CiC distributions. The best-fit GQED $b$ parameter as an indicator of the degree of clustering is found to be closer to unity for larger cells. Because $b$ is an integral of the two-point correlation function within the scale of the cell, the values of $b$ closer to 1 in larger cells are expected. The absolute difference of the best-fit $b$ between $w$CDM and $\Lambda$CDM is very small across all redshifts and cell radii. The largest percentage difference of $b$ between $w$CDM and $\Lambda$CDM is found to be about 2\% with a cell radius of $2h^{-1}$Mpc at $z=4$. Therefore, $b$ is not very useful for distinguishing $w$CDM and $\Lambda$CDM. The difference in $b$ between RPCDM and $\Lambda$CDM is more obvious. The absolute difference between $b_{RPCDM}$ and $b_{\Lambda CDM}$ is larger with smaller cells and at higher redshifts. It is up to about 18\%. It is interesting to note that the percentage differences between $b_{RPCDM}$ or $b_{wCDM}$ compared to $b_{\Lambda CDM}$ change sign at $z\sim 1.7$, (except for RPCDM with $R=2h^{-1}$Mpc cells). In other words, at the turning point $z\sim 1.7$, all three dark energy models have nearly identical best-fit values of $b$ for all CiC with $R=4$ to $25h^{-1}$Mpc. At lower $z<1.7$, $b_{RPCDM}$ is most distinguishable at $z=0.65$ in cells $R=4h^{-1}$Mpc with a percentage difference of about 2\%. In Figure \ref{fig:gqed} it is worth noting that the turning point for $b$ occurs at a higher redshift for $w$CDM and at a lower redshift for RPCDM compared to $\Lambda$CDM. Our method makes a clear prediction for the redshift at which $b$ reaches the maximum for different dark energy models. CiC in galaxy surveys should be able to test our prediction.

The NBD parameter $g$ can be interpreted as another clustering parameter that is approximately equivalent to the two-point correlation \citep{yang11}. The absolute difference between $g_{RPCDM}$ and $g_{\Lambda CDM}$ is larger than the absolute difference between $g_{wCDM}$ and $g_{\Lambda CDM}$ for a given cell radius and redshift. Both tend to be more significant for smaller cell radii across all redshifts. In the last panel of Figure \ref{fig:nbd} the percentage differences of $g_{RPCDM}$ and $g_{wCDM}$ compared to $g_{\Lambda CDM}$ are both larger for large cell radii and increase to their maxima at $z\sim2.3$ to about 48\% difference for RPCDM and  18\% for $w$CDM. The substantial differences in the evolution history of clustering indicated by $b$ and $g$ as a result of various dark energy models suggest that the clustering parameters as a function of redshift are promising probes for distinguishing RPCDM and $w$CDM from $\Lambda$CDM.

The comparison between PLN and PLNB clearly shows that if the matter density is modeled by a log-normal distribution, bias must be considered in order to model the CiC accurately. The bias parameter in PLNB does not take the scale dependence of bias and nonlinear bias into account, so a better model of bias would be necessary to improve the accuracy of the PLN or the PLNB. The PLNB, being the best-fit model to the CiC in large cells, shows about 40\% differences in $b$ and larger than 100\% differences in $C_{b}$ between RPCDM and $\Lambda$CDM. The variance $C_{b}$ in the PLNB increases from high redshift to low redshift as matter density fluctuations grow. The bias parameter clearly shows a scale dependence, but does not vary much as a function of redshift at small scales. The bias at large scales tends to decrease towards lower redshifts. The PLN and the PLNB both show larger percentage differences in larger cells at higher redshifts. The trends of the parameters in PLNB are not as smooth as in the other models, which may make clearer predictions. 

\subsection{Relations to Observations}
Our CiC methods and model fitting can be applied to current and future galaxy surveys. The trends of CiC and their best-fit models are expected to be similar for galaxies, but whether the large percentage differences would be measured in the CiC of galaxies might be less certain for a number of reasons. Firstly, the galaxy surveys may not have large enough samples of galaxies at $z>2.3$ or $z>4$ to measure the very large percentage differences between dark energy models precisely. At very high redshifts, the galaxy samples are not complete samples because very faint galaxies are below detection limits. 

Secondly, the percentage differences between dark energy models may be smaller when cylindrical cells are used for galaxy surveys, which are subject to photometric redshift uncertainties and redshift-space distortions. These cylindrical cells are circular cells elongated in the redshift direction. The CiC measured with the cylindrical cells are redshift-averaged, whereas our measurements of simulations are snapshots at particular redshifts. CiC measurements done in 3D cells and 2D projected cells are not directly comparable. We will explore 2D CiC in redshift-space in further detail in future work. The redshift-averaged CiC will only be close to our simulation measurements if the redshift thickness is small. However, the galaxies in a thin redshift bin may not be a large enough sample to give reliable results. Furthermore, there are larger uncertainties associated the photometric redshifts that make galaxy samples in very thin redshift bins even more uncertain. The count of objects per steradian per unit increment in redshift can be approximated by $\frac{dN}{dz}=\frac{n_0}{H_0}\frac{r^2(z)}{\sqrt{\Omega_m(1+z)^3+\Omega_{\Lambda}}}$, where $r(z)=\frac{1}{H_0}\int_0^z \frac{dz^{\prime}}{\sqrt{\Omega_m(1+z^{\prime})^3+\Omega_{\Lambda}}}$ \citep{peebles93}. Assuming that photometric redshift uncertainties and redshift-space distortions produce $\sigma_z=0.002(1+z)$, the fractional galaxy count uncertainty $dN/N$ for a redshift bin $0.4<z<0.6$  is about 2.8\%. Assuming a larger $\sigma_z=0.02(1+z)$ for a redshift bin $0.8<z<1.2$, $dN/N\approx20\%$. Assuming a small $\sigma_z=0.002(1+z)$ at high redshift $3.8<z<4.2$, $dN/N\approx5\%$. The relative uncertainty in the galaxy count for the DES Science Verification catalog at $0.1<z<0.5$ is about 5\% \citep{clerkin17}. The requirement for Year 10 LSST large-scale structure analysis is that the systematic uncertainty in the mean redshift of each tomographic bin shall not exceed $0.003(1 +z)$ and the systematic uncertainty in the photometric redshift scatter $\sigma_z$ shall not exceed $0.03(1 +z)$ \citep{lsstdesc18}. The average number of galaxies in our fitting models should have similar fractional uncertainty as the above values of $dN/N$. The uncertainties and percentage differences of the clustering parameters of our fitting models for different cosmologies must be found by fitting the CiC of galaxies. Our predicted large differences between dark energy models at $z>1$ should still be detectable if the galaxy counts have 5\% uncertainties due to photometric redshift uncertainty. Further redshift-space CiC studies with simulations that sample the cosmological parameter space more densely would better forecast the detectability of the differences from different dark energy models in future surveys and are a target of our future research.

Lastly, the process of removing satellite galaxies may introduce more uncertainties. Once a halo merges into a larger halo, its count in our measurement will become zero. This means that for galaxy CiC, satellite galaxies in galaxy clusters that reside in smaller halos that previously merged into larger halos would not be counted. From a theoretical standpoint, not considering subhalos or satellite galaxies is easier, because the stochasticity of the halo occupation number is no longer a concern. However, for observations, it is harder to decide if a galaxy belongs to a galaxy cluster, especially without the spectroscopic redshifts of the galaxies in the cluster.

CiC for galaxy surveys do not need accurate mass determination, especially when compared to the CiC distribution of galaxy-sized halos in simulations, because the observable galaxies are above the minimum observable mass for galaxy-mass halos and dwarf galaxies. The large sample of galaxies and the easily obtained discrete number of galaxies make CiC measurements easier to implement than other probes of dark energy mentioned in Section \ref{sec:intro}.

\section{Conclusions} \label{sec:conclusions}
We have developed a new technique using the counts-in-cells (CiC) probability distribution functions (PDFs) of dark matter halos to distinguish dark energy models. We have compared the CiC statistics and their models on $2-25h^{-1}$Mpc scales at $0<z<4$ for dark matter halos more massive than $2.4\times10^{11} M_{\odot}$ in the DEUS simulations. In comparison to $\Lambda$CDM, the RPCDM and $w$CDM show larger percentage differences in the best-fit model parameters of the CiC PDFs at high redshifts up to $z=4$. These are consistent with the trends for the CiC mean, variance, skewness and kurtosis. We advocate comparing the best-fit analytical models of the CiC PDFs, because they inherently contain moments of all orders and shed light on the stochastic processes underlying the clustering statistics. This also provides hints about the physical processes behind structure formation and clustering. The physics behind the analytical models needs to be better understood in order to improve the models of the CiC. The analytical models of the CiC fit best at $15-25h^{-1}$Mpc scales between $1<z<2.3$. The best-fit model of the CiC is the NBD for $2-6h^{-1}$Mpc scales and the PLNB for $8-25h^{-1}$Mpc scales. The clustering parameter $b$ in the GQED shows larger percentage differences between the RPCDM and $\Lambda$CDM models for small scales down to $2h^{-1}$Mpc at $z>2.3$. The clustering parameter $g$ in the NBD shows larger percentage differences between the RPCDM or $w$CDM model and $\Lambda$CDM for large scales up to $25h^{-1}$Mpc at high redshifts up to $z=4$. The bias and the variance parameters in the PLNB show larger percentage differences for different dark energy cosmologies on large scales up to $25h^{-1}$Mpc at $1<z<4$.

The percentage differences in mean and variance of the CiC PDFs do not depend strongly on scale between $2-25h^{-1}$Mpc. The percentage difference in skewness is a smooth function of redshift on $2-10h^{-1}$Mpc scales and increases as a function of scale. The percentage differences in kurtosis are more pronounced on small scales between $2-6h^{-1}$Mpc. Quantifying the differences in CiC PDFs with the residuals $\triangle f(N)_{RP-\Lambda}$ and $\triangle f(N)_{w-\Lambda}$, the variance, skewness and kurtosis of these residuals on $2h^{-1}$Mpc scale at $z>2.3$ are the largest and the most different. The skewness and kurtosis of the residuals on $4-6h^{-1}$Mpc scales are larger at low redshift of $z<0.65$. At scales above $8h^{-1}$Mpc at $0<z<4$, the moments of the residuals are too close to zero for distinguishing dark energy models. 

For the cosmologies explored in the DEUS simulations, which are consistent with the CMB and Type Ia supernovaents at the time of the simulation, the mean number of dark matter halos more massive than $2.4\times10^{11} M_{\odot}$ is up to 20\% more at $z=4$ in the quintessence dark energy model and up to 70\% less than in the phantom dark energy model compared to a constant $\Lambda$ dark energy model (Figure \ref{fig:residual_moments}). Comparing to $\Lambda$CDM at $z=4$, the clustering parameter g in the NBD is up to 80\% different in RPCDM and up to 20\% different in $w$CDM (Figure \ref{fig:nbd}). The quantitative conclusions comparing RPCDM and wCDM to $\Lambda$CDM are only for the cosmologies explored in the DEUS simulations, but the methodology is generally applicable to cosmological simulations and galaxy surveys. Our CiC analyses of the simulations provide very useful hints on the scales and redshifts to probe for deviations caused by different dark energy. The moments and model parameters of the CiC PDFs for galaxies on $2-25h^{-1}$Mpc scales at $0.65<z<4$ are more likely to show detectable differences among various dark energy models for future galaxy surveys like the LSST. The methods of CiC analyses that we develop here, using available models and simulations, can be extended to gain considerable insights into future models and observations. Our methodology applied to a suite of simulations that systematically explores a wider range of cosmological parameters would make useful predictions about the minimum detectable deviations and possible degeneracies.

\acknowledgments
We would like to thank the anonymous referee for helpful insights and comments. This research is part of the Blue Waters sustained-petascale computing project, which is supported by the National Science Foundation (awards OCI-0725070 and ACI-1238993) and the state of Illinois. Blue Waters is a joint effort of the University of Illinois at Urbana-Champaign and its National Center for Supercomputing Applications.

\software{MPFIT \citep{markwardt09}, MPFR \citep{fousse07}, GSL \citep{galassi09}, MPI \citep{gropp14}, FOFReaderLib \citep{pasdeloup16}, Matplotlib \citep{hunter07}, Numpy \citep{vanderwalt11}}

\appendix\section{CiC Resolution Study\label{appendix}}
We perform a resolution study of the CiC methodology to understand the effects of uncertain counts due to the floating point number precision limit of the halo positions on the resulting CiC probability distribution function and to choose an appropriate CiC resolution. Cell radii of 10 and 25 $h^{-1}$Mpc are selected for comparison. The resolution of the CiC measurement is increased by increasing the number of grids on each side of the simulation box, hence increasing the number of cells.

We divide the cubic simulation box into a cubic grid with equal cells. Every cubic grid's center is the center of one cell. The total number of cells is given by the cubic power of the number of divisions on one edge of the simulation box. In every iteration, the square of the distance between a halo center and a cell center is computed. If the square of the ratio between the distance and the cell radius, $r$, is smaller than 1, the halo is considered inside the cell. In practice, infinitely increasing the number of cells does not necessarily yield lower measurement errors due to the limited precision of the halo positions. Because the halo position catalogs from DEUS store single-precision floating point numbers for the spatial coordinates of halos, the coordinates of halo centers have only 6 or 7 significant figures. When the ratio $r$ is in close proximity to 1, whether the halo is in the cell is uncertain. To reduce uncertain counts, double-precision floating numbers are used for variables and operations. In our measurement, eps (epsilon), the smallest positive number that, added to 1.0, is not equal to 1.0 within machine precision, is found to be $1.19209\times10^{-7}$; negeps (negative epsilon), the smallest positive number that, subtracted from 1.0, is not equal to 1.0, is found to be $5.96046\times10^{-8}$ for our algorithm \citep{press02}. When $r$ falls into the range (1-negeps, 1+eps), an uncertain count is recorded. After determining how many halo centers fall in each cell, the number of occurrences of each number count, starting from 0, is counted, which gives a discrete number distribution, which is then normalized to give a count probability distribution known as the CiC distribution. The number of occurrences of each number count is first computed by considering all uncertain cases as ‘not-inside-the-cells’, then is computed again by considering all uncertain cases as ‘inside-the-cells’. By comparing the two, the uncertainties of the CiC distribution due to the halo catalog’s limited precision level are calculated for every number count. When an uncertain count is considered ‘inside’ instead of ‘not-inside’, the occurrence of a particular count $N$ will decrease by 1, whereas the occurrence of the count $N+1$ will increase by 1. The lower limit to a count $N$ corresponds to the case that uncertain cases increase the occurrence of $N+1$ at the expense of the occurrence of $N$ and no uncertain cases increase the occurrence of N at the expense of the occurrence of $N-1$. The upper limit to a count $N$ corresponds to the case that uncertain cases increase the occurrence of $N$ at the expense of the occurrence of $N-1$ and no uncertain cases increase the occurrence of $N+1$ at the expense of the occurrence of $N$.

\begin{figure*}
\plottwo{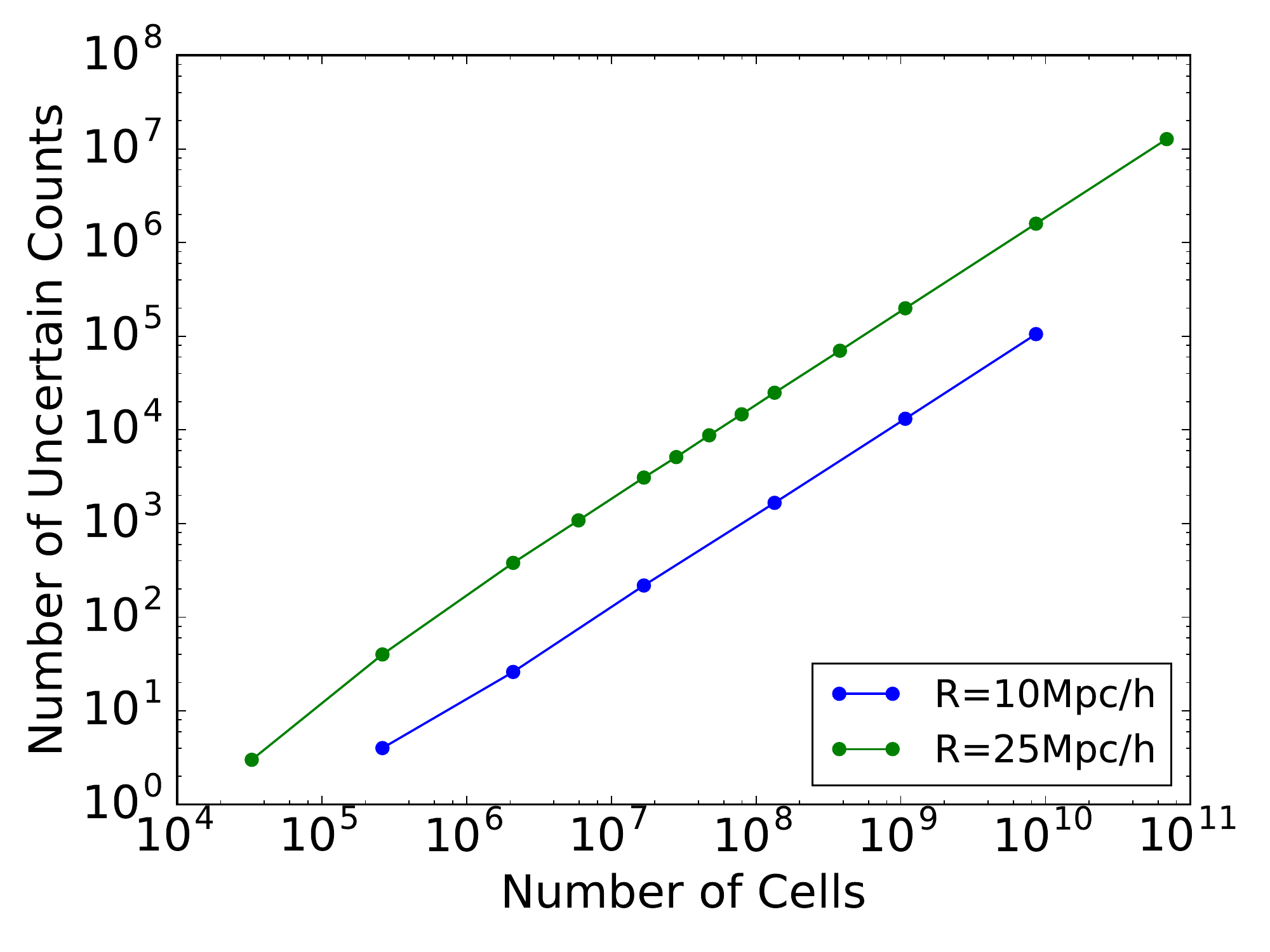}{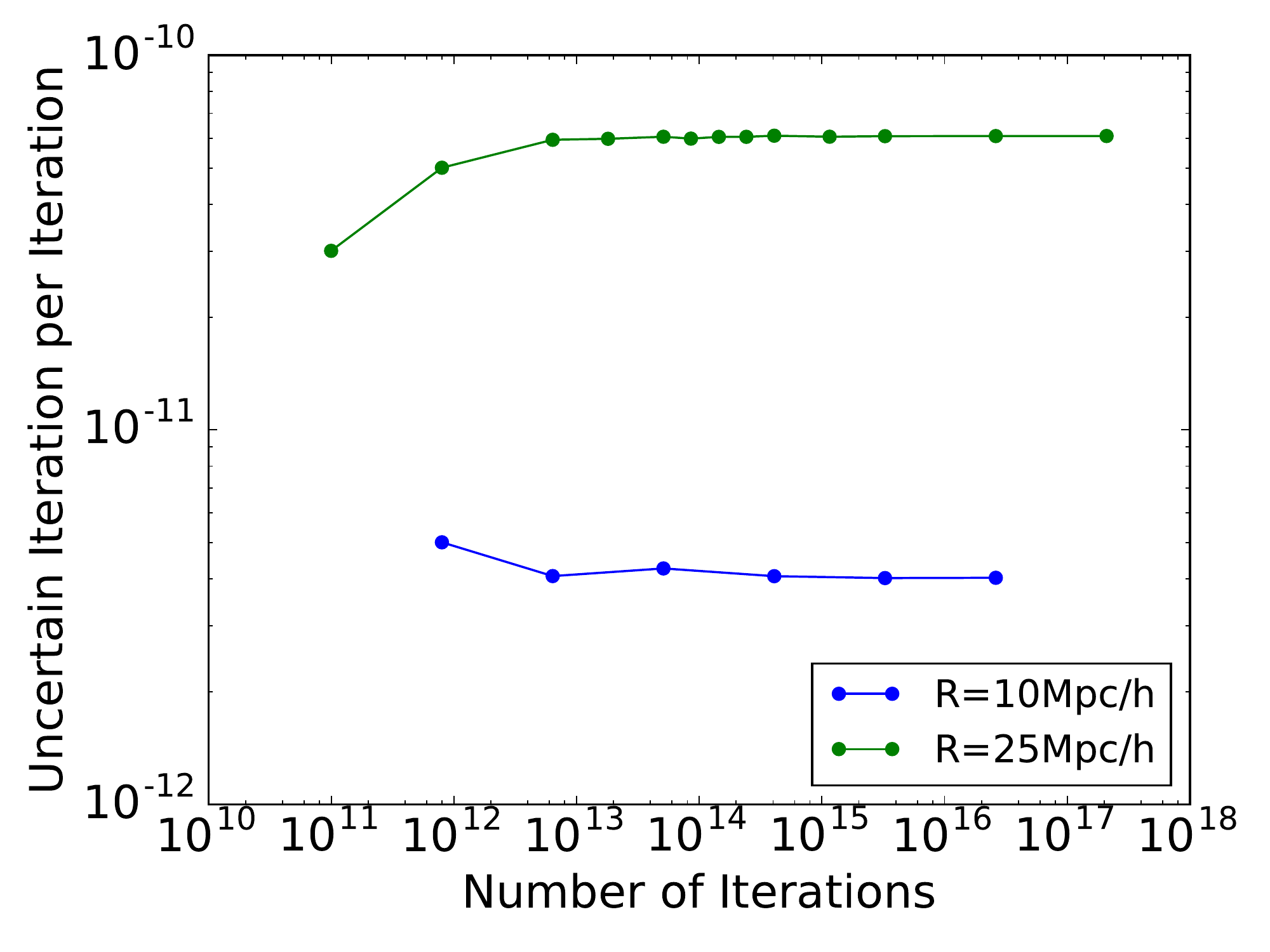}
\caption{On the left: Number of uncertain cases as a function of number cells. On the right: Number of uncertain iterations as a function of total number of iterations.}
\label{fig:uncertain}
\end{figure*}

\begin{figure*}
\plottwo{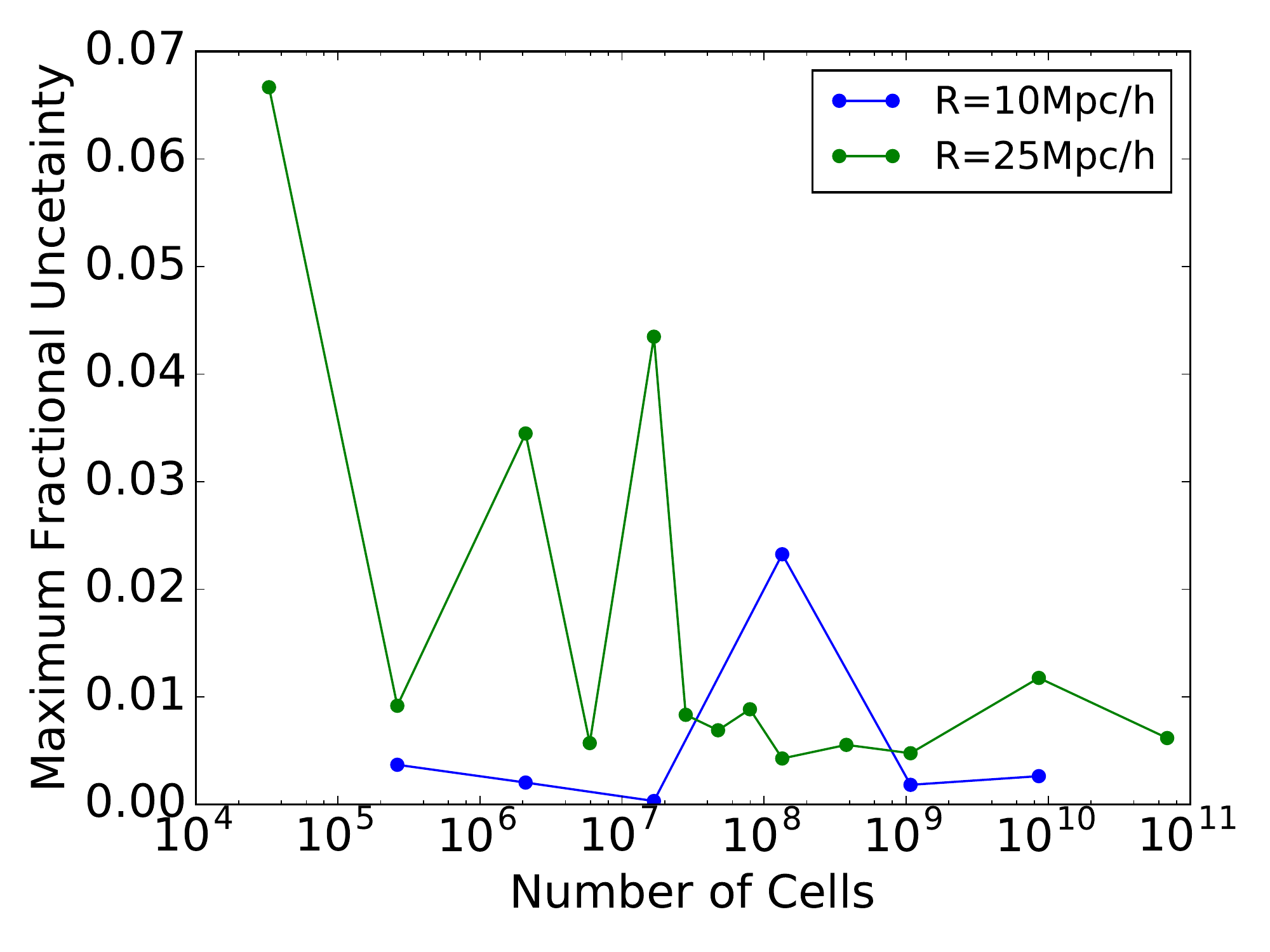}{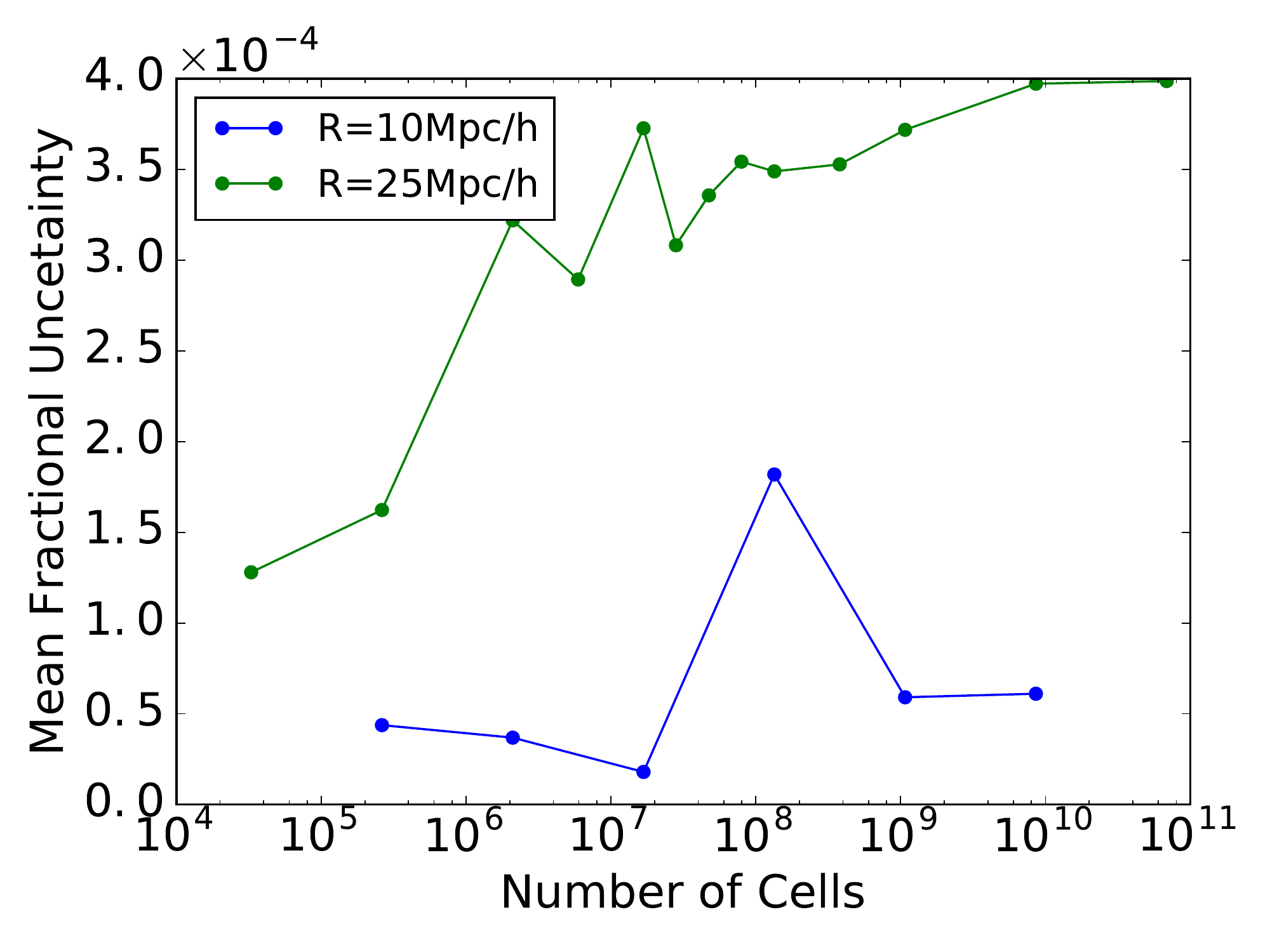}
\caption{On the left: Maximum fractional uncertainty in halo count as a function of total number of cells. On the right: Mean fractional uncertainty as a function of total number of cells.}
\label{fig:max_uncertain}
\end{figure*}

Figure \ref{fig:uncertain} shows that the number of uncertain cases increases linearly with the number of cells for both cell sizes. The total number of iterations is given by the number of cells multiplied by the number of halos, $N_{halo}=3045305$. The fraction of uncertain cases in all the iterations converges to $\sim4\times10^{-12}$ for cell radius 10 $h^{-1}$Mpc and $\sim6\times10^{-11}$ for cell radius 25 $h^{-1}$Mpc after the number of cells reaches $128^{3}$. The fraction of uncertain cases for cell radius 25 $h^{-1}$Mpc is about 15 times that for cell radius 10 $h^{-1}$Mpc, which is close to the expected ratio of $(25/10)^{3}=15.625$. For cell radius of 10 $h^{-1}$Mpc the occurrence count $N$ ranges from 0 to 273 and for cell radius of 25 $h^{-1}$Mpc the occurrence count $N$ ranges from 0 to 1602. For each count $N$, the difference between the upper limit and the lower limit of its number of occurrences, divided by $N$, gives a fractional uncertainty of that count. Figure \ref{fig:max_uncertain} (left) shows how the uncertain cases affect the maximum fractional uncertainty differently as a function of the number of cells. It shows that the largest fractional uncertainty is below 7\% for a cell radius of 25 $h^{-1}$Mpc and below 2.5\% for a cell radius of 10 $h^{-1}$Mpc. Figure \ref{fig:max_uncertain} (right) shows the sum of fractional uncertainty averaged over all $N$ as a function of the number of cells. For a cell radius of 25 $h^{-1}$Mpc, the mean fractional uncertainty converges to 0.04\%. For a cell radius of 10 $h^{-1}$Mpc, the mean fractional uncertainty does not show a clear trend of convergence within the range of resolution tested. The maximum fractional uncertainty and the mean fractional uncertainty for a cell radius of 10 $h^{-1}$Mpc show possible convergences after the number of cells reaches $1024^3$. The maximum fractional uncertainty and the mean fractional uncertainty for a cell radius of 25 $h^{-1}$Mpc show that fluctuations become small after the number of cells reaches $512^3$. 

\begin{figure*}
\plottwo{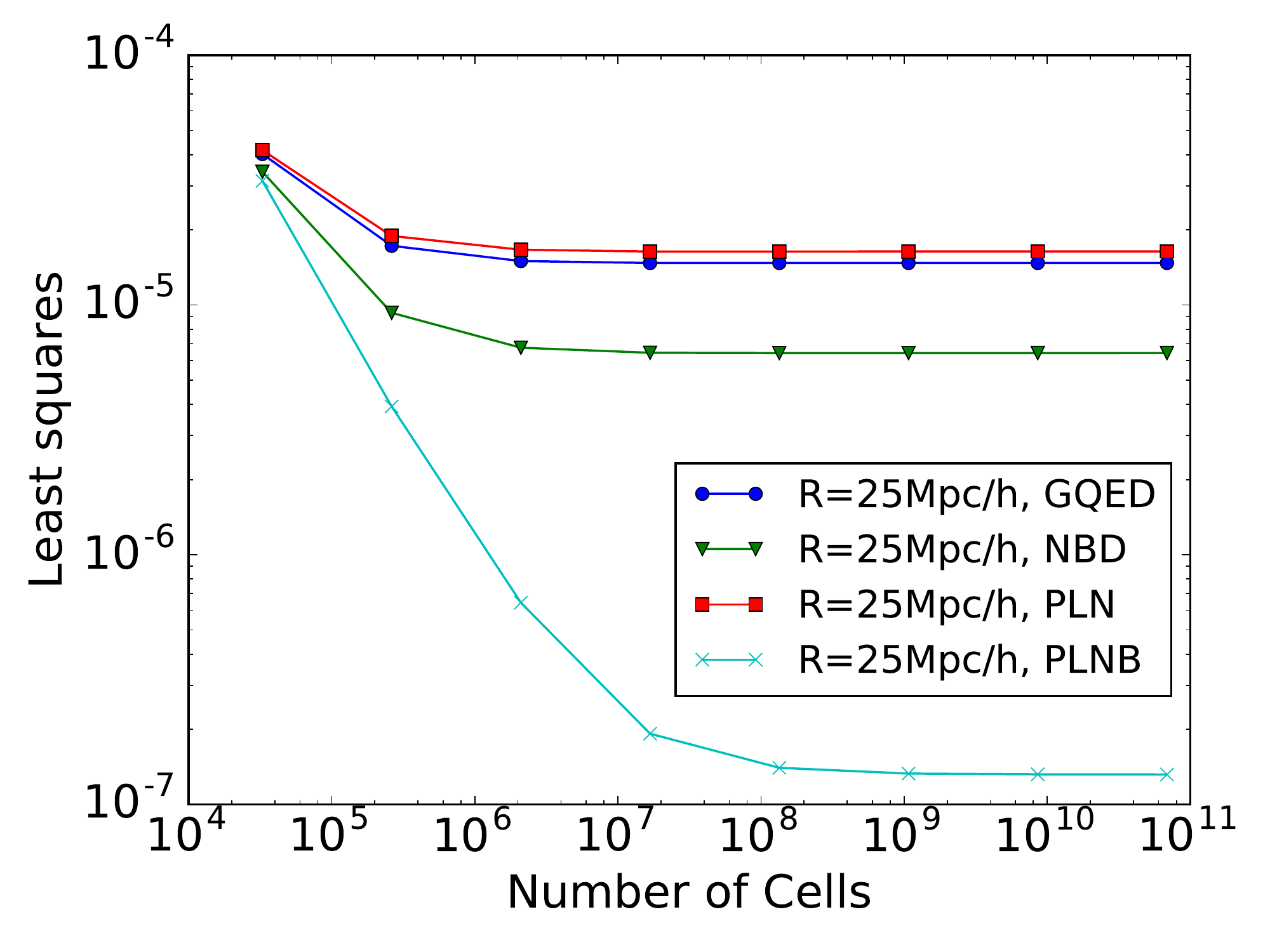}{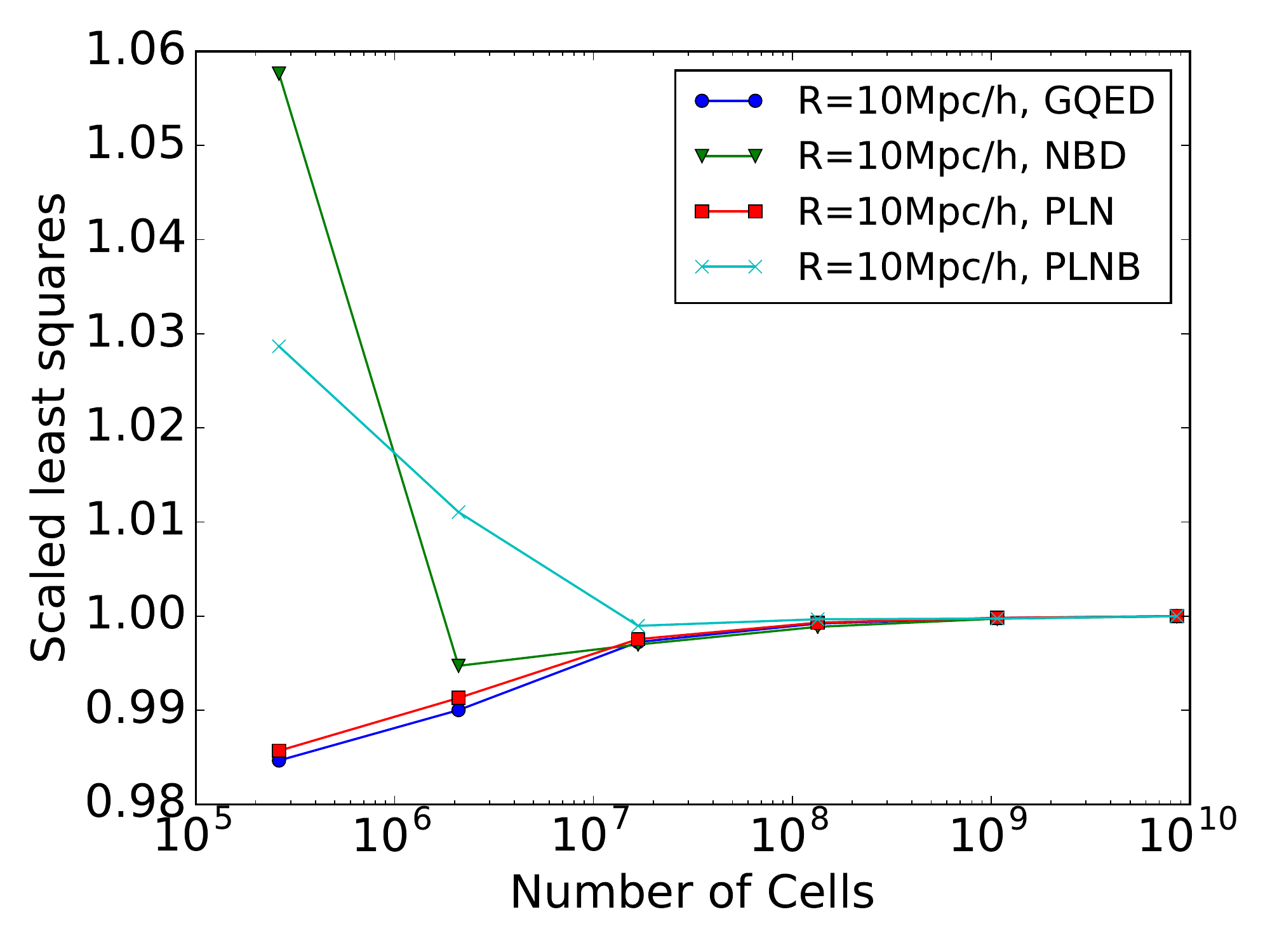}
\caption{On the left: Minimized least squares by fitting four distribution functions to counts-in-cells distribution as a function of total number of cells for a cell radius of $25h^{-1}$Mpc. On the right: Minimized least squares by fitting four distribution functions to counts-in-cells distribution as a function of total number of cells for a cell radius of $10h^{-1}$Mpc, scaled to PLNB with $2048^3$ cells.}
\label{fig:mls}
\end{figure*}

\begin{figure*}
\plottwo{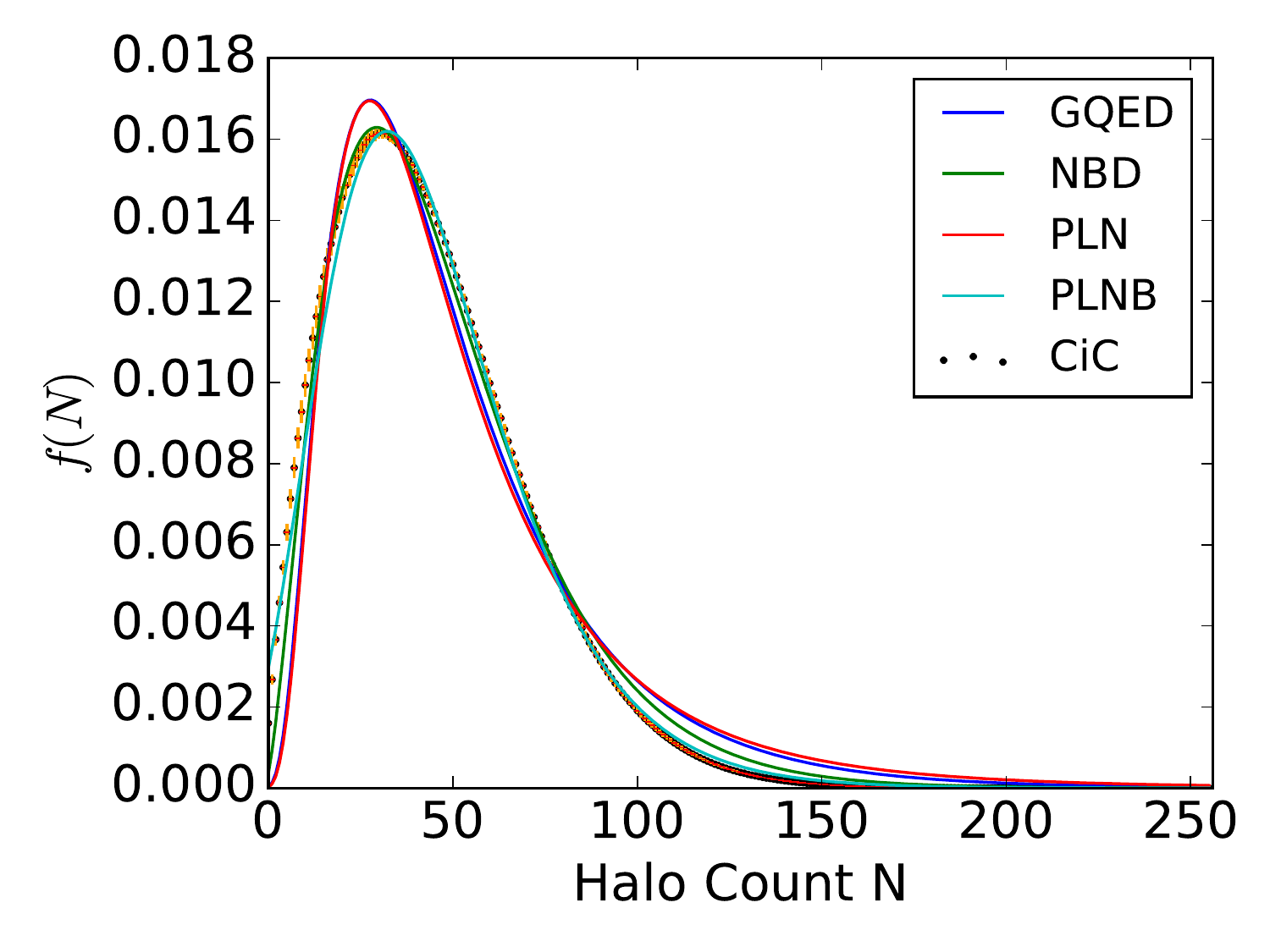}{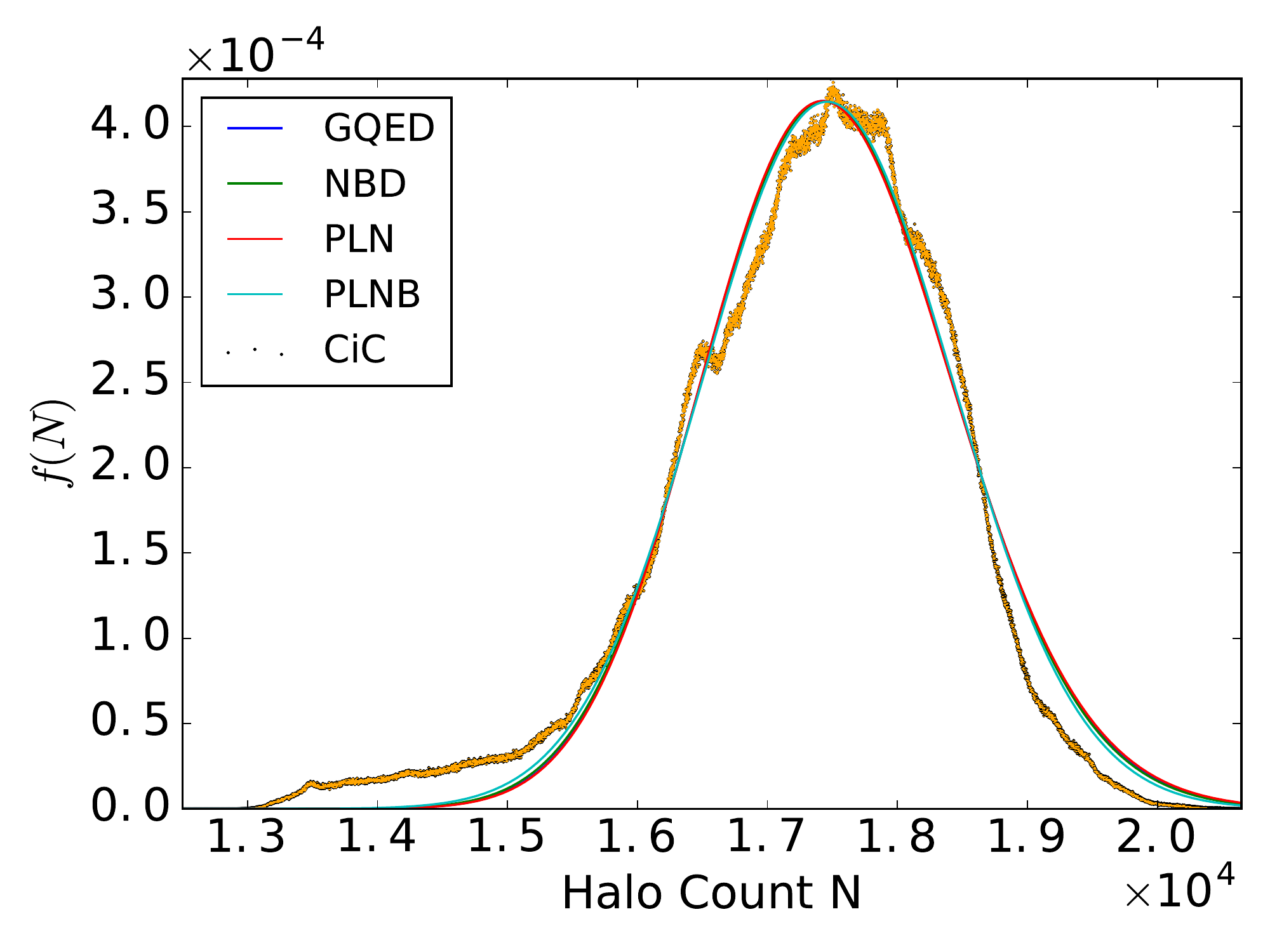}
\caption{The counts-in-cells distributions of halos $M>2.4\times10^{11} M_{\odot}$ for cell radii of $R=10h^{-1}$Mpc in $\Lambda CDM$ ({\it left}) and $R=25h^{-1}$Mpc in $w$CDM ({\it right}) at $z=4$, fitted with four models. The orange error bars on the left are jackknife errors obtained with 1/8 deleted fraction (see Section \ref{jackknife}). The orange error bars on the right are uncertainties due to uncertain counts near the boundaries of cells (see text of this Appendix).}
\label{fig:example}
\end{figure*}

Four analytical models are used to fit the CiC distribution. Figure \ref{fig:mls} (left) shows the least squares of the best-fits of the four models to the CiC distributions of dark matter halos. We use the least squares as an indicator for the closeness of the fit. The quality of fit of all models clearly improves as the number of cells increases. After the number of cells reaches $512^3$, the least squares of all models converges. Figure \ref{fig:mls} (right) shows the least squares for the best-fit models normalized by that of the highest resolution for each model, scaled least squares, as a function of the number of cells. The scaled least squares make comparison between least squares that differ by orders of magnitude easy to visualize. The quality of fits clearly converges after the number of cells reaches $512^3$ and no longer improves significantly with more cells. The convergence of the least squares for a cell radius of 25 $h^{-1}$Mpc and the scaled least squares for a cell radius of 10 $h^{-1}$Mpc suggest that for the purpose of measuring and fitting the CiC distribution, $512^3$ cells are sufficient to obtain the best-fit models given the precision level of the halo catalog.

\end{document}